\documentclass[11pt,a4paper]{article}
\pdfoutput=1
\usepackage{jheppub}
\usepackage{amsthm}
\usepackage{physics}
\usepackage{xcolor}
\usepackage{mathtools}
\usepackage{tensor}
\usepackage{empheq}

\usepackage{tikz}
\usepackage{graphicx}
\usepackage{subfigure}
\usetikzlibrary{calc,fadings,decorations.pathreplacing}
\usepackage{verbatim}

\usepackage{ulem} % defines \uline and replaces \emphasis

\usepackage{url}
\DeclareMathOperator{\MyProd}{\scalebox{1.4}{$\mathrm{I\kern-0.2ex I}$}}
\newcommand\fft[2]{\frac{#1}{#2}}

\newcommand\nn{\nonumber}
\DeclareMathOperator{\Li}{Li}
\newcommand\pgfmathsinandcos[3]{%
  \pgfmathsetmacro#1{sin(#3)}%
  \pgfmathsetmacro#2{cos(#3)}%
}
\newcommand\LongitudePlane[3][current plane]{%
  \pgfmathsinandcos\sinEl\cosEl{#2} % elevation
  \pgfmathsinandcos\sint\cost{#3} % azimuth
  \tikzset{#1/.style={cm={\cost,\sint*\sinEl,0,\cosEl,(0,0)}}}
}
\newcommand\LatitudePlane[3][current plane]{%
  \pgfmathsinandcos\sinEl\cosEl{#2} % elevation
  \pgfmathsinandcos\sint\cost{#3} % latitude
  \pgfmathsetmacro\yshift{\cosEl*\sint}
  \tikzset{#1/.style={cm={\cost,0,0,\cost*\sinEl,(0,\yshift)}}} %
}

\newcommand\DrawLatitudeCircle[2][1]{
  \LatitudePlane{\angEl}{#2}
  \tikzset{current plane/.prefix style={scale=#1}}
  \pgfmathsetmacro\sinVis{sin(#2)/cos(#2)*sin(\angEl)/cos(\angEl)}
  \pgfmathsetmacro\angVis{asin(min(1,max(\sinVis,-1)))}
  \draw[current plane] (\angVis:1) arc (\angVis:-\angVis-180:1);
  \draw[current plane,dashed] (180-\angVis:1) arc (180-\angVis:\angVis:1);
}

\tikzset{%
  >=latex, % option for nice arrows
  inner sep=0pt,%
  outer sep=2pt,%
  mark coordinate/.style={inner sep=0pt,outer sep=0pt,minimum size=3pt,
    fill=black,circle}%
}

\preprint{LCTP-20-16}

\title{Sub-leading Structures in Superconformal Indices:
Subdominant Saddles and Logarithmic Contributions\\
}

\author[a,b,c]{\small{Alfredo Gonz\'alez Lezcano}}

\emailAdd{agonzale@ictp.it}

\affiliation[a]{The Abdus Salam International Centre for Theoretical Physics, 34014 Trieste, Italy}

\affiliation[b]{SISSA International School for Advanced Studies\\
Via Bonomea 265, 34136 Trieste\\
\hspace*{15mm}  and\\
INFN, sezione di Trieste}
\affiliation[c]{Departamento de F\'isica, Universidad de Pinar del R\'io
Avenida Jos\'e Mart\'i No. 270,\\ CP 20100, Pinar del R\'io, Cuba }

\author[d]{\small{Junho Hong}}
\emailAdd{junhoh@umich.edu}

\author[d]{\small{James T. Liu}}

\emailAdd{jimliu@umich.edu}

\affiliation[d]{Leinweber Center for Theoretical Physics, University of Michigan, Ann Arbor, MI 48109, U.S.A.}

\author[a, d]{\small{Leopoldo A. Pando Zayas}}

\emailAdd{lpandoz@umich.edu}

\abstract{We systematically study various sub-leading structures in the superconformal index of ${\cal N}=4$ supersymmetric Yang-Mills theory with SU($N$) gauge group. We  concentrate in the superconformal index  description as a matrix model of elliptic gamma functions and in  the Bethe-Ansatz presentation. Our saddle-point approximation goes beyond the Cardy-like limit and we uncover various saddles governed by a matrix model corresponding to SU($N$) Chern-Simons theory. The dominant saddle, however, leads to perfect agreement with the Bethe-Ansatz approach. We also determine the logarithmic correction to the superconformal index to be   $\log N$, finding precise agreement between  the saddle-point and Bethe-Ansatz approaches in their respective approximations. We generalize the two approaches to cover a  large class of  4d ${\cal N}=1$  superconformal theories.  We find that also in this case both approximations agree all the way down to a universal contribution of the form  $\log N$.  The universality of this last result constitutes a robust signature of this ultraviolet description of asymptotically AdS$_5$ black holes and could be tested by low-energy IIB supergravity. }

\keywords{}

\arxivnumber{}

\begin{document}

\maketitle

\newpage
%%%%%%%%%%%%%%%%%%%%%%%%%%%%
\section{Introduction and Summary}

One of the most remarkable results in the context of the AdS/CFT correspondence has been the microscopic explanation of the entropy of electrically charged, rotating black holes in AdS$_5$ using the superconformal index (SCI) of ${\cal N}=4$ SYM theory. Three recent works have provided microscopic foundations for the black hole entropy using the dual supersymmetric field theory 
 \cite{Cabo-Bizet:2018ehj, Choi:2018hmj, Benini:2018ywd}. The answer was obtained more or less simultaneously by three groups using slightly different starting points. Initially, there were  three approaches  to the question of AdS$_5$ black hole entropy:  (i) The collaboration in  \cite{Cabo-Bizet:2018ehj} exploited a supersymmetric localization argument; (ii) The work \cite{Choi:2018hmj} started from the physical partition function at weak coupling; (iii) The authors of \cite{Benini:2018ywd} started from a Bethe-Ansatz (BA) presentation of the SCI.   Soon after these original works, it became evident that  these groups basically proposed different approaches to the SCI whose leading term was first conjectured in \cite{Hosseini:2017mds}. It is worth noting that (i) and (ii) relied on a Cardy-like limit while (iii) did not require such restrictions. Naturally, the ideas put forward in those works have inspired similar computations that  have been carried out in various field theories with the resounding outcome of providing microscopic foundations for the entropy of rotating, electrically charged, asymptotically AdS black holes  in various dimensions including AdS$_4$, AdS$_6$ and AdS$_7$, see for example,  \cite{Choi:2019miv,Choi:2019zpz, Kantor:2019lfo, Nahmgoong:2019hko, Nian:2019pxj,Bobev:2019zmz,Benini:2019dyp,Crichigno:2020ouj}; this collective body of work reinforces the original intuition.

One question that  follows from this embarrassment of richness  is to determine the precise relation between the different approaches. This situation motivates us to embark on a systematic study of those presentations at leading and sub-leading orders. We demonstrate explicitly that the two main presentations are different approximations schemes to the index which result, nevertheless, in the same answer including sub-leading terms all the way down to a universal logarithmic correction. This process helps us clarify a number of central elements and provides a glimpse into an effective matrix model theory governed by SU($N$) Chern-Simons theory. 

Let us briefly describe some of our main results.  Recall that the index counts (with sign) $\frac{1}{16}$-BPS states and depends on the fugacity $\tau$ and chemical potentials $\Delta_a$. When written as a matrix model, we discuss the saddle-point approach to the SCI. In this approach we consider a Cardy-like limit but extend it to include all terms up to exponentially suppressed ones, ${\cal O}(e^{-1/|\tau|})$. We compute leading and sub-leading terms of the SCI explicitly in this Cardy-like expansion, based on the large-$N$ analysis of saddle points. Our main result is a computation of the SCI of $\mathcal N=4$ SYM that goes beyond the leading $\fft{1}{|\tau|^2}$ order in the Cardy-like limit and takes the following form  

\begin{empheq}[box=\fbox]{equation}
\begin{split}\label{Eq:ResultSP}
	\mathcal{I}(\tau;\Delta) & =\mathcal I(\tau;\Delta)\big|_\text{Main Saddle Point} + \text{(contribution from other saddles)}\\
	\log\mathcal I(\tau;\Delta)\big|_\text{Main Saddle Point} &=-\fft{\pi i(N^2-1)}{\tau^2}\prod_{a=1}^3\left(\{\Delta_a\}_\tau-\fft{1+\eta}{2}\right)+\log N+\mathcal O(e^{-1/|\tau|}).
\end{split}
\end{empheq}
The value of $\eta=\pm 1$ is determined by the domain of $\Delta_a$ in (\ref{eq:eta}). We are also able to compute explicitly the contribution from other saddles in Appendix \ref{App:C-center:saddle} and in Appendix \ref{App:saddle:sublead}, where the latter is in particular subdominant compared to (\ref{Eq:ResultSP}) by an $N^2$-leading order term independent of chemical potentials. We obtain an analogous expression for a wide class of ${\cal N}=1$ 4d SCFT's in section \ref{Sec:SaddleGen}.

The other approach to the index that we scrutinize in this manuscript is the BA  approach.  In this approximation the index is written as the sum of contributions from BA solutions and we focus on the contribution of the so-called basic solution.  This solution to the eigenvalues first appeared in the high temperature limit of the topologically twisted index of ${\cal N}=4$ SYM on $T^2\times S^2$  \cite{Hosseini:2016cyf}. It was later  shown in \cite{Hong:2018viz} that it provides an exact solution for the topologically twisted index without the need of the high temperature approximation. This solution was also used by Benini and Milan in their discussion of the SCI \cite{Benini:2018ywd} and it was further  extended in \cite{Lezcano:2019pae,Lanir:2019abx} where it was shown that it furnishes a solution for a generic class of ${\cal N}=1$ supersymmetric field theories. More recently, the BA approach to the index based on the basic solution  was extended to include  two different angular velocities \cite{Benini:2020gjh}, thus covering the  most general type of asymptotically   AdS$_5\times S^5$ black holes.  Our results for the BA approach goes beyond the leading $N^2$ order in the large-$N$ limit and takes the form 

\begin{empheq}[box=\fbox]{equation}
\begin{split}\label{Eq:ResultBA}
	\mathcal I(\tau;\Delta)& =\mathcal I(\tau;\Delta)\big|_\text{Basic BA} + \text{(contribution from other BA solutions)}\\
		\log\mathcal I(\tau;\Delta)\big|_\text{Basic BA}&=-\fft{\pi i(N^2-1)}{\tau^2}\prod_{a=1}^3\left(\{\Delta_a\}_\tau-\fft{1+\eta}{2}\right)+\log N+\mathcal O(N^0),
\end{split}
\end{empheq}
where the last term $\mathcal O(N^0)$ can be replaced with $\mathcal O(e^{-1/|\tau|})$ in the Cardy-like limit. We are also able to compute explicitly the contribution from other BA solutions in Appendix \ref{App:C-center:BA}. We obtain an analogous expression for a wide class of ${\cal N}=1$ 4d SCFT's in section \ref{Sec:BAGen}. 

There are two important lessons that we provide:

\begin{itemize}
\item The expressions (\ref{Eq:ResultSP}) and (\ref{Eq:ResultBA}) explicitly demonstrate that both approximations yield the same contribution to the SCI up to exponentially suppressed terms of the form $e^{-1/|\tau|}$ in the Cardy-like limit, filling a gap in the literature regarding sub-leading corrections, namely $o(|\tau|^{-2})$ in (\ref{Eq:ResultSP}) and $o(N^2)$ in (\ref{Eq:ResultBA}). 

\item One of our main results is finding the logarithmic corrections which required control beyond the Cardy-like limit. In both approximations we find the same term, $\log N$.  The $\log N$ terms constitute, as remarked by Ashoke Sen \cite{Sen:2011ba}, a litmus test for any theory aspiring to be the ultraviolet complete description of gravity and such term should match the corresponding supergravity one-loop computation, presenting a unique UV/IR connection. The robustness of this term in the two approaches to the index is an important UV signature that we derive. 

\end{itemize}
The rest of the manuscript is organized as follows.  We start with a brief review of the SCI in section \ref{Sec:IndexGen}. The ${\cal N}=4$ SCI and its large $N$ saddle point approximation  is presented in section \ref{Sec:Saddle}.  The results of the saddle-point approach to arbitrary 4d ${\cal N}=1$ SCI are discussed in section \ref{Sec:SaddleGen}. In section \ref{Sec:BA} we study the BA approach to the SCI; we extend this approach to arbitrary 4d ${\cal N}=1$ theories in section \ref{Sec:BAGen} finding perfect agreement with the results based on the saddle point approximation. We conclude in section \ref{Sec:Conclusions} where we discuss a number of open problems that naturally follow from the work presented here. Given the technical nature of our investigation we relegate a number of important tests and results to a series of appendices. In Appendix~\ref{App:elliptic:functions} we clarify our notation and the definitions of the functions used in the main body of the manuscript. Appendix~\ref{App:C-center} investigates contributions to the SCI from $C$-center saddle-points and BA-solutions, respectively. These $C$-center solutions describe particular eigenvalue configurations that can be dominant over  those studied in the main  sections \ref{Sec:Saddle} and \ref{Sec:BA} in certain domain of chemical potentials. Appendix~\ref{App:saddle} presents various intuition-building facts, including details of the matrix model solution and the nature of the sub-leading saddles. Part of Appendix~\ref{App:saddle} describes our numerical treatment of the full ${\cal N}=4$ SCI and the level of compatibility with the analytical results described using the Cardy-like expansion in the main body up to and including the sub-leading saddles. Appendix~\ref{App:CS:partition:function} reviews the partition function of SU($N$) Chern-Simons theory which is quite relevant to our computations.

%\end{document}
%%%%%%%%%%%%%%%%%%%%%%%%%%%%%%%%%%%%%%%%%%%%%%%
%%%%%
\section{The Superconformal Index}\label{Sec:IndexGen}
%%%%%%
The SCI counts (with sign) BPS states that can not combine to form long representations of the superconformal algebra. For $\mathcal{N} =1$ theories on $S^1 \times S^3$, the SCI was defined in \cite{Romelsberger2005, Kinney:2005ej} and takes the form: 
\begin{equation}
\mathcal{I}(p, q; v) = \text{Tr}_{\mathcal{H}(S^1\times S^3)}\left[\left(-1\right)^{F}e^{- \beta\{\mathcal{Q}, \mathcal{Q}^{\dagger}\}}v_a^{Q_a}p^{J_1 + \fft{r}{2}} q^{J_2 + \fft{r}{2}}\right], \label{Eq:TheSCI}
\end{equation}
where $Q_a$ are the charges of states that commute with the super charge $\mathcal{Q}$ and $r$ is the R-charge. The fugacities $p$ and  $q$ are associated to the two angular momenta $J_{1,2}$ of $S^3$. We have that, $\mathcal{I}(p, q; v)$ counts $\fft{1}{16}-$BPS states for $\mathcal{N} =4$ SYM theory and $\fft14-$BPS states for more generic $\mathcal{N}= 1$ SCFT's. It was shown in \cite{Romelsberger:2007ec} that the SCI \eqref{Eq:TheSCI} for $\mathcal{N}= 1$ Superconformal gauge theories can be expressed as a complex integral:
\begin{eqnarray}
    \mathcal{I}(p, q; v) & = & \int \left[\mathcal{D}U\right] \exp \left(\sum_{n}^{\infty}\sum_{\text{R}\in\mathfrak{R}}\frac{1}{n} f(p^n, q^n; v^n)\chi_{\text{R}}(U^n)\right), \label{Eq:TrU}
\end{eqnarray}
where $U$ is the holonomy of the gauge field around $S^1$, $[\mathcal{D}U]$ is the invariant group measure and $\text{R}$ runs over all the representations $\mathfrak{R}$ in which the matter fields of the theory transform. The character of such representation is denoted as $\chi_{\text{R}}(\cdot)$. Equation \eqref{Eq:TrU} becomes an integral over complex eigenvalues $z_i$ ($|z_i| =1$) upon diagonalization of the unitary matrix $U$. 
The function $f(p, q; v)$ has the interpretation of the single-letter index of the supersymmetric gauge theory.  Specifically, for chiral matter, the single letter index has the form  \cite{Romelsberger:2007ec}
\begin{equation}
    i_{\Phi} (p, q; v) = \fft{v - pq /v}{(1- q)(1-p)}, \label{I-chiral}
\end{equation} 
whereas the vector multiplet single letter index is given by:
\begin{equation}
    i_{\text{V}}(p, q) = 1 - \fft{1- pq}{(1-p)(1-q)}. \label{I-vector}
\end{equation}
Inserting \eqref{I-chiral} and \eqref{I-vector} in \eqref{Eq:TrU} will generate terms of the form:
\begin{equation}
\begin{split}
    \exp \left(\sum_{n=1}^{\infty}\fft1n i_{\Phi}(p^n, q^n; v^n)\right)& = \prod_{j,k \geq 0}\fft{1 - v^{-1}p^{j+1}q^{k+1}}{1- vp^j q^k} = \Gamma \left(v; p , q\right) \\ 
    \exp \left[\sum_{n= 1}^{\infty}\fft1n i_{\text{V}}(p^n, q^n)\left(z^{n} + z^{-n}\right)\right] & = \fft{1}{(1- z)(1 -z^{-1})}\fft{1}{\Gamma(z;p ,q)\Gamma(z^{-1}; p, q)} \\
    \exp \left(\sum_{n= 1}^{\infty}\fft1n i_{\text{V}}(p^n, q^n)\right) & = \left(p; p\right)_{\infty}\left(q; q\right)_{\infty}, \label{Eq:Prod}
\end{split}
\end{equation} 
which hold in the domain where $|p|, |q|<1$ for $p, q \in \mathbb{R}$. The function $\Gamma(x; y, z)$ is the elliptic gamma function and $\left(q;q\right)_{\infty}$ is the $q$-Pochhammer symbol, both of which we define in appendix \ref{App:elliptic:functions} together with some  of the properties that will be useful for our study. One main ingredient in the successful account for the asymptotic growth of the SCI in the large-$N$ limit has been to allow the chemical potentials to be complex \cite{Choi:2018hmj, Cabo-Bizet:2018ehj, Benini:2018ywd}, therefore, we consider the analytic continuation of \eqref{Eq:indexToric} below  in the fugacities $p, q, v$. \par 
Consider  now a generic $\mathcal{N}=1$ theory with semi-simple gauge group $G$ with rank $\text{rk}(G)$, flavor symmetry $G_F$ and non-anomalous $U\left(1\right)_R$ R-symmetry. The matter content of this theory is taken to be $n_{\chi}$ chiral multiplets $\Phi_a$ in representations $\mathfrak{R}_a$ of $G$ having weight $\rho_a$, with flavor weights $\omega_a$ in some representation $\mathfrak{R}_F$ of $G_F$ and superconformal R-charge $r_a$. 

Using \eqref{Eq:Prod} the complex integral for the SCI \eqref{Eq:TrU} can be written as \cite{Romelsberger:2007ec,Dolan:2008qi} %Assel:2014paa}:
\begin{eqnarray}
    \mathcal{I} \left(p , q ; v\right) & = & \bar{\kappa}_G\oint_{\mathbb{T}^{\text{rk}\left(G\right)}}\frac{\prod_{a=1}^{n_{\chi}}\prod_{\rho_a \in \mathfrak{R}_a}\Gamma \big(\left(p q\right)^{r_a/2}z^{\rho_a}v^{\omega_a};p,q\big)}{\prod_{\alpha \in D}\Gamma \left(z^{\alpha}; p, q\right)}\prod_{i =1}^{\text{rk}\left(G\right)}\frac{d z_i}{2 \pi i z_i}, \label{Eq:indexToric} \\ \nonumber
    \bar{\kappa}_G & \equiv& \frac{\left(p ; p \right)_{\infty}^{\text{rk}\left(G\right)}\left(q ; q \right)_{\infty}^{\text{rk}\left(G\right)}}{|\mathcal{W}_G|}.
\end{eqnarray}
In \eqref{Eq:indexToric} we have adopted the notation of \cite{Benini:2018mlo} in which $z^{\rho_a} =\prod_{i=1}^{\text{rk}(G)}z_i^{\rho^i_a}$ and $v^{\omega_a} = \prod_{l =1}^{\text{rk}(G_F
)}v_{a}^{\omega_{l}}$. With $D$ we denote the set of all simple roots of the Lie algebra of $G$. The integration contour is the product of $\text{rk}(G)$ unit circles $|z_i| =1, \hspace{1.5mm} i = 1, \cdots \text{rk}(G)$.  The order of the Weyl group is denoted as $|W_G|$. To evaluate \eqref{Eq:indexToric} we can follow different paths which can be divided into two classes: 
\begin{itemize}
\item[\textit{a})] The saddle point method can be used to approximate $\mathcal{I}(p=q; v)$ provided we have a large control parameter. This is, in fact, the method pursued in various works \cite{Choi:2018hmj, Cabo-Bizet:2019osg, Honda:2019cio, ArabiArdehali:2019tdm, Kim:2019yrz, Amariti:2019mgp}, where the evaluation was performed in the Cardy-like limit $|q|\rightarrow 1$. A different  version of the saddle-point approach was applied in references \cite{ Cabo-Bizet:2019eaf, Cabo-Bizet:2020nkr} where an Elliptic extension of the integrand in  \eqref{Eq:indexToric} was proposed as an alternative to the more common analytic extension.

\item[\textit{b})] One can evaluate the complex integral using the residue theorem and  exploiting the properties of the  pole structure of the integrand. This is the so-called BA approach which has been followed by Benini and Milan \cite{Benini:2018mlo, Benini:2018ywd}.
\end{itemize}

Approach $a)$ provides, by definition,  an approximate answer while approach $b)$ is designed to yield an exact evaluation of the integral \eqref{Eq:indexToric}. There is, however, a catch in using the BA approach. As we will review later, in section~\ref{ssec:BAder}, the BA approach reduces the problem of evaluating \eqref{Eq:indexToric} to the problem of finding \textit{all solutions} of the Bethe-Ansatz Equations (BAEs). For the important question of matching the black hole entropy it has been sufficient to utilize a particular set of solutions to the BAEs. It is precisely in this sense that not {\it all BA solutions} have been used to evaluate $\mathcal{I}(p,q;v)$ that we talk about a BA approximation.   \par 
For later convenience, we introduce the following quantities:
\begin{eqnarray}
    p =e^{2 \pi i  \tau}, \hspace{3mm} q =e^{2 \pi i \sigma}, \hspace{3mm} v_{a} = e^{2 \pi i \xi_{a}}, \hspace{3mm} z_i = e^{2 \pi i u_i}\label{fugacities}
\end{eqnarray}
and the R-charge chemical potential which is fixed by supersymmetry to:
\begin{eqnarray}
    \nu_R = \frac{1}{2}\left(\tau + \sigma\right). \label{nur}
\end{eqnarray}
In terms of these quantities we use a modified version of the elliptic gamma function $\widetilde{\Gamma}(u;\tau, \sigma)$ defined in appendix \ref{App:elliptic:functions}.
We can further define
\begin{equation}
    y_a \equiv e^{2 \pi i \Delta_a} \equiv v^{\omega_a}(pq)^{\fft{r_a}{2}}\quad\Rightarrow\quad\Delta_a = \xi_a +r_a\nu_R\label{DeltaA},
\end{equation}
which allows to write \eqref{Eq:indexToric} as 
\begin{eqnarray}
    \mathcal{I} \left(\tau , \sigma ; \Delta\right) & = & \bar{\kappa}_G\int_{\mathcal{C}}\frac{\prod_{a=1}^{n_{\chi}}\prod_{\rho_a \in \mathfrak{R}_a}\widetilde{\Gamma} \left(\rho^a(u)+\Delta_a;\tau,\sigma\right)}{\prod_{\alpha \in D}\widetilde{\Gamma} \left(\rho^{\alpha}(u); \tau, \sigma\right)}\prod_{i =1}^{\text{rk}\left(G\right)}d u_i, \label{Eq:indexuts} \\ \nonumber
\end{eqnarray}
where $\mathcal{C} = \bigcup_{i =1}^{\text{rk}(G)}(0,1]$ and we have defined $\rho^a(u)$ such that $z^{\rho_a} = e^{2 \pi i \rho^{a}(u)}$. We shall be interested only in the case of equal angular momenta $J_1 = J_2 =J$, thus we set $\sigma = \tau$, which yields:
\begin{eqnarray}
\mathcal{I} \left(\tau; \Delta\right) & = & \bar{\kappa}_G\int_{\mathcal{C}}\frac{\prod_{a=1}^{n_{\chi}}\prod_{\rho_a \in \mathfrak{R}_a}\widetilde{\Gamma} \left(\rho^a(u)+\Delta_a;\tau\right)}{\prod_{\alpha \in D}\widetilde{\Gamma} \left(\rho^{\alpha}(u); \tau\right)}\prod_{i =1}^{\text{rk}\left(G\right)}d u_i, \label{Eq:indexutau} \\ \nonumber
\end{eqnarray}
where we have replaced $\mathcal I(\tau,\tau;\Delta)$ and $\widetilde{\Gamma}(u; \tau,\tau)$ with $\mathcal I(\tau;\Delta)$ and $\widetilde{\Gamma}(u; \tau)$ respectively for notational convenience. Particularizing for the $\mathcal{N} =4$ SYM theory in which $\rho^{a}(u) =u_{ij} \equiv u_i - u_j$, equation \eqref{Eq:indexutau} takes the form:
\begin{eqnarray}
\mathcal{I} \left(\tau; \Delta\right) & = & \kappa_N\int_{\mathcal{C}}\prod_{\mu =1}^{N-1}d u_{\mu}\frac{\prod_{a=1}^{3}\prod_{i \neq j}\widetilde{\Gamma} \left(u_{ij}+\Delta_a;\tau\right)}{\prod_{i \neq j}\widetilde{\Gamma} \left(u_{ij}; \tau\right)}, \label{Eq:indexN4} \\ \nonumber
\kappa_{N} & = & \bar{\kappa}_{SU(N)}\prod_{a=1}^{3}\left(\widetilde{\Gamma}(\Delta_a; \tau)\right)^{N-1}.
\end{eqnarray}

%%%%%%%%%%%%%%%%%%%%%%%%%%%%%%%%%%%%%%%%%%
\subsection{The structure of poles in the Superconformal Index }
\begin{figure}[t]

\centering 
\begin{tikzpicture} % MERC
\def\C{-8}; %Complex origin

%% some definitions

\def\R{1.5} %  radius
\def\angEl{25} % elevation angle
\def\angAz{-100} % azimuth angle
\def\angPhiOne{-50} % longitude of point P
\def\angPhiTwo{-35} % longitude of point Q
\def\angBeta{33} % latitude of point P and Q

%% working planes

\pgfmathsetmacro\H{2*\R*cos(\angEl)} % distance to north pole
\LongitudePlane[xzplane]{\angEl}{\angAz}
\LongitudePlane[pzplane]{\angEl}{\angPhiOne}
\LongitudePlane[qzplane]{\angEl}{\angPhiTwo}
\LatitudePlane[equator]{\angEl}{0}

%% characteristic points

\coordinate (O) at (8,3);

\path[xzplane] (\R,0) coordinate (XE);
\path[pzplane] (\angBeta:\R) coordinate (P);
\path[pzplane] (\R,0) coordinate (PE);
\path[qzplane] (\angBeta:\R) coordinate (Q);
\path[qzplane] (\R,0) coordinate (QE);

\draw[dashed, thin] (-3.5, -\H)--(-3.5,4);
\draw[->,thick] (1.375*\C, 0) -- (.625*\C,0) node[above]{Re$(z)$};
\draw[->, thick](\C,-.85*\H) -- (\C, -.375*\C) node[right]{Im$(z)$};
\fill[gray, opacity=0.2] (\C,0) circle (1cm);
\draw[thick, blue ] (\C,0) circle (1cm);
\draw[->,thick, blue](\C,-.125*\C) arc (90:60:1cm);
\fill (\C,0) circle (1pt) ;

%\DrawLatitudeCircle[\R]{0} ; % equator

\node[blue, thick] at (0,0) {\tikz{\DrawLatitudeCircle[\R]{0}}};
\node at (0,2*\R) { \tikz{\DrawLatitudeCircle[\R]{0}} };

\node at (0,-\H) { \tikz{\DrawLatitudeCircle[\R]{0}} };
\fill (0,-.42*\R) circle (1pt)  node[above=10]{Re$(u)\in (0,1]$};

%% draw lines and put labels
\draw[ thick] (-1.1*\R,0) --(-1.1*\R,.5*\H) node[left]{Poles};
\draw[->,thick] (-1.1*\R,.5*\H) -- (-1.1*\R,\H);
\draw (-\R,-\H) -- (-\R,2*\R);
\draw (\R,-\H) --(\R,0)node[right=2]{Im$(u)=0$};
\draw (\R,0)-- (\R,2*\R) ;
\draw (XE) -- +(0,2*\R) ;
\draw[thin] (-\R,-\H)--(-\R,-1.001*\H)  node[right] {$\hspace{2mm}$Im$(u) \in \mathbb{R}$};

\path[pzplane] (0.5*\angBeta:\R);% node[right] {$\hat{1}$};
\path[qzplane] (0.5*\angBeta:\R) ;%node[right] {$\hat{2}$};
\draw[equator, blue,->] (-90:\R) arc (-90:-50:\R) ;
\end{tikzpicture}
%\end{subfigure}

\caption{This figure shows the two complex domains for the holonomies related through the map $z = e^{ 2 \pi i u}$. The $z$ plane is represented such that the unit circle over which the integration is originally performed is the boundary between the gray and white regions. The complex variable $u$ lives on a cylinder. The unit circle on the $z$ plane is mapped to the circle in the middle of the cylinder (both represented in blue) where Re$ (u) \in [0,1]$ and $0 \sim 1$.}
\label{lb}
\end{figure}
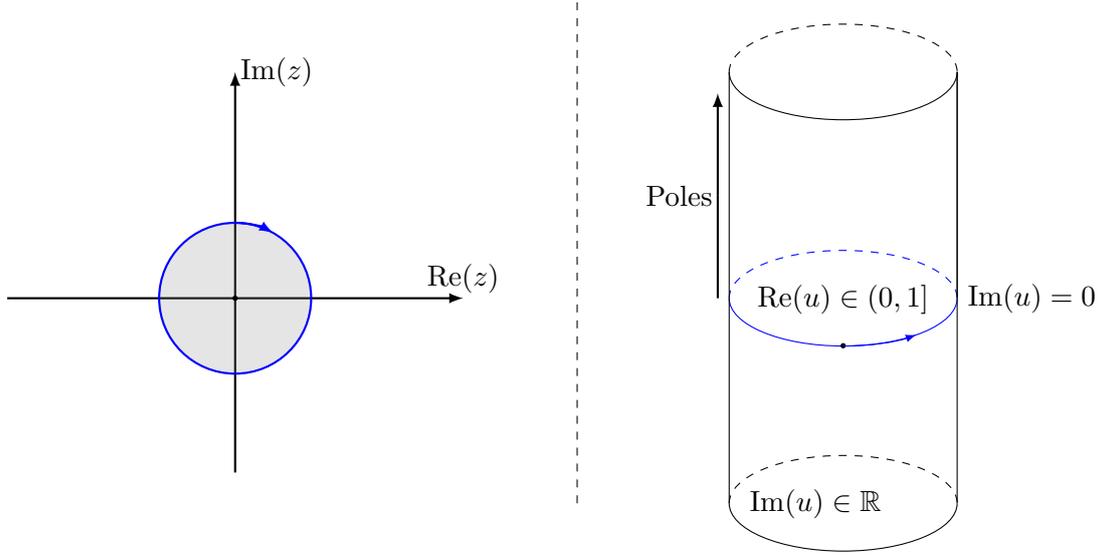

As emphasized already in \cite{Benini:2018mlo}, the only singularities of the integrand of \eqref{Eq:indexToric} come from the elliptic gamma functions associated to the chiral multiplets and in the $z_i$ variables take the form:
\begin{equation}
    z^{\rho_a} = v^{-\omega_a}q^{-r_a - k}, \hspace{2mm} k  \in \mathbb{Z}^{\geq 0}.
\end{equation}
The map $z = e^{ 2 \pi i u}$ preserves the singularity structure of the integrand in \eqref{Eq:indexutau}, therefore any deformation of the contour even in the variables $u$ has to keep track of possible poles being crossed while deforming the contour. 
In Figure~\ref{lb} we illustrate how the domains transform under $z=e^{2 \pi i u}$, where the periodicity of $u$ implies it takes values on a cylinder. With the $u$-variables is easier to visualize the location of the poles: for a fixed value of $\rho^a(u)$, the poles are separated from each other by $\tau$ translations on the surface of the $u-$cylinder as \begin{equation}
\rho^a(u) + \Delta_a + k \tau =0, \label{PolesT}
\end{equation}
which can be read from the integrand of \eqref{Eq:indexutau}.

%%%%%%%%%%%%%%%%%%%%%%%%%%%%%%%%%%%%%%%%%%%%%%%%%%%%%%%%%%%%%%%%%%

%%%%%%%%%%%%%%%%%%%%%%%%%%%%%%%%%%%%%%%%%%%%%%%%%%%%%%%%%%%%%%%%%%%

We will study $\mathcal{N} =4$ SYM theory, thus we can write:
\begin{equation}
u_{ij} + \Delta_a + k \tau =0. \label{PolesTn4}
\end{equation} 
Note that, even when applying the saddle point method we will eventually have to deform the contours, therefore we want to make sure not to cross non-trivial poles in this process. By non-trivial pole we mean a point $P=\{u_1, \cdots, u_N\} \in \mathbb{C}^N$ whose residue contribution to \eqref{Eq:indexutau} is different from $0$. Given a point $P \in \mathbb{C}^N$, if there is at least one coordinate $u_i$ not satisfying \eqref{PolesTn4}, then the integral over that coordinate in \eqref{Eq:indexutau} vanishes. Thus we call non-trivial poles those satisfying that $\forall \hspace{1.5mm} i=1, \cdots, N$ there is at least one point of the form \eqref{PolesTn4} through which the holonomy $u_i$ passes.

%%%%%%%%%%%%%%%%%%%%%%%%%%%%%%%%%%%%%%%%%%%%%%%%%%%%%%%%%
\subsection{The Bethe-Ansatz formulation of the SCI}
\label{ssec:BAder}
Benini and Milan in \cite{Benini:2018ywd}, represented the SCI using a BA approach developed in \cite{Benini:2018mlo}. The conceptual basis for writing the SCI as a sum over solutions to BAEs were originally clarified in \cite{Closset:2017bse} based on interesting relations between observables on manifolds of different topologies developed in \cite{Closset:2017zgf}. For completeness, we present a heuristic derivation of the BAEs in which they arise as the outcome of properly organizing the residues contributing to \eqref{Eq:indexutau}. Let us define the integrand of \eqref{Eq:indexN4} such that we can write the integral as:
\begin{equation}
\begin{split}
\mathcal{I}(\tau; \Delta) & = \kappa_N \int_{\mathcal{C}} \prod_{\mu =1}^{N-1}du_{\mu} \mathcal{Z}(u; \Delta, \tau),\label{Eq:IndexZij}\\ 
\mathcal{Z}(u; \Delta, \tau) &\equiv \prod_{i \neq j}\mathcal{Z}_{ij}(u_{ij}; \Delta, \tau). \\
\end{split}
\end{equation}
Under shifting by $\tau$  the argument of $\mathcal{Z}(u; \Delta, \tau)$ we have the following property:
\begin{equation}
\begin{split}
\mathcal{Z}(u - \delta_{k} \tau; \Delta, \tau)& = Q_{k}(u; \Delta, \tau) \mathcal{Z}(u; \Delta, \tau)\label{Eq:shift}\\
\delta_k &\equiv \left(\delta_{kl}\right)_{l=1}^{N-1},
\end{split}
\end{equation}
where the function $Q_k(u; \Delta, \tau)$ measures the lack of periodicity of $\mathcal{Z}(u; \Delta, \tau)$ in the variable $u_k$ under shifting by $\tau$ and is defined in \cite{Benini:2018mlo} as:
\begin{equation}
Q_k(u; \Delta, \tau) = e^{2 \pi i \lambda} \prod_{l=1 ( \neq k)}^{N}\prod_{a =1}^3 \fft{ \theta_1(- u_{kl}+\Delta_a; \tau )}{\theta_1(u_{kl}+\Delta_a; \tau)}, \label{Eq:BAOperator}
\end{equation}
where $\lambda$ is a Lagrange multiplier implementing the SU($N$) constraint on the holonomies and $\theta_1(u;\tau)$ is the elliptic theta function defined in appendix \ref{App:elliptic:functions}. These functions are called BA operators and have the crucial property of being doubly periodic with periods $1$ and $\tau$ as proved in \cite{Benini:2018mlo}, namely
\begin{equation}
Q_k(u+n+m\tau; \Delta, \tau) =Q_k(u; \Delta, \tau). \label{Period}
\end{equation}

Another important property of the BA operator is that, wherever $\mathcal{Z}(u;\Delta, \tau)$ has a pole, it has a pole of higher order. Specifically, as demonstrated in \cite{Benini:2018mlo}, the points \eqref{PolesTn4} are such that $Q_{k}(u; \Delta, \tau)$ have stronger singularities than the integrand $\mathcal{Z}(u; \Delta,\tau)$.
Using the change of variable $u_k \rightarrow u_k +\tau$, a contour $\mathcal{C}^0_k = (0,1] $ transforms into $\mathcal{C}^1_k = (\tau, \tau+1]$. Defining $\mathcal{C}_{i_1 \cdots i_n}$ as
\begin{equation}
    \mathcal{C}_{i_1 \cdots i_n} =\left( \bigcup_{i=1}^{N-1-n}\mathcal{C}^0_i\right) \times \left(\bigcup_{j=1}^n\mathcal{C}^1_{i_j}\right), \label{Ciii}
\end{equation}
the following relation holds:
\begin{equation}
    \sum_{n=1}^{N-1}\fft{(-1)^n}{n!}\bigcup_{i_1 \neq \cdots \neq i_n}\mathcal C_{i_1, \cdots, i_n}= - \bigcup_{k =1}^{N-1}\mathcal{C}^1_{k}\equiv \mathcal{C}^1. \label{Eq:C1}
\end{equation}
Using \eqref{Eq:BAOperator} with the corresponding change of variables and \eqref{Eq:C1}, we can define the shifting operator:
\begin{equation}
    Q(u; \Delta,\tau) = \sum_{n=1}^{N-1}\fft{(-1)^n}{n!}Q_{i_1}(u; \Delta,\tau)\cdots Q_{i_n}(u; \Delta,\tau).  \label{Eq:Q}
\end{equation}
Therefore we can write:
\begin{equation}
\begin{split}
    \mathcal{Z}(u -  \tau; \Delta, \tau)& = Q(u; \Delta, \tau) \mathcal{Z}(u; \Delta, \tau)\label{Eq:Qshift}\\ 
    & \Downarrow \\ 
    \mathcal{Z}(u - m \tau; \Delta, \tau)& = (Q(u; \Delta, \tau))^m \mathcal{Z}(u; \Delta, \tau).
\end{split}
\end{equation}

%%%%%%%%%%%%%%%%%%%%%%%%%%%%%%%%%%%%%%%%%%%%%%%%%%%%%%%%%%%%%%%%%%%%%%%%%%%%%%%%FIGURE
\begin{figure}[t]

%\begin{subfigure}
  \centering 
\begin{tikzpicture} % MERC
\def\C{-8}; %Complex origin

%% some definitions

\def\R{1.9} % sphere radius
\def\angEl{25} % elevation angle
\def\angAz{-100} % azimuth angle
\def\angPhiOne{-50} % longitude of point P
\def\angPhiTwo{-35} % longitude of point Q
\def\angBeta{33} % latitude of point P and Q

%% working planes

\pgfmathsetmacro\H{2*\R*cos(\angEl)} 
\LongitudePlane[xzplane]{\angEl}{\angAz}
\LongitudePlane[pzplane]{\angEl}{\angPhiOne}
\LongitudePlane[qzplane]{\angEl}{\angPhiTwo}
\LatitudePlane[equator]{\angEl}{0}

%% characteristic points

\coordinate (O) at (8,3);

\path[xzplane] (\R,0) coordinate (XE);
\path[pzplane] (\angBeta:\R) coordinate (P);
\path[pzplane] (\R,0) coordinate (PE);
\path[qzplane] (\angBeta:\R) coordinate (Q);
\path[qzplane] (\R,0) coordinate (QE);

\DrawLatitudeCircle[\R]{0} ; % equator

\node at (0,2*\R) { \tikz{\DrawLatitudeCircle[\R]{0}} };
\node at (0,0.35*\H) { \tikz{\DrawLatitudeCircle[\R]{0}} };
\node at (0,0.7*\H) { \tikz{\DrawLatitudeCircle[\R]{0}} };
\node at (0,-\H) { \tikz{\DrawLatitudeCircle[\R]{0}} };
\fill (0,-.42*\R) circle (2pt) ;
\draw[->,thick, red] (0,-0.42*\R) --(0.42*\R, .26*\R) node[below=9]{$\tau$};
\fill[red] (.42*\R,.26*\R) circle (2pt) ;

%% draw lines and put labels
\draw[ thick] (-1.1*\R,0) --(-1.1*\R,.5*\H) node[left]{Poles};
\draw[->,thick] (-1.1*\R,.5*\H) -- (-1.1*\R,\H);
\draw (-\R,-\H) -- (-\R,2*\R) (\R,-\H) -- (\R,2*\R);
\draw (XE) -- +(0,2*\R) ;

\path[pzplane] (0.5*\angBeta:\R);
\path[qzplane] (0.5*\angBeta:\R) ;%node[right] {$\hat{2}$};
%*********************************************************************
\draw[equator,->] (-90:\R) arc (-90:-50:\R) ;
\draw[equator,->] (-90:-0.500*\R) arc (-90:-140:\R) node[above=2]{$\mathcal{C}^1$};
\draw[equator,->] (-90:-.5*\R) arc (-90:-50:\R)   node[above=2]{$-\mathcal{C}^1$};
\draw[equator,->] (-90:-2*\R) arc (-90:-50:\R)   node[above=2]{$-\mathcal{C}^2$};
\draw[equator,->] (-90:-2*\R) arc (-90:-140:\R)   node[above=2.5]{$\mathcal{C}^2$};
\draw[thick,dotted](0,1.3*\R)--(0,1.5*\R);
\draw[equator,->] (-90:-3.76*\R) arc (-90:-50:\R)   node[above=2]{$-\mathcal{C}^m$};
\draw[equator,->] (-90:-3.76*\R) arc (-90:-140:\R)   node[above=2.5]{$\mathcal{C}^m$};

\end{tikzpicture}
%\end{subfigure}

\caption{The figure shows the pairs of contours added and subtracted in order to obtain the final form of integration contour and the integrand for the SCI using the BA approach. The final integration contour is simply $\mathcal{C}\bigcup \mathcal{C}^1$.  }
\label{fig:ring}
\end{figure}
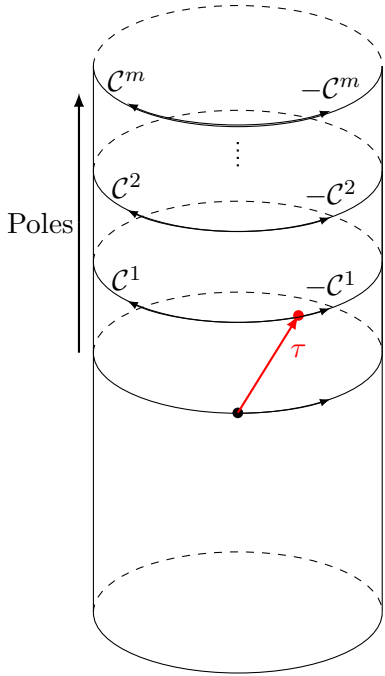

%%%%%%%%%%%%%%%%%%%%%%%%%%%%%%%%%%%%%%%%%%%%%%%%%%%%%%%%%%%%%%%%%%%%%%%%%%%%%%%%%%%%%
A way to systematically collect all non-trivial poles is to sum the contribution of poles located in strips slicing up the upper half cylinder in Figure~\ref{lb}. Using properties \eqref{Period} and \eqref{Eq:Qshift}, one can add and subtract infinitely many times the same integral taken over contours successively $\tau$-shifted as shown in Figure~\ref{fig:ring}. This yields 
\begin{equation}
\begin{split}
\mathcal{I}(\tau; \Delta)& = \kappa_N\int_{\mathcal{C}\bigcup\mathcal{C}^{1}}\prod_{\mu =1}^{N-1}du_{\mu} \sum_{m=0}^{\infty}(Q(u; \Delta, \tau))^m \mathcal{Z}(u; \Delta, \tau) \label{Eq:BAder} \\ 
& = \kappa_N\int_{\mathcal{C}\bigcup\mathcal{C}^{1}}\prod_{\mu =1}^{N-1}du_{\mu} \fft{1}{1-Q(u; \Delta, \tau)}\mathcal{Z}(u; \Delta, \tau)\\ 
& = \kappa_N\int_{\mathcal{C}\bigcup\mathcal{C}^{1}}\prod_{\mu =1}^{N-1}du_{\mu} \fft{\mathcal{Z}(u; \Delta, \tau)}{\prod_{k=1}^{N-1}\left(1-Q_k(u; \Delta, \tau)\right)}.
\end{split}
\end{equation}
 Since points of the form \eqref{PolesTn4} are stronger singularities for $Q_k(u; \Delta, \tau)$, then the only singularities contributing to $\mathcal{I}(\tau; \Delta)$ are those satisfying the BAEs which take the form:
\begin{equation}
\begin{split}
Q_k(\hat{u}; \Delta, \tau) =1, \hspace{2mm} \forall \hspace{1.5mm} k =1, \cdots, N \label{BAEs} 
\end{split}
\end{equation}
The values $\hat{u}$ satisfying \eqref{BAEs} are called BA solutions.
Upon direct application of the residue theorem, $\mathcal{I}(\tau, \Delta)$ can be rewritten in terms of a discrete sum as: 
\begin{equation}
\begin{split}
	\mathcal I(\tau; \Delta)&=\kappa_N\sum_{\hat u\in\mathrm{BA}}\mathcal Z(\hat u;\Delta,\tau)H(\hat u;\Delta,\tau)^{-1},\label{eq:index:BA}\\
H\left(\hat{u}; \Delta, \tau\right)& = \det \left[\frac{1}{2 \pi i} \frac{\partial \left(Q_1, \cdots, Q_N\right)}{\partial \left(u_1, \cdots , u_{N-1}, \lambda\right)}\right].
	\end{split}
\end{equation}
%

% Emphasize the approximation used; contour deformation, etc. 

% Applied to arbitrary 4d ${\cal N}=1$ SCFT \cite{Lezcano:2019pae} and \cite{Lanir:2019abx}. More recently extended to account for different angular velocities \cite{Benini:2020gjh}.

%\textbf{Working............................................................}\\ \par 

%%%%%%%%%%%%%%%%%%%%%%%%%%%%%%%%%%%%%
\section{Saddle-point approach to the SCI}\label{Sec:Saddle}

The classical gravity regime where the Bekenstein-Hawking entropy of the rotating, electrically charged, asymptotically AdS$_5$ black hole is known to correspond, on the field theory side, to the large-$N$ regime. This situation motivates the study of the SCI in the large-$N$ limit. Having an integral expression for the SCI of the form $\mathcal{I} \sim \int [d u]\exp (N^2 S_{\text{eff}}(u_{ij}))$ (see \eqref{Eq:TrU} and \eqref{Eq:indexutau}) makes it suitable for a saddle point evaluation. The pairwise nature of the full effective action, however,  prevents us from directly applying standard matrix models techniques. Recall that standard matrix model effective actions have a typically attractive potential depending only on the matrix eigenvalues, thus playing the role of an external source and a  Vandermonde-like term which is pairwise, specifically: $S_{\text{eff}}(u) = V_{\text{external}}(u_i)+W_{\text{pairs}}(u_{ij})$, such that the two terms $V_{\text{external}}$ and $W_{\text{pairs}}$ compete until the eigenvalues $u_i$ stabilize in the equilibrium configuration  \cite{Eynard:2015aea}. In contrast, for the SCI we have an effective action where $V_{\text{external}}$ is absent, thus it is purely pairwise interaction $W_{\text{pairs}}(u_{ij})$. This structure resembles the so-called frustrated systems appearing in condensed matter. For these systems the building blocks of the full interaction term compete among themselves yielding structurally rich set of  vacua and, consequently, a plethora of new phenomena \cite{2004fss..book.....D}. Precisely because such frustrated systems have several equilibrium configurations beyond the dominant one,  the application of saddle-point approaches becomes inefficient. Indeed, we found various such subdominant configurations when analyzing the SCI numerically in terms of elliptic gamma functions (see appendix \ref{App:saddle} for more details). It would be interesting to understand if there is a deeper and more explicit connection between the SCI and frustrated systems.

In \cite{ Cabo-Bizet:2019eaf, Cabo-Bizet:2020nkr}, the authors proposed to circumvent the difficulties of having only  pairwise interaction by introducing an elliptic  extension of the SCI. Such extensions exploit the central fact that $W_{\text{pairs}}$ have saddle points configurations consisting of eigenvalues $u_i$ uniformly distributed along the periodic directions of the interaction term. 

The Cardy-like limit has resolved the question of saddle-points by simplifying the analysis of the SCI to a limit where it is easy to find the dominant saddle-point configuration. In our systematic Cardy-like expansion, we effectively depart from the leading Cardy-like limit in a way that automatically eliminates the pairwise nature of the effective potential. In this process we uncover an interesting connection with an effective SU($N$) Chern-Simons theory on $S^3$.

With these ideas in mind, we proceed to compute the index using the conventional saddle-point approach. For simplicity, in section~\ref{Sec:Saddle:N=4} we start with $\mathcal N=4$ SYM and compute the corresponding index (\ref{Eq:indexN4}). Then we move on to a generic $\mathcal N=1$ SCFT and compute the corresponding index (\ref{Eq:indexutau}) in section~\ref{Sec:SaddleGen}.

%%%%%
\subsection{Saddle point approximation for \texorpdfstring{$\mathcal N=4$}{N=4} SYM}\label{Sec:Saddle:N=4}
%%%%%
To compute the integral in (\ref{Eq:indexN4}) using the conventional saddle point approach, we introduce an effective action $S_\text{eff}(\hat u;\Delta,\tau)$ as
\begin{equation}
\begin{split}
	N^2S_\text{eff}(\hat u;\Delta,\tau)&=\sum_{i\neq j}\left(\sum_{a=1}^3\log\widetilde\Gamma(u_{ij}+\Delta_a;\tau)+\log\theta_0(u_{ij};\tau)\right)\\
	&\quad+(N-1)\sum_{a=1}^3\log\widetilde\Gamma(\Delta_a;\tau)+2(N-1)\log(q;q)_{\infty},\label{eq:Seff:1}
\end{split}
\end{equation}
such that the index (\ref{Eq:indexN4}) can be rewritten simply as
\begin{equation}
	\mathcal I(\tau; \Delta)=\fft{1}{N!}\int_{-\fft{1}{2N}}^{1-\fft{1}{2N}}\prod_{\mu=1}^{N-1}du_\mu\,\exp[N^2S_\text{eff}(\hat u;\Delta,\tau)].\label{eq:index:1}
\end{equation}
Here $\hat u$ denotes a set of holonomies $\hat u=\{u_j|j=1,\cdots,N\}$ and we have chosen the above integration range for later convenience. Note that we have replaced $-\sum_{i\neq j}\log\widetilde\Gamma(u_{ij};\tau)$ with $\sum_{i\neq j}\log\theta_0(u_{ij};\tau)$ to get (\ref{eq:Seff:1}) and (\ref{eq:index:1}) from (\ref{Eq:indexN4}), using the quasi-double-periodicity (\ref{theta0:periodic}), (\ref{Gamma:periodic}) and the inversion formula (\ref{theta0:inversion}), (\ref{Gamma:inversion}) of the elliptic functions. 

Given the effective action (\ref{eq:Seff:1}) and the integral form of the index (\ref{eq:index:1}), we can now apply the saddle-point approach. First, we find solutions to the saddle point equations
\begin{equation}
	0=\fft{\partial}{\partial u_\mu}S_\text{eff}(\hat u;\Delta,\tau)\bigg|_{\hat u=\hat u_\ast}\qquad(\mu=1,\cdots,N-1).\label{eq:saddle:1}
\end{equation}
Then the index (\ref{eq:index:1}) can be approximated as
\begin{equation}
	\mathcal I(\tau; \Delta)\sim\sum_{\hat u_\ast\in\mathcal C'}\fft{1}{N!}\int_{D_{\hat u_\ast}}\prod_{\mu=1}^{N-1}du_\mu\,\exp[N^2S_\text{eff}(\hat u;\Delta,\tau)],
\label{eq:index:1:saddle}
\end{equation}
where the integration is along the steepest descent contour $\mathcal C'$ passing through one or more saddle points.  For each saddle point, $D_{\hat u_\ast}$ is a neighborhood of the corresponding saddle point solution $\hat u_\ast$.  For a real saddle point, where $\hat u_\ast$ lies
on the original contour $\mathcal C$ of (\ref{eq:index:1}), we have
\begin{equation}
	\hat u_\ast\in D_{\hat u_\ast}\subseteq\mathcal C=\bigcup_{\mu=1}^{N-1}[-\fft{1}{2N},1-\fft{1}{2N}].
\label{contour}
\end{equation}
However, in general, we may expect the saddle point to be complex, in which case the original contour $\mathcal C$ will have to be deformed to pass through the saddle point.  Here we assume this to be the case, but will further comment on the contour deformation in section~\ref{Cardy:finite}.

Note that (assuming contour deformation is possible) if we did not restrict the integral in (\ref{eq:index:1:saddle}) to the neighborhoods of the saddle points, but kept the full integration contour $\mathcal C'$, then we would still have an exact expression for the index.  The approximation comes from integrating only near the saddle points, and this needs to be controlled by a large parameter.  Such a parameter would naturally be $N^2$ in the 't~Hooft expansion.  But in the Cardy-like limit, $1/|\tau|$ can also play the role of a large parameter.  In either case, the saddle point result (\ref{eq:index:1:saddle}) is valid up to exponentially suppressed terms in the large parameter. From here on, we choose $1/|\tau|$ as a large control parameter.

To make contact with the results in the literature we take the Cardy-like limit that imposes $|\tau|\ll1$ from here on. In section~\ref{Cardy:infinite}, we revisit the  leading term in the Cardy-like limit $|\tau|\to0$ \cite{Choi:2018hmj,Honda:2019cio}. In section~\ref{Cardy:finite}, we keep track of sub-leading corrections in the finite Cardy-like expansion with $|\tau|\ll 1$. In both sections, our goal is to obtain an explicit expression for the SCI using the saddle-point approximation (\ref{eq:index:1:saddle}).

%%%%%
\subsubsection{Leading term in the Cardy-like limit}\label{Cardy:infinite}
%%%%%
In the Cardy-like limit, $|\tau|\to0$, we substitute the asymptotic formulas of the Pochhammer symbol (\ref{pochhammer:asymp}), the elliptic theta function (\ref{elliptic:theta:0:asymp}), and the elliptic gamma function (\ref{elliptic:Gamma:asymp}) into the effective action (\ref{eq:Seff:1}).  The leading order term then scales as $\mathcal O(1/\tau^2)$, and we find
\begin{equation}
	N^2S_\text{eff}(\hat u;\Delta,\tau)=-\fft{\pi i}{3\tau^2}\sum_{a=1}^3\left(\sum_{i\neq j}B_3(\{u_{ij}+\Delta_a\}_\tau)+(N-1)B_3(\{\Delta_a\}_\tau)\right)+\mathcal O(|\tau|^{-1}),
\label{eq:Seff:2:infinite}
\end{equation}
where $B_3(x)$ is the third Bernoulli polynomial.
The definition of a $\tau$-modded value $\{\cdot\}_\tau$ is given in (\ref{tau-modded}). Here we assumed
\begin{equation}
	\{\tilde u_{ij}+\tilde\Delta_a\}\not\to0~\text{or}~1\label{assumption:leading}
\end{equation}
for any $u_i$'s and $\Delta_a$'s to use the asymptotic formula of the elliptic gamma function (\ref{elliptic:Gamma:asymp}). The `tilde' values $\tilde u_i$ and $\tilde\Delta_a$ are defined following (\ref{u:component}) and the curly bracket $\{\cdot\}$ is defined in (\ref{modded}). 

The saddle point equation (\ref{eq:saddle:1}) is given from the effective action (\ref{eq:Seff:2:infinite}) as
\begin{equation}
\begin{split}
	0&=-\fft{\pi i}{\tau^2}\sum_{a=1}^3\sum_{j=1}^N\Big(B_2(\{u_{\mu j}+\Delta_a\}_\tau)-B_2(\{u_{Nj}+\Delta_a\}_\tau)\\
	&\kern7em~-B_2(\{-u_{\mu j}+\Delta_a\}_\tau)+B_2(\{-u_{Nj}+\Delta_a\}_\tau)\Big)+\mathcal O(|\tau|^{-1}),\label{eq:saddle:2:infinite}
\end{split}
\end{equation}
under the assumption (\ref{assumption:leading}). As we have commented in the beginning of this section, the pairwise saddle point equation (\ref{eq:saddle:2:infinite}) yields a rich set of solutions and we expect that one or a handful of solutions yields a dominant contribution to the index in the saddle point approximation (\ref{eq:index:1:saddle}). One of the most well known solutions is the one with all identical holonomies, namely $u_i=u_j$ for all $i,j\in\{1,\cdots,N\}$ \cite{Choi:2018hmj,Honda:2019cio}. The effective action at this saddle point successfully counted the dual AdS$_5$ black hole microstates \cite{Choi:2018hmj}. In the main text, we focus on the case where this particular saddle point with identical holonomies is dominant over the other saddle points and therefore this black hole microstate counting is valid. We put off the discussion on other types of saddle points, in particular the ones dubbed as $C$-center solutions\footnote{The $C$-center solution is related to the $\{C,N/C,0\}$ BA solution in \cite{Hong:2018viz} and the $(C,N/C)$ saddle in \cite{Cabo-Bizet:2019eaf}.} in \cite{ArabiArdehali:2019orz}, to Appendix \ref{App:C-center}. 

On the integration contour (\ref{contour}), there are $N$ distinct sets of identical holonomies satisfying the SU($N$) constraint $\sum_{i=1}^Nu_i\in\mathbb Z$, namely
\begin{equation}
	\hat u^{(m)}=\left\{u_j^{(m)}=\fft{m}{N}\,\Big|\,j=1,\cdots,N\right\}\quad(m=0,1,\cdots,N-1).
\label{eq:saddle:sol:infinite}
\end{equation}
We can compute the effective action (\ref{eq:Seff:2:infinite}) at this saddle point (\ref{eq:saddle:sol:infinite}) as 
\begin{equation}
	N^2S_\text{eff}(\hat u^{(m)};\Delta,\tau)=-\fft{\pi i(N^2-1)}{\tau^2}\prod_{a=1}^3\left(\{\tilde\Delta_a\}-\fft{1+\eta}{2}\right)+\mathcal O(|\tau|^{-1}),\label{eq:Seff:2:infinite:saddle}
\end{equation}
where we have introduced $\eta\in\{\pm 1\}$ as
\begin{equation}
	\sum_{a=1}^3\{\Delta_a\}_\tau=2\tau+\sum_{a=1}^3\{\tilde\Delta_a\}=2\tau+\fft{3+\eta}{2},
\label{eq:eta}
\end{equation}
from the constraint $\sum_{a=1}^3\Delta_a-2\tau\in\mathbb Z$ and the assumption (\ref{assumption:leading}). The SCI is then given by substituting (\ref{eq:Seff:2:infinite:saddle}) into the saddle point approximation (\ref{eq:index:1:saddle}) as
\begin{equation}
\begin{split}
	\mathcal I(\tau; \Delta)&=N\exp[-\fft{\pi i(N^2-1)}{\tau^2}\prod_{a=1}^3\left(\{\tilde\Delta_a\}-\fft{1+\eta}{2}\right)+o\left(|\tau|^{-2}\right)]\\
	&\quad+(\text{contribution from other saddles}).\label{eq:index:2:saddle:infinite}
\end{split}
\end{equation}
This reproduces the result of \cite{Choi:2018hmj,Honda:2019cio,ArabiArdehali:2019tdm}. In the context of a 4d Cardy formula, (\ref{eq:index:2:saddle:infinite}) is also consistent with the result of \cite{Kim:2019yrz} and closely related to the supersymmetric Casimir energy \cite{Assel:2015nca,Bobev:2015kza} used to count the dual black hole microstates \cite{Cabo-Bizet:2018ehj}. The factor of $N!$ in the denominator of (\ref{eq:index:1:saddle}) is removed by the degeneracy from permuting $N$ holonomies within the saddle point (\ref{eq:saddle:sol:infinite}).

%%%%%
\subsubsection{Sub-leading terms in the Cardy-like expansion}\label{Cardy:finite}
%%%%%
The fact that the $|\tau|^{-2}$-leading term in the Cardy-like limit (\ref{eq:index:2:saddle:infinite}) also captures the $N^2$-leading term in the large-$N$ limit is not clear {\it a priori}, since (\ref{eq:index:2:saddle:infinite}) could have terms of order $N^2$ but sub-leading in the Cardy-like expansion such as $\mathcal O(N^2|\tau|^{-1})$. In this subsection we clarify that such a correction does \emph{not} show up in fact and therefore (\ref{eq:index:2:saddle:infinite}) captures the $N^2$-leading term in the large-$N$ limit correctly, by keeping track of all the sub-leading terms up to exponentially suppressed ones in the Cardy-like expansion. 

To go beyond the leading term in the Cardy-like limit, we have to expand the special functions to higher order.  In particular, we substitute the asymptotic formulas of the Pochhammer symbol (\ref{pochhammer:asymp}), the elliptic theta function (\ref{elliptic:theta:0:asymp}), and the elliptic gamma function (\ref{elliptic:Gamma:asymp}) into (\ref{eq:Seff:1}) and keep track of sub-leading terms in the finite Cardy-like expansion. The result is given in terms of Bernoulli polynomials as
\begin{equation}
\begin{split}
	N^2S_\text{eff}(\hat u;\Delta,\tau)&=-\fft{\pi i}{3\tau^2}\sum_{a=1}^3\left(\sum_{i\neq j}B_3(\{u_{ij}+\Delta_a\}_\tau)+(N-1)B_3(\{\Delta_a\}_\tau)\right)\\
	&\quad+\fft{\pi i}{\tau}\left(\sum_{a=1}^3\sum_{i\neq j}B_2(\{u_{ij}+\Delta_a\}_\tau)+(N-1)\sum_{a=1}^3B_2(\{\Delta_a\}_\tau)+\sum_{i\neq j}\{u_{ij}\}_\tau(1-\{u_{ij}\}_\tau)\right)\\
	&\quad-\fft{5\pi i}{6}\sum_{a=1}^3\left(\sum_{i\neq j}B_1(\{u_{ij}+\Delta_a\}_\tau)+(N-1)B_1(\{\Delta_a\}_\tau)\right)\\
	&\quad+\pi i\sum_{i\neq j}\{u_{ij}\}_\tau+\fft{\pi i(2\tau^2-3\tau-1)N^2}{6\tau}+\pi iN-\fft{\pi i(2\tau^2+3\tau-1)}{6\tau}\\
	&\quad-(N-1)\log\tau+\sum_{i\neq j}\log(1-e^{-\fft{2\pi i}{\tau}(1-\{u_{ij}\}_\tau)})\left(1-e^{-\fft{2\pi i}{\tau}\{u_{ij}\}_\tau}\right)\\
	&\quad+\mathcal O\left(|\tau|^{-1}e^{\fft{2\pi\sin(\arg\tau)}{|\tau|}X}\right),
\label{eq:Seff:2:finite}
\end{split}
\end{equation}
where the first line above is just the leading order term (\ref{eq:Seff:2:infinite}).  As in the previous subsection, we follow the conventions in (\ref{tau-modded}), (\ref{u:component}), (\ref{modded}) and the assumption (\ref{assumption:leading}). The higher order terms are of $\mathcal O(|\tau|^{-1}e^{\fft{2\pi\sin(\arg\tau)}{|\tau|}X})$ where $X$ is defined as
\begin{equation}
	X=\min(\{\tilde u_{ij}+\tilde\Delta_a\},~1-\{\tilde u_{ij}+\tilde\Delta_a\}:a=1,2,3,~i,j=1,\cdots,N).
\end{equation}
This is exponentially suppressed under the assumption (\ref{assumption:leading}). Thus, we are treating the SCI in all powers of $\tau$ up to exponentially suppressed terms.

Using this finite Cardy-like expansion of the effective action (\ref{eq:Seff:2:finite}), we would like to  evaluate  sub-leading corrections to the saddle point solution (\ref{eq:saddle:sol:infinite}) and the index (\ref{eq:index:2:saddle:infinite}) obtained in the infinite Cardy-like limit. For that purpose, it suffices to focus on the effective action (\ref{eq:Seff:2:finite}) \emph{near} the leading saddle point solution (\ref{eq:saddle:sol:infinite}). To be specific, we make the ansatz for saddle point solutions in the finite Cardy-like expansion,
\begin{equation}
	\hat u^{(m)}=\left\{u_j^{(m)}=\fft{m}{N}+v_j\tau\,\Big|~v_j\sim\mathcal O(|\tau|^0),~\sum_{j=1}^Nv_j=0\right\}\quad(m=0,1,\cdots,N-1),\label{eq:saddle:ansatz:finite}
\end{equation}
and investigate the effective action (\ref{eq:Seff:2:finite}) around this ansatz. This ansatz is natural as it is equivalent to the leading order solution (\ref{eq:saddle:sol:infinite}) up to sub-leading corrections given by $v_j$. Note that $\sum_{j=1}^Nv_j=0$ is required to satisfy the SU($N$) constraint.

The effective action (\ref{eq:Seff:2:finite}) \emph{near} the saddle point ansatz (\ref{eq:saddle:ansatz:finite}) can be simplified using 
\begin{equation}
\begin{split}
	\{u_{ij}+\Delta_a\}_\tau&=u_{ij}+\{\Delta_a\}_\tau,\\
	\{u_{ij}\}_\tau&=\begin{cases}
	u_{ij} & (\tilde u_i\geq\tilde u_j)\\
	1+u_{ij} & (\tilde u_i<\tilde u_j),
\end{cases}\label{assumption:sub-leading}
\end{split}
\end{equation}
since $u_{ij}=v_{ij}\tau$ is at most order $\mathcal O(|\tau|)$ and therefore we can factor it out from the modded values carefully. The resulting simplified effective action is given as
\begin{equation}
\begin{split}
	N^2S_\text{eff}(\hat u;\Delta,\tau)&=-\fft{\eta \pi i}{\tau^2}N\sum_{j=1}^N\left(u_j-\fft{\sum_{k=1}^Nu_k}{N}\right)^2+\sum_{j\neq k}\log(2\sin\fft{\pi u_{jk}}{\tau})\\
	&\quad-\fft{\pi i}{\tau^2}(N^2-1)\prod_{a=1}^3\left(\{\Delta_a\}_\tau-\fft{1+\eta}{2}\right)+\fft{\pi i(6-5\eta)(N^2-1)}{12}\\
	&\quad-\fft{\pi iN(N-1)}{2}-(N-1)\log\tau+\mathcal O(|\tau|^{-1}e^{\fft{2\pi\sin(\arg\tau)}{|\tau|}X}),\label{eq:Seff:2:finite:near-ansatz}
\end{split}
\end{equation}
where we have used the same $\eta$ introduced in (\ref{eq:eta}).

The saddle point equation (\ref{eq:saddle:1}) is given from the effective action (\ref{eq:Seff:2:finite:near-ansatz}) and the ansatz (\ref{eq:saddle:ansatz:finite}) as
\begin{equation}
	i\eta\,v_j=\fft{1}{N}\sum_{k=1\,(\neq j)}^N\cot\pi v_{jk}\quad(i=1,\cdots,N)
\label{eq:saddle:2:finite}
\end{equation}
and is valid up to exponentially suppressed terms. Note that the system of equations is $\tau$-independent, thus justifying our assumption $v_j\sim\mathcal O(|\tau|^0)$.  In addition, the log term in the first line of (\ref{eq:Seff:2:finite:near-ansatz}) leads to a repulsion between pairs of eigenvalues.  It is this term that shows up away from the strict Cardy-like limit that pushes the eigenvalues apart and modifies the leading order solution, (\ref{eq:saddle:sol:infinite}), of condensed eigenvalues.
In fact, as will be highlighted below, this set of equations closely resemble those of an SU($N$) Chern-Simons model.

%%%%%
\subsubsection*{The steepest descent contour}

%%%%%

At leading order in the Cardy-like limit, we found $N$ distinct real saddle points (\ref{eq:saddle:sol:infinite}).  However, at sub-leading order, while there are still $N$ distinct saddle points, each one is now complex, as the solutions to (\ref{eq:saddle:2:finite}) are complex.  As a result, we seek to deform the original contour (\ref{contour}) to a new contour $\mathcal C'$ that passes through these $N$ saddles.

%%%%%
\begin{figure}[t]
	\centering
	\includegraphics[scale=.65]{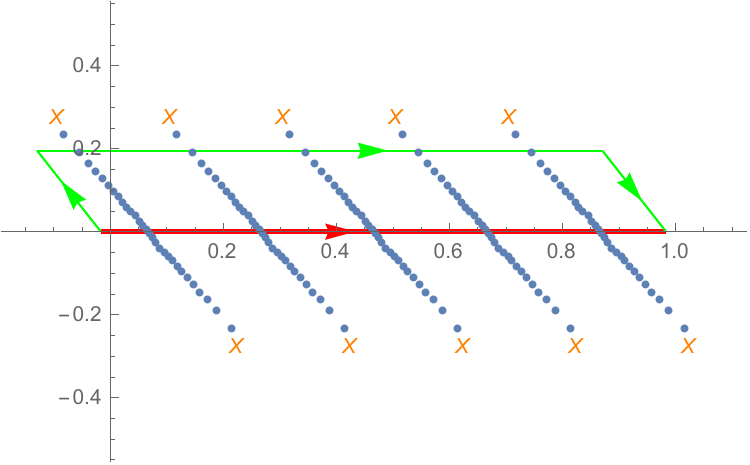}
	\caption{Numerical leading saddle points\,(blue dots) discussed in Appendix~\ref{App:saddle:compare} with $N=30$ and $\tau=\fft{ie^{\pi i/6}}{\pi}$. There must be $N=30$ distinct sets of holonomies in the above figure but here only 5 copies of them are shown for presentation. Orange crosses denote $\pm\tau+\fft{m}{N}~(m=2,8,14,20,26)$ and therefore it is straightforward to see that each set of holonomies collapses to $\fft{m}{N}$ as $|\tau|\to0$. \label{Contour-deform}}
\end{figure}
%%%%%

To be more specific, we show a typical complex saddle point solution in Figure~\ref{Contour-deform}.  The original contour integrates all eigenvalues along the real line, as shown by the red path.  The first step is then to deform the contour so that the integration path of each eigenvalue $u_\mu$ passes through the corresponding saddle point solution as indicated by the green path in the figure.  Since the contributions from the left and the right ends of green contours cancel each other, the deformed contour can be written simply as
\begin{equation}
	\mathcal C'=\bigcup_{\mu=1}^{N-1}(v_\mu\tau-\fft{1}{2N},v_\mu\tau+1-\fft{1}{2N}],
\label{contour:deformed}
\end{equation}
where $\{v_\mu\}$ is a solution to the saddle-point equation (\ref{eq:saddle:2:finite}).  Note that we are implicitly assuming that the effective action is analytic in this region so that the deformation is valid.

The saddle-point approximation to the SCI is then obtained from the effective action (\ref{eq:Seff:2:finite:near-ansatz}) as
\begin{equation}
\begin{split}
	\mathcal I(\tau;\Delta)&\sim\sum_{m=0}^{N-1}\fft{\mathcal A}{N!}\int_{D_{\hat u^{(m)}}}\prod_{\mu=1}^{N-1}du_\mu\,\exp[-\fft{\eta\pi i}{\tau^2}N\sum_{j=1}^N\Big(u_j-\fft{\sum_{k=1}^Nu_k}{N}\Big)^2+\sum_{j\neq k}\log(2\sin\fft{\pi u_{jk}}{\tau})]\\
	&\quad+(\text{contribution from other saddles}),\label{eq:index:2:saddle:finite:1}
\end{split}
\end{equation}
up to exponentially suppressed terms, where $D_{\hat u^{(m)}}$ denotes a small neighborhood of a saddle point solution $\hat u^{(m)}$ on the deformed contour (\ref{contour:deformed}), namely
\begin{equation}
	D_{\hat u^{(m)}}=\bigcup_{\mu=1}^{N-1}(v_\mu\tau+\fft{m}{N}-\epsilon,v_\mu\tau+\fft{m}{N}+\epsilon]\subseteq\mathcal C',
\label{contour:saddle:u}
\end{equation}
for some small positive number $\epsilon$. The prefactor $\mathcal A$ is defined as
\begin{equation}
\begin{split}
	\mathcal A&=\exp\left[-\fft{\pi i}{\tau^2}(N^2-1)\prod_{a=1}^3\left(\{\Delta_a\}_\tau-\fft{1+\eta}{2}\right)+\fft{\pi i(6-5\eta)(N^2-1)}{12}\right.\\
	&\kern3em~\left.-\fft{\pi iN(N-1)}{2}-(N-1)\log\tau+\mathcal O(|\tau^{-1}|e^{\fft{2\pi\sin(\arg\tau)}{|\tau|}X})\right].\label{eq:A}
\end{split}
\end{equation}
Finally, it is convenient to introduce new integration variables $\lambda_j$ with the constraint $\sum_{j=1}^N\lambda_j=0$ as
\begin{equation}
	u_j=u_j^{(m)}-(i\lambda_j+v_j)\tau=\fft{m}{N}-i\lambda_j\tau.
\label{u:to:lambda}
\end{equation}
This allows us to rewrite (\ref{eq:index:2:saddle:finite:1}) as
\begin{equation}
\begin{split}
	\mathcal I(\tau; \Delta)&\sim N\tau^{N-1}e^{-\fft{\pi i(N^2-1)}{2}}\fft{\mathcal A}{N!}\int_{D_{\hat\lambda}}\prod_{\mu=1}^{N-1}d\lambda_\mu\,\exp[\eta\pi iN\sum_{j=1}^N\lambda_j^2+\sum_{j\neq k}\log(2\sinh\pi\lambda_{jk})]\\
	&\quad+(\text{contribution from other saddles}),\label{eq:index:2:saddle:finite:2}
\end{split}
\end{equation}
where the integration contour $D_{\hat\lambda}$ is given from the contour (\ref{contour:saddle:u}) and the change of variables (\ref{u:to:lambda}) as
\begin{equation}
	D_{\hat\lambda}=\bigcup_{\mu=1}^{N-1}(iv_\mu-\fft{i\epsilon}{\tau},iv_\mu+\fft{i\epsilon}{\tau}].\label{contour:saddle:lambda}
\end{equation}

Remarkably, the steepest descent integral in (\ref{eq:index:2:saddle:finite:2}) is identical to that used to evaluate the $S^3$ sphere partition function of supersymmetric $\text{SU}(N)_k$ Chern-Simons theory
\begin{equation}
	Z_{\text{SU}(N)_k}^{CS}=\fft{1}{N!}\int_{-\infty}^\infty\prod_{\mu=1}^{N-1}d\lambda_\mu\,\exp[-\pi ik\sum_{j=1}^N\lambda_j^2+\sum_{j\neq k}\log(2\sinh\pi\lambda_{jk})],
\end{equation}
under the constraint $\sum_{j=1}^N\lambda_j=0$, provided we make the identification $k=-\eta N$.  This does depend on the ability to deform the contour of the Chern-Simons theory to pass through the $D_{\hat\lambda}$ contour, which we assume to be the case.  We investigate this further in Appendix~\ref{App:saddle}.  The final result is that the index can be written as
\begin{equation}
\begin{split}
	\mathcal I(\tau; \Delta)&\sim N\tau^{N-1}e^{-\fft{\pi i(N^2-1)}{2}}\mathcal A\,Z_{\text{SU}(N)_{k=-\eta N}}^{CS}\\
	&\quad+(\text{contribution from other saddles}).
\label{eq:index:2:saddle:finite:3}
\end{split}
\end{equation}
We have computed this SU($N$) partition function in Appendix \ref{App:CS:partition:function} based on the U($N$) partition function from \cite{Kapustin:2009kz}. Substituting the result (\ref{eq:Z:SU(N)}) into (\ref{eq:index:2:saddle:finite:3}), we get
\begin{empheq}[box=\fbox]{equation}
\begin{split}
	\mathcal I(\tau; \Delta)&\sim N\exp[-\fft{\pi i(N^2-1)}{\tau^2}\prod_{a=1}^3\left(\{\Delta_a\}_\tau-\fft{1+\eta}{2}\right)+\mathcal O(e^{-1/|\tau|})]\\
	&\quad+(\text{contribution from other saddles}).\label{eq:logI1}
\end{split}
\end{empheq}
The leading Cardy-like limit of \eqref{eq:logI1} reproduces the result obtained in the leading Cardy-like limit by \cite{ Cabo-Bizet:2018ehj, Choi:2018hmj, Cabo-Bizet:2019osg, Honda:2019cio, ArabiArdehali:2019tdm}. One of the main results of this paper is that there are no sub-leading $\tau$-corrections besides exponentially suppressed contributions. We also obtain a $\log N$ contribution to the logarithm of the SCI, which comes directly from the degeneracy of $N$ different saddle points contributing equally to the superconformal index
\footnote{Note that we do not include contributions to $\log N$ that are present in the full quantum Chern-Simons theory \cite{Ooguri:2002gx}. These contributions arise from integration near the trivial connection which has a residual global gauge symmetry corresponding to SU($N$) gauge transformations and therefore contributes a factor of inverse volume of SU($N$) to the path integral. Our computation connects to  the matrix model of SU($N$) Chern-Simons theory, not to the quantum path integral of Chern-Simons theory.}.
This is in fact an important lesson we learn, and it will ensure the universality of the logarithmic correction for a large class of $\mathcal{N}=1$ 4d SCFT's as we will see in subsequent sections.

%%%%%
\subsection{Saddle point approximation for generic \texorpdfstring{${\cal N}=1$}{N=1} SCFT }\label{Sec:SaddleGen}
%%%%%

We now generalize the previous set of results to the case of $\mathcal{N} =1$ toric quiver gauge theories. 
Toric quiver gauge theories describe the low energy dynamics of a stack of $N$ D3 branes probing the tip of a toric Calabi-Yau singularity;  there is by now a vast literature detailing how to construct a supersymmetric field theory given toric data (see, for example, \cite{Benvenuti:2004dy,Franco:2005sm}). Consider a toric quiver gauge theory whose gauge group $G$ has $n_{\text{v}}$ simple factors (in all the $\mathcal{N} = 1$ quiver gauge theories we will deal with, the number of simple factors coincides with the number of vector multiplets). We focus, for concreteness, on the case in which all the gauge group factors are SU($N_a$), $a$ goes from $1$ to $n_{\text{v}}$, with $N_a=N \hspace{2mm} \forall \hspace{1mm} a$. In these theories the weight vectors $\rho$ are such that for any bi-fundamental field $\Phi_{ab}$ (notice that in the more generic notation used in \cite{Benini:2018mlo}, the index $a$ of $\Phi_a$ would now split into $ab$):
\begin{eqnarray}
\rho_{ij}^{\Phi_{ab}}\left(u\right) \equiv u_{ij}^{ab} \equiv u_i^{a} - u_j^b.
\end{eqnarray} 
There are $d - 1$ fugacities corresponding to flavor symmetries appearing in the generic toric gauge theories that we will study,  $d$ is the number of external points of the toric diagram that are related to the quivers defining the theory (see for example \cite{Amariti:2019mgp}).
The integrand of \eqref{Eq:indexToric} can be now exponentiated and treated like an effective action:
\begin{equation}
\begin{split}
    S_{\text{eff}}(u; \Delta,\tau) & = \sum_{\Phi_{ab}}\sum_{i_a\neq j_b}\log \widetilde{\Gamma} \left(u_{ij}^{ab}+ \Delta_{ab}; \tau\right)-\sum_{a =1}^{n_{\text{v}}}\sum_{i_a \neq j_a}\widetilde{\Gamma} \left(u^a_{i j}, \tau\right) \\ 
&\quad +2 n_{\text{v}}(N-1)\log \left(q ; q\right)_{\infty}+ (N-1)\sum_{\Phi_{ab}}\widetilde{\Gamma}(\Delta_{ab}; \tau)\label{Eq:logIgen}
\end{split}
\end{equation}
where we have denoted $u^{ab}_{ij} \equiv u_{i_a} - u_{j_b}$ for the holonomies associated to the chiral multiplets $\Phi_{ab}$ and $u^a_{ij} \equiv u_{i_a} - u_{j_a}$ for the vector multiplets. 
Making use of the expression for the elliptic gamma function in the $|\tau | \ll 1$ limit, the effective action can be expressed as:
\begin{equation}
\begin{split}
	S_\text{eff}(u;\Delta,\tau)&=-\fft{\pi i}{3\tau^2}\sum_{\Phi_{ab}}\left(\sum_{i_a\neq j_b}\{u^{ab}_{ij}+\Delta_{ab}\}_\tau(\{u^{ab}_{ij}+\Delta_{ab}\}_\tau-\fft12)(\{u^{ab}_{ij}+\Delta_{ab}\}_\tau-1)\right. \label{Eq:sefftoric}\\
	&\kern6em~\left.+(N-1)\{\Delta_{ab}\}_\tau(\{\Delta_{ab}\}_\tau-\fft12)(\{\Delta_{ab}\}_\tau-1)\right)\\
	&\quad+\fft{\pi i}{\tau}\left(\sum_{\Phi_{ab}}\sum_{i_a\neq j_b}(\{u^{ab}_{ij}+\Delta_{ab}\}_\tau^2-\{u^{ab}_{ij}+\Delta_{ab}\}_\tau+\fft16)+\sum_{a =1}^{n_{\text{v}}}\sum_{i_a\neq j_a}\{u^{a}_{ij}\}_\tau(1-\{u^{a}_{ij}\}_\tau)\right.\\
	&\kern4em~\left.+(N-1)\sum_{\Phi_{ab}}(\{\Delta_{ab}\}_\tau^2-\{\Delta_{ab}\}_\tau+\fft16)\right)\\
	&\quad-\fft{5\pi i}{6}\sum_{\Phi_{ab}}\left(\sum_{i_a \neq j_b}(\{u^{ab}_{ij}+\Delta_{ab}\}_\tau-\fft12)+(N-1)(\{\Delta_{ab}\}_\tau-\fft12)\right)\\
	&\quad+\pi i\sum_{a= 1}^{n_{\text{v}}}\sum_{i_a\neq j_a}\{u^{a}_{ij}\}_\tau +\sum_{a =1}^{n_{\text{v}}}\sum_{i_a \neq j_a}\log(1-e^{-\fft{2\pi i}{\tau}(1-\{u^{a}_{ij}\}_\tau)})\left(1-e^{-\fft{2\pi i}{\tau}\{u^{a}_{ij}\}_\tau}\right)\\
	&\quad+\fft{i \pi N^2}{6 \tau}\left((n_{\chi}- n_{\text{v}})\tau^2-  n_{\text{v}}\right)+\fft{i \pi N n_{\text{v}}}{6 \tau}\left(\tau^2  +1\right)-\fft{i \pi n_{\chi}\tau}{6}\\
	&\quad + 2 n_{\text{v}}(N-1)\log \left(q ; q\right)_{\infty}+\mathcal O(e^{-1/|\tau|}).
\end{split}
\end{equation}
Using the asymptotic formula \eqref{pochhammer:asymp} valid for $|\tau|\ll 1$ and  following section~\ref{Cardy:finite} we generalize the statement that $u_{ij}^{ab}$ satisfy the ansatz for the saddles \eqref{eq:saddle:ansatz:finite}, which allows us to rewrite the effective action as
\begin{equation}
\begin{split}
	S_\text{eff}(u;\Delta,\tau)&=-\fft{\pi i}{\tau^2}\sum_{\Phi_{ab}}(\{\Delta_{ab}\}_\tau-\tau-\fft12)\sum_{i_a\neq j_b}(u^{ab}_{ij})^2- \fft{\pi i}{\tau}\sum_{a=1}^{n_{\text{v}}}\sum_{i_a \neq j_a}(u^a_{ij})^2+\fft{2\pi i}{\tau}\sum_{a=1}^{n_{\text{v}}}\sum_{i_a>j_a}u^a_{ij}\\
	&\quad +2\sum_{a=1}^{n_{\text{v}}}\sum_{i_a>j_a}\log(1-e^{-\fft{2\pi i}{\tau}u^a_{ij}})\\
	&\quad-\fft{\pi i}{3\tau^2}(N^2-1)\sum_{\Phi_{ab}}\left( B_3\left(\{\Delta_{ab}\}_{\tau}\right)- 3 \tau B_2\left(\{\Delta_{ab}\}_{\tau}\right)+ \frac{5\tau^2}{2}B_1\left(\{\Delta_{ab}\}_{\tau}\right)\right)\\
	&\quad+\fft{i \pi N^2}{6 \tau}\left((n_{\chi}- n_{\text{v}})\tau^2 - n_{\text{v}}\right)+ \fft{i \pi n_{\text{v}}}{2}N \\
	&\quad -\fft{i \pi}{6 \tau}\left((n_{\chi} - n_{\text{v}})\tau^2+ 3 n_{\text{v}}\tau - n_{\text{v}}\right)- n_{\text{v}}(N-1)\log \tau+\mathcal O(e^{-1/|\tau|}).
\label{Eq:eff-toric}
\end{split}
\end{equation}
We have used Bernoulli polynomials for the sake of compactness. To simplify the terms depending on $u_{ij}$ in  \eqref{Eq:eff-toric}, it will be convenient to rewrite the sum over the chiral multiplets $\sum_{\Phi_{ab}}$ as $\fft12\sum_{a=1}^{n_{\text{v}}}\sum_{\phi_a = 1}^{n_{\phi_a}}$, where $\phi_a$ labels an arrow connected to $a$ and $n_{\phi_a}$ is the degree of incidence the node $a$ in the graph representing the quiver.  The effective action then simplifies to
\begin{equation}
\begin{split}
	S_\text{eff}(u;\Delta,\tau)
    & =	\sum_{a=1}^{n_{\text{v}}}F_a(u;\eta_a, \tau) -\fft{\pi i}{3\tau^2}(N^2-1)\sum_{\Phi_{ab}}\left[ B_3\left(\{\Delta_{ab}\}_{\tau}\right)- 3 \tau \left(B_2\left(\{\Delta_{ab}\}_{\tau}\right)- \fft16\right)\right.\\ 
    &\quad- \left. \fft{\tau}{2}+\frac{5\tau^2}{2}B_1\left(\{\Delta_{ab}\}_{\tau}\right)\right]+\fft{i \pi N^2}{6 \tau}\left((n_{\chi}- n_{\text{v}})\tau^2- n_{\text{v}}\right)+ \fft{i \pi n_{\text{v}}}{2}N\\
	&\quad -\fft{i \pi}{6 \tau}\left((n_{\chi} - n_{\text{v}})\tau^2+ 3 n_{\text{v}}\tau - n_{\text{v}}\right)- n_{\text{v}}(N-1)\log \tau+\mathcal O(e^{-1/|\tau|}),
\label{Eq:eff-toric1.1}
\end{split}
\end{equation}
where we have defined
\begin{equation}
\begin{split}
	F_a(u;\eta_a, \tau) & \equiv -\fft{ \pi i \eta_a}{\tau^2} N\sum_{i_a}u_{i_a}^2+ \sum_{i_a \neq j_a} \log \left(2 \sin \fft{ \pi u^a_{ij}}{\tau}\right). \label{FA}
\end{split}
\end{equation}
 Here the factors $\eta_a$ are given by
\begin{equation}
    \sum_{\phi_a =1}^{n_{\phi_a}}\{\Delta_{\phi_a}\}_\tau\equiv\tau(n_{\phi_a} -2)+\eta_a +\fft{n_{\phi_a}}{2}. \label{eq:constraintba}
\end{equation}
We have written $\Delta_{\phi_a}$ to emphasize that we are summing over a very specific subset of $\Delta_{ab}$, that is, for a fixed value of $a$, we sum only over chiral multiplets connected to $a$ in the quiver. Let us further note that, upon the shifting \eqref{shift}, conservation of $U(1)$ global charge at each node of the quiver yields $\eta_a =1$ as long as we remain within the domain of chemical potentials specified by:
\begin{eqnarray}
 \text{Im}\left(-\frac{1}{\tau}\right) & > &\text{Im}\left(\frac{\sum_{i =1}^{d-1}\left[\Delta_i\right]_{\tau}}{\tau}\right)>0.\label{domain}
\end{eqnarray}

It is remarkable that the holonomy-dependent part of (\ref{Eq:eff-toric1.1}), namely (\ref{FA}), takes exactly the same form with the effective action in (\ref{eq:index:2:saddle:finite:1}) for the $\mathcal N=4$ SYM case. Hence the arguments of section~\ref{Cardy:finite} can be applied here to solve the matrix model with the effective action \eqref{FA}. In the physics point of view, this implies that the relation between the sub-leading structures of the $\mathcal N=4$ SCI in the Cardy-like limit and the 3D Chern-Simons theory studied in the previous subsection \ref{Cardy:finite} can be extended to more general 4D $\mathcal N=1$ quiver gauge theories. To be specific, we can use the expression for $Z_{\text{SU}(N)}$ shown in Appendix \ref{App:CS:partition:function} to compute the first term of (\ref{Eq:eff-toric1.1}):
\begin{equation}
\begin{split}
	\sum_{a=1}^{n_{\text{v}}}\log Z^{(a)}_{SU(N)} & = \fft{i n_{\text{v}} \pi N(N-1)}{2}+\fft{5 \pi i (N^2 -1)}{12}\sum_{a=1}^{n_{\text{v}}}\eta_{a} .\label{Za}
\end{split}
\end{equation}
Even though we are restricting ourselves to the special case with $\eta_a=1$ in this subsection, we do not replace $\eta_a$ by $1$ to be able to compare with the $\mathcal{N}=4$ case. 

Before substituting (\ref{Za}) into (\ref{Eq:eff-toric1.1}), however, let us first simplify the remaining terms of (\ref{Eq:eff-toric1.1}). We rewrite \eqref{Eq:eff-toric1.1} as
\begin{equation}
\begin{split}
	S_\text{eff}(u;\Delta,\tau)&=	\sum_{a=1}^{n_{\text{v}}}F_a(u;\eta_a, \tau) -\fft{\pi i}{3\tau^2}(N^2-1)\sum_{\Phi_{ab}}\left[ B_3\left(\left[\Delta_{ab}\right]_{\tau}+1\right)- 3 \tau \left[\Delta_{ab}\right]_{\tau} \left(\left[\Delta_{ab}\right]_{\tau}+1\right)\right. \\ 
&\quad- \left. \fft{\tau}{2}+\frac{5\tau^2}{2}\left(\left[\Delta_{ab}\right]_{\tau}+\fft12\right)\right]+\fft{i \pi N^2}{6 \tau}\left((n_{\chi}- n_{\text{v}})\tau^2 - n_{\text{v}}\right)+ \fft{i \pi n_{\text{v}}}{2}N \\
	&\quad -\fft{i \pi}{6 \tau}\left((n_{\chi} - n_{\text{v}})\tau^2+ 3 n_{\text{v}}\tau - n_{\text{v}}\right)- n_{\text{v}}(N-1)\log \tau+\mathcal O(e^{-1/|\tau|}).
\label{Eq:eff-toric2}
\end{split}
\end{equation}
For the sake of compactness, we have used the relation $\{x\}_{\tau} =[x]_{\tau}+1$ thus we will express our final answer in terms of $[ \Delta]_{\tau}$. Let us now analyze separately the second term in \eqref{Eq:eff-toric2}:
\begin{equation}
\begin{split}
 \mathcal{K}(\left[\Delta_{ab}\right]_{\tau}; \tau) & \equiv B_3\left(\left[\Delta_{ab}\right]_{\tau}+1\right)- 3 \tau \left[\Delta_{ab}\right]_{\tau} \left(\left[\Delta_{ab}\right]_{\tau}+1\right) \\
 &=  B_3\left(\left[\Delta_{ab}\right]_{\tau}- \tau+1\right) - 3 \tau^2\left[\Delta_{ab}\right]_{\tau}+B_3\left(\tau\right),\\ 
\sum_{\Phi_{ab}} \mathcal{K}(\left[\Delta_{ab}\right]_{\tau}; \tau) & = \sum_{\Phi_{ab}}B_3\left(\left[\Delta_{ab}\right]_{\tau}- \tau+1\right) - 3 \tau^2\sum_{\Phi_{ab}}\left[\Delta_{ab}\right]_{\tau} + n_{\chi}B_{3}(\tau).
\label{Eq:eff-toric21}
\end{split}
\end{equation}

To simplify (\ref{Eq:eff-toric21}), first let us analyze the shifting of the chemical potentials 
\begin{equation}
 \Delta_l \rightarrow \Delta_l - \fft{2 \tau}{d}. \label{shift} 
\end{equation}
For the chemical potentials defined in \eqref{DeltaA} we have:
\begin{equation}
\Delta_{ab} = \xi_{ab}+ r_{ab}\tau=\sum_{l=1}^{d-1}q_{ab}^{l}\Delta_{l} + r_{ab} \tau,
\end{equation}
where $r_{ab}$ is the R-charge assigned to $a$-maximization and $q^{l}_{ab}$, $\Delta_{l}$ are the global charge, and chemical potentials respectively. As discussed in \cite{Lezcano:2019pae}, the shifting \eqref{shift} ensures optimal obstruction of bosonic-fermionic cancellation in the SCI by canceling the $(-1)^F$ in \eqref{Eq:TheSCI} and can be interpreted as a redefinition of R-charges such that only those chiral multiplets with negative $U(1)$ global charge acquire R-charge equals 2. Concretely, upon implementation of the shifting \eqref{shift}, we obtain:
\begin{equation}
\Delta_{ab} \rightarrow \bar{\xi}_{ab} + 2 \tau d_{ab}, \qquad \bar{\xi}_{ab} \equiv \sum_{l=1}^{d-1}q_{ab}^{l}\Delta_{l},
\end{equation}
where
\begin{equation}
d_{ab} = \begin{cases}
1 & \text{for} \hspace{2mm} \sum_{l=1}^{d-1}q^l_{ab} <0 \\
0 & \text{for} \hspace{2mm}\sum_{l=1}^{d-1}q^l_{ab} \geq 0. \\
\end{cases}
\end{equation}
To define $d_{ab}$ we are using the fact that, for a given chiral multiplet $\Phi_{ab}$,  $\text{sign}(q^{l}_{ab}) = \text{sign}(q^{m}_{ab}), \hspace{1.5mm} \forall \hspace{1.5mm}l, m= 1, \cdots, d-1$ which is the main property of the global charges that we will need during our calculations. Detailed tables with the values for the global charges $q^{l}_{ab}$ are presented in \cite{Lezcano:2019pae}. 

The properties of the Bernoulli polynomials allows to write $\mathcal{K}(\left[\Delta_{ab}\right]_{\tau}; \tau) $ after the shifting \eqref{shift} in the following way:
\begin{equation}
\begin{split}
    \sum_{\Phi_{ab}} \mathcal{K}(\left[\Delta_{ab}\right]_{\tau}; \tau) & \rightarrow \sum_{\Phi_{ab}}\left(-1\right)^{d_{ab}}B_3\left(\left(-1\right)^{d_{ab}}\left[\bar{\xi}_{ab}\right]_{\tau}- \tau+1\right) - 3 \tau^2\sum_{\Phi_{ab}}\left[\bar{\xi}_{ab}\right]_{\tau}\\
    &\quad- 6\tau^3 n_s + n_{\chi}B_{3}(\tau), \\ 
    n_s & \equiv \sum_{\Phi_{ab}}d_{ab}, \\ 
    \sum_{\Phi_{ab}} \mathcal{K}(\left[\bar{\xi}_{ab}\right]_{\tau}; \tau) &=\sum_{\Phi_{ab}}\left(-1\right)^{d_{ab}}B_3\left(\left(-1\right)^{d_{ab}}\left[\bar{\xi}_{ab}\right]_{\tau}+1\right) - 3\tau \sum_{\Phi_{ab}}\left[\bar{\xi}_{ab}\right]^2_{\tau}\\
    &\quad-6\tau^3 n_s + (n_{\chi}- n_{\text{v}})B_{3}(\tau).
\end{split}
\end{equation}
Using the fact that the quiver diagrams analyzed here can be drawn on a torus, and that $2 n_s = n_{F}=n_{\chi} - n_{\text{v}}$ is the number of monomial terms in the superpotential (which corresponds to the number of faces of the quiver representing the theory, see for example \cite{Lezcano:2019pae} and references therein), we obtain
\begin{equation}
\begin{split}
\sum_{\Phi_{ab}} \mathcal{K}(\left[\bar{\xi}_{ab}\right]_{\tau}; \tau) & = \sum_{\Phi_{ab}}\left(-1\right)^{d_{ab}}B_3\left(\left(-1\right)^{d_{ab}}\left[\bar{\xi}_{ab}\right]_{\tau}+1\right) - 3\tau \sum_{\Phi_{ab}}\left[\bar{\xi}_{ab}\right]^2_{\tau}\label{Eq:Kappa}\\
&\quad- (n_{\chi}- n_{\text{v}})\left(3\tau^3-B_{3}(\tau)\right).
\end{split}
\end{equation}
Let us now analyze separately the following term:
\begin{equation}
\begin{split}
\fft{5 \tau^2}{2}\sum_{\Phi_{ab}}\left([\Delta_{ab}]_{\tau} + \fft12\right) & = \frac{5\tau^2}{2}\sum_{a=1}^{n_{\text{v}}}\left(\fft12\sum_{\phi_a}^{n_{\phi_a}}\left[\Delta_{\phi_a}\right]_{\tau} + \fft{n_{\phi_a}}{4}\right)\\
&=\frac{5\tau^2}{2}\sum_{a=1}^{n_{\text{v}}}\fft{\eta_a}{2}+ \fft{5 \tau^3}{2}(n_{\chi} - n_{\text{v}}),\label{half}
\end{split}
\end{equation}
where we have organized the sum over chiral multiplets $\sum_{\Phi_{ab}} \rightarrow \fft12\sum_{a=1}^{n_{\text{v}}}\sum_{\phi_a=1}^{n_{\phi_a}}$  and in the 2nd line we have used \eqref{eq:constraintba}. Inserting \eqref{Eq:Kappa} and \eqref{half} back into \eqref{Eq:eff-toric2} yields:
\begin{equation}
\begin{split}
	S_\text{eff}(u;\Delta,\tau)&=	\sum_{a=1}^{n_{\text{v}}}F_a(u;\eta_a, \tau)\\
	&\quad- \fft{i  \pi (N^2-1)}{3\tau^2}\left\{ \sum_{\Phi_{ab}}\left[K\left(\left[\bar{\xi}_{ab}\right]_{\tau}, \tau\right)\right] - \fft{n_{\chi}\tau}{2}- (n_{\chi} - n_{\text{v}})\left(\fft12\tau^3- B_3(\tau)\right)\right.\\ &\quad \left.+\frac{5\tau^2}{2}\sum_{a=1}^{n_{\text{v}}}\fft{\eta_a}{2}\right\} +\fft{i \pi N^2}{6 \tau}\left((n_{\chi}- n_{\text{v}})\tau^2 - n_{\text{v}}\right)+ \fft{i \pi n_{\text{v}}}{2}N \\
	&\quad -\fft{i \pi}{6 \tau}\left((n_{\chi} - n_{\text{v}})\tau^2+ 3 n_{\text{v}}\tau - n_{\text{v}}\right)- n_{\text{v}}(N-1)\log \tau+\mathcal O(e^{-1/|\tau|}). \\ 
	&\quad 
\label{Eq:eff-toric4}
\end{split}
\end{equation}
Here we have defined
\begin{equation}
\begin{split}
 K(\Delta , \tau) & \equiv  B_3\left(\Delta+1\right)- 3 \tau \Delta^2  =  \frac{1}{2}\left(2 \Delta^3 - 3 |\Delta | \Delta +\Delta - 6 \tau |\Delta | \Delta \right)  \label{KK}
\end{split}
\end{equation}
and the final equality holds when $|\Delta | < 1$.
Upon simplification of \eqref{Eq:eff-toric4} we obtain:
\begin{equation}
\begin{split}
	S_\text{eff}(u;\Delta,\tau)&=	\sum_{a=1}^{n_{\text{v}}}F_a(u;\eta_a, \tau)- \fft{i  \pi (N^2-1)}{3\tau^2} \sum_{\Phi_{ab}}K\left(\left[\bar{\xi}_{ab}\right]_{\tau}, \tau\right) \\&\quad - \fft{i \pi (N^2 -1)}{12 }5  \left(\sum_{a=1}^{n_{\text{v}}}\eta_a\right) - \fft{\pi i n_{\text{v}}N(N-1)}{2} + \fft{i \pi n_{\chi}}{2}(N^2 -1)\\
	&\quad- n_{\text{v}}(N-1)\log \tau+\mathcal O(e^{-1/|\tau|}).  
\label{Eq:eff-toric5}
\end{split}
\end{equation}

The term $- \fft{i  \pi (N^2-1)}{3\tau^2} \sum_{\Phi_{ab}}K\left(\left[\bar{\xi}_{ab}\right]_{\tau}, \tau\right)$ gives the leading entropy function:
\begin{equation} 
\begin{split}
 S_E &=-\fft{i \pi (N^2 -1)}{6 \tau^2}C_{IJK}\left[\Delta_I\right]_{\tau}\left[\Delta_J\right]_{\tau}\left[\Delta_K\right]_{\tau} \quad (I = 1, \cdots, d), \label{Eq:SE}\\ 
\sum_{I=1}^d \Delta_I - 2 \tau &= 1,
 \end{split}
\end{equation}
as shown in \cite{Amariti:2019mgp, Lezcano:2019pae, Lanir:2019abx}. The coefficients $C_{IJK}$ are the triangular 't Hooft anomaly coefficients which correspond, as pointed out originally in \cite{Hosseini:2018dob} and later in  \cite{Amariti:2019mgp}, to the Chern-Simons couplings of the holographic dual gravitational description as elucidated in \cite{Benvenuti:2006xg}. Note that, as already pointed out in the previous subsection \ref{Cardy:infinite}, \eqref{Eq:SE} is closely related to the supersymmetric Casimir energy \cite{Assel:2015nca,Bobev:2015kza,Cabo-Bizet:2018ehj}. The structure \eqref{Eq:SE} was derived on a case by case basis for a large class of toric quiver gauge theories in the following domain of chemical potentials \eqref{domain}.

Inserting \eqref{Za} back into \eqref{Eq:eff-toric5}, we have
\begin{equation}
\begin{split}
	S_\text{eff}(u;\Delta,\tau)&=	-\fft{i \pi (N^2 -1)}{6 \tau^2}C_{IJK}\left[\Delta_I\right]_{\tau}\left[\Delta_J\right]_{\tau}\left[\Delta_K\right]_{\tau}\\
	&\quad+ \fft{i \pi n_{\chi}}{2}(N^2 -1)- n_{\text{v}}(N-1)\log \tau+\mathcal O(e^{-1/|\tau|}). 
\label{Eq:entropyfunct}
\end{split}
\end{equation}
Therefore, we can write $\mathcal{I}(\tau;\Delta)$ by generalizing the expression \eqref{eq:index:2:saddle:finite:2} which yields
\begin{equation}
\begin{split}
	\mathcal{I}(\tau; \Delta ) & = N\tau^{n_{\text{v}}(N-1)} e^{- \fft{n_{\text{v}}i \pi(N^2-1)}{2}}e^{S_\text{eff}(u;\Delta,\tau)}\\ 
	& =n_vN\exp\left[-\fft{i \pi (N^2 -1)}{6 \tau^2}C_{IJK}\left[\Delta_I\right]_{\tau}\left[\Delta_J\right]_{\tau}\left[\Delta_K\right]_{\tau}+\fft{i \pi }{2}(n_{\chi} - n_{\text{v}})(N^2 -1)+\mathcal O(e^{-1/|\tau|})\right],
\end{split}
\end{equation}
up to contribution from other saddles. Using the fact that $n_{\chi} - n_{\text{v}} =2 n_s$ we can ignore the term $\frac{i \pi }{2}(n_{\chi} -n_{\text{v}})(N^2 -1)$ because it is a multiple of $2 \pi i$ which does not affect the value of $\mathcal{I}(\tau; \Delta)$.  Finally we can write
\begin{empheq}[box=\fbox]{equation}
\begin{split}
	\mathcal I(\tau; \Delta)& =n_vN\exp\left[-\fft{i \pi (N^2 -1)}{6 \tau^2}C_{IJK}\left[\Delta_I\right]_{\tau}\left[\Delta_J\right]_{\tau}\left[\Delta_K\right]_{\tau}+\mathcal O(e^{-1/|\tau|})\right]\\
	&\quad+(\text{contribution from other saddles}). \label{eq:logI2}
\end{split}
\end{empheq}
The Cardy-like leading term in \eqref{eq:logI2} reproduces the results of \cite{ Cabo-Bizet:2019osg, Kim:2019yrz, Amariti:2019mgp, Lezcano:2019pae, Lanir:2019abx}. More importantly, the answer in \eqref{eq:logI2} contains no sub-leading contributions in 
$\tau$ up to exponentially suppressed terms, justifying {\it a posteriori} the efficacy of the Cardy-like limit. The logarithmic correction has the same origin, arising from a normalization, as it did in the  $\mathcal{N}=4$ case. Note that we can recover the case $\eta =1$ of \eqref{eq:logI1} by setting $n_{\chi} = 3$ and $n_{\text{v}} =1$ in \eqref{eq:logI2}.

%%%%%
\section{Bethe Ansatz approach}\label{Sec:BA}
%%%%%
%%%%%
\subsection{Bethe Ansatz approximation for \texorpdfstring{$\mathcal N=4$}{N=4} SYM}\label{Sec:BA:N=4}
%%%%%
According to the BA formula \cite{Benini:2018mlo}, the integral representation of the SCI of the $\mathcal N=4$ SU($N$) SYM theory (\ref{Eq:indexN4}) can be rewritten in terms of a discrete sum \eqref{eq:index:BA} which we reproduce here for convenience:
\begin{equation}
	\mathcal I(\tau; \Delta)=\kappa_N\sum_{\hat u\in\mathrm{BA}}\mathcal Z(\hat u;\Delta,\tau)H(\hat u;\Delta,\tau)^{-1},\label{eq:index:BA:N=4}
\end{equation}
where the building blocks are given as
\begin{subequations}
\begin{align}
	\kappa_N & = \frac{1}{N!}\left(\left(q; q\right)_{\infty}^2\prod_{a=1}^3\widetilde {\Gamma}(\Delta_a;\tau)\right)^{N-1}\label{def:kappa}\\ 
	\mathcal{Z}\left(\hat{u};\Delta,\tau\right) & = \prod_{i \neq j}^N \frac{\prod_{a=1}^3\widetilde\Gamma(u_{ij}+\Delta_a;\tau)}{\widetilde{\Gamma}\left(u_{ij};\tau\right)}\label{def:Z}\\ 
	H\left(\hat{u}; \Delta, \tau\right)& = \det \left[\frac{1}{2 \pi i} \frac{\partial \left(Q_1, \cdots, Q_N\right)}{\partial \left(u_1, \cdots , u_{N-1}, \lambda\right)}\right],\label{def:H}
\end{align}
\end{subequations}
and the BA operator $Q_i$ \eqref{Eq:BAOperator} takes the explicit form: 
\begin{equation}
	Q_i(\hat u;\Delta,\tau)\equiv e^{2\pi i\lambda}\prod_{j=1}^N\fft{\theta_1(u_{ji}+\Delta_1;\tau)\theta_1(u_{ji}+\Delta_2;\tau)\theta_1(u_{ji}-\Delta_1-\Delta_2;\tau)}{\theta_1(u_{ij}+\Delta_1;\tau)\theta_1(u_{ij}+\Delta_2;\tau)\theta_1(u_{ij}-\Delta_1-\Delta_2;\tau)}.\label{eq:Q}
\end{equation}
recalling that the BAEs are given as
\begin{equation}
	Q_i(\hat u;\Delta,\tau)=1.\label{eq:BAE}
\end{equation}

As we reviewed in section \ref{Sec:IndexGen}, the BA operator has a double-periodicity, namely 
\begin{equation}
	Q_i(\hat u;\Delta,\tau)=Q_i(\hat u';\Delta,\tau)\label{Q:double-period}
\end{equation}
for two sets of holonomies
\begin{equation}
\begin{split}
	\hat u&=\left\{u_i|i=1,\cdots,N\right\},\\
	\hat u'&=\left\{u_i+m_i+n_i\tau|\,m_i,n_i\in\mathbb Z,~i=1,\cdots,N\right\}.\label{u:u'}
\end{split}
\end{equation}
Hence, if we find one solution $\hat u$ to the BAEs (\ref{eq:BAE}), we can generate infinitely many solutions $\hat u'$ with different sets of integers $m_i$'s and $n_i$'s. Since the building blocks $\mathcal{Z}(\hat{u};\Delta,\tau)$ and $H(\hat{u}; \Delta,\tau)$ are invariant under $\hat u\to\hat u'$ (\ref{u:u'}) provided $\hat u$ is a solution to the BAEs (\ref{eq:BAE}) \cite{Benini:2018ywd}, the contributions from these infinitely many BAE solutions to the index through (\ref{eq:index:BA:N=4}) are all identical.

One of the simplest solution to the BAEs (\ref{eq:BAE}) is a so-called `basic' solution, namely
\begin{equation}
	\hat u_{\text{basic}}=\left\{u_i=\bar u+\fft{i}{N}\tau\,\Big|\,i=1,2,\cdots,N-1\right\}\bigcup\left\{u_N=\bar u\right\}\label{BAE:sol:basic}
\end{equation}
where $\bar u$ is determined as
\begin{equation}
	N\bar u+\fft{N(N-1)}{2}\tau\in\mathbb Z\label{ubar:general}
\end{equation}
from the SU($N$) constraint $\sum_{i=1}^Nu_i\in\mathbb Z$. Due to the double-periodicity of the BA operator (\ref{Q:double-period}), there are infinitely many basic solutions as
\begin{equation}
	\hat u_{\text{basic}}=\left\{u_i=\bar u+\fft{i}{N}\tau+m_i+n_i\tau\,\Big|\,i=1,2,\cdots,N-1\right\}\bigcup\left\{u_N=\bar u+m_N+n_N\tau\right\}\label{BAE:sol:basic:many}
\end{equation}
with arbitrary integers $m_i$'s and $n_i$'s. Note that $m_i$'s are redundant since $u_i$'s are defined modulo integers in the first place in (\ref{Eq:indexN4}) due to periodicity (see the cylinder in Figure~\ref{lb}).

In this section, we will compute the contribution from these basic solutions to the SCI through the BA formula (\ref{eq:index:BA:N=4}) in the large-$N$ limit, assuming that it dominates the other contributions.

%%%%%
\subsubsection*{Degeneracy}
%%%%%
To determine the contribution from infinitely many basic solutions (\ref{BAE:sol:basic:many}) to the index through the BA formula (\ref{eq:index:BA:N=4}), first we must figure out how many times we should add the contribution from a single basic solution: in short, we need the `relevant degeneracy' of basic solutions. There are two possible origins of degeneracies:
\begin{subequations}
	\begin{align}
	&i)~\text{The $\#$ of permuting }N\text{ holonomies within a given basic solution }\hat u_{\text{basic}}\\
	&ii)~\text{The $\#$ of }\bar u\text{ and }n_i's\text{ which yield inequivalent basic solutions }\hat u_{\text{basic}}.
	\end{align}
\end{subequations}
The $i)$ factor is obviously $N!$. The $ii)$ factor should be treated more carefully. First, the number of $\bar u$ that yields inequivalent basic solutions is given from (\ref{ubar:general}) as $N$, namely
\begin{equation}
	\bar u\in\left\{\fft{k}{N}-\fft{N-1}{2}\tau\,\Big|\,k=0,1,\cdots,N-1\right\}.
\end{equation}
The infinitely many inequivalent basic solutions obtained by manipulating $n_i$'s, however, must \emph{not} be taken into the `relevant degeneracy' when we compute the index using (\ref{eq:index:BA:N=4}). This is because, the equation (3.9) of \cite{Benini:2018mlo} restricts $u_i$'s to be inside the fundamental domain. To be specific, it requires
\begin{equation}
	0\leq\mathrm{Im}[u_i]<\mathrm{Im}\tau,
\end{equation}
which uniquely fixes every $n_i$. Hence the relevant degeneracy of a basic solution is $N\times N!$. Consequently, the BA formula (\ref{eq:index:BA:N=4}) reads
\begin{equation}
\begin{split}
	\mathcal I(\tau; \Delta)&=N\times N!\times \kappa_N\mathcal Z(\hat u_{\mathrm{basic}};\Delta,\tau)H(\hat u_{\mathrm{basic}};\Delta,\tau)^{-1}\\
	&\quad+(\mathrm{from~other~BA~solutions}).\label{BA:formula}
\end{split}
\end{equation}

We focus on the logarithm of the basic contribution, namely the first line of (\ref{BA:formula}):
\begin{equation}
\begin{split}
	\log\mathcal I(\tau; \Delta)\big|_\text{Basic BA}&=\log N!+\log N+\log\kappa_N\\
	&\quad+\log\mathcal Z(\hat u_{\mathrm{basic}};\Delta,\tau)-\log H(\hat u_{\mathrm{basic}};\Delta,\tau).\label{eq:index:basic}
\end{split}
\end{equation}
The contribution $\log\kappa_N$ can be written explicitly from the definition (\ref{def:kappa}) as
\begin{equation}
\begin{split}
	\log\kappa_N&=-\log N!+(N-1)\left(\sum_{a=1}^3\log\widetilde\Gamma(\Delta_a;\tau)+2\log(q;q)_{\infty}\right).\label{eq:kappa}
\end{split}
\end{equation}
In the remaining part of this section, we compute the remaining two contributions in the second line of (\ref{eq:index:basic}) in order, mainly following the results of \cite{Benini:2018ywd}. We omit the subscript `basic' of $\hat u_{\text{basic}}$ for notational convenience from here on.

%%%%%
\subsubsection*{The contribution from $\log\mathcal Z(\hat u;\Delta,\tau)$}
%%%%%
The contribution $\log\mathcal Z(\hat u;\Delta;\tau)$ to the index (\ref{eq:index:basic}) can be written explicitly as
\begin{equation}
	\log\mathcal Z(\hat u;\Delta;\tau)=\sum_{i\neq j}^N\left(\sum_{a=1}^3\log\widetilde\Gamma(\fft{i-j}{N}\tau+\Delta_a;\tau)-\log\widetilde\Gamma(\fft{i-j}{N}\tau;\tau)\right).\label{eq:gamma}
\end{equation}
To simplify the expression (\ref{eq:gamma}) further, recall that section 4 of \cite{Benini:2018ywd} yields
\begin{subequations}
	\begin{align}
	\sum_{i\neq j}^N\log\widetilde\Gamma(\fft{i-j}{N}\tau+\Delta_a;\tau)&=2\pi i\sum_{i\neq j}^NQ(\fft{i-j}{N}\tau+\{\Delta_a\}_\tau;\tau)-N\log\fft{\theta_0(\fft{N(\{\Delta_a\}_\tau-1)}{\tau};-\fft{N}{\tau})}{\theta_0(\fft{\{\Delta_a\}_\tau-1}{\tau};-\fft{1}{\tau})}\nn\\
	&\quad+\sum_{k=0}^\infty\bigg(\log\fft{\psi(\fft{N(k+\{\Delta_a\}_\tau)}{\tau})}{\psi(\fft{N(k+1-\{\Delta_a\}_\tau)}{\tau})}-N\log\fft{\psi(\fft{k+\{\Delta_a\}_\tau}{\tau})}{\psi(\fft{k+1-\{\Delta_a\}_\tau}{\tau})}\bigg),\\
	\sum_{i\neq j}^N\log\widetilde\Gamma(\fft{i-j}{N}\tau;\tau)&=2\pi i\sum_{i\neq j}^NQ(\fft{i-j}{N}\tau+1;\tau)-N\log N-2N\log\fft{(\tilde q^N;\tilde q^N)_{\infty}}{(\tilde q;\tilde q)_{\infty}}+\fft{\pi i}{12}(N-1),
	\end{align}\label{eq:gamma:1}%
\end{subequations}
where we have followed the conventions of $\theta_0(u;\tau)$ and $\psi(u)$ in Appendix \ref{App:elliptic:functions} and also defined $\tilde q \equiv e^{-\fft{2\pi i}{\tau}}$ and 
\begin{equation}
	Q(u;\tau)\equiv-\fft{B_3(u)}{6\tau^2}+\fft{B_2(u)}{2\tau}-\fft{5}{12}B_1(u)+\fft{\tau}{12}\label{eq:Q:Gamma}
\end{equation}
in terms of Bernoulli polynomials $B_n(x)$. Then, following the conventions introduced in Appendix \ref{App:elliptic:functions}, we can show that some of the contributions in (\ref{eq:gamma:1}) are exponentially suppressed in the large-$N$ limit as
\begin{subequations}
\begin{align}
	\log(\tilde q^N;\tilde q^N)_\infty&=\mathcal O(e^{-\fft{2\pi N\sin(\arg\tau)}{|\tau|}}),\\
	\theta_0(\fft{N(\{\Delta_a\}_\tau-1)}{\tau};-\fft{N}{\tau})&=\mathcal O(e^{-\fft{2\pi N\sin(\arg\tau)}{|\tau|}\min(\{\tilde\Delta_a\},1-\{\tilde\Delta_a\})}),\\
	\sum_{k=0}^\infty\log\psi(\fft{N(k+1-\{\Delta_a\}_\tau)}{\tau})&=\mathcal O(Ne^{-\fft{2\pi N\sin(\arg\tau)}{|\tau|}(1-\{\tilde\Delta_a\})}),\\
	\sum_{k=0}^\infty\log\psi(\fft{N(k+\{\Delta_a\}_\tau)}{\tau})&=\mathcal O(Ne^{-\fft{2\pi N\sin(\arg\tau)}{|\tau|}\{\tilde\Delta_a\}}),
\end{align}
\end{subequations}
where we have assumed
\begin{equation}
	\{\tilde\Delta_a\}\not\to 0,1.\label{assumption:BA}
\end{equation}
Ignoring the above exponentially suppressed terms and using the identity (\ref{Gamma:S-transform}), we can simplify (\ref{eq:gamma:1}) as
\begin{subequations}
	\begin{align}
	\sum_{i\neq j}^N\log\widetilde\Gamma(\fft{i-j}{N}\tau+\Delta_a;\tau)&=-\fft{\pi iN^2(\{\Delta_a\}_\tau-\tau)(\{\Delta_a\}_\tau-\tau-\fft12)(\{\Delta_a\}_\tau-\tau-1)}{3\tau^2}\nn\\&\quad+\fft{\pi i(\{\Delta_a\}_\tau-\tau-\fft12)}{6}-N\log\widetilde\Gamma(\Delta_a;\tau)\nn\\
	&\quad+\mathcal O(Ne^{-\fft{2\pi N\sin(\arg\tau)}{|\tau|}\min(\{\tilde\Delta_a\},1-\{\tilde\Delta_a\})}),\\
	\sum_{i\neq j}^N\log\widetilde\Gamma(\fft{i-j}{N}\tau;\tau)&=\fft{\pi iN^2(\tau-\fft12)(\tau-1)}{3\tau}-\fft{\pi iN(\tau^2-3\tau+1)}{6\tau}-\fft{\pi i\tau}{6}\nn\\&\quad-N\log N+2N\log(\tilde q;\tilde q)_{\infty}+\mathcal O(e^{-\fft{2\pi N\sin(\arg\tau)}{|\tau|}}).
	\end{align}\label{eq:gamma:2}%
\end{subequations}
Finally, substituting (\ref{eq:gamma:2}) into (\ref{eq:gamma}) and introducing $\eta\in\{\pm1\}$ as (\ref{eq:eta}), we obtain
\begin{equation}
\begin{split}
	\log\mathcal Z(\hat u;\Delta_a,\tau)&=-\fft{\pi iN^2}{\tau^2}\prod_{a=1}^3\left(\{\Delta_a\}_\tau-\fft{1+\eta}{2}\right)+\fft{(1-\eta)\pi i}{2}N^2+\fft{\eta\pi i}{12}\\
	&\quad+N\log N-N\sum_{a=1}^3\log\widetilde\Gamma(\Delta_a;\tau,\tau)-2N\log(\tilde q;\tilde q)_{\infty}\\
	&\quad+\fft{\pi iN(\tau^2-3\tau+1)}{6\tau}+\mathcal O(Ne^{-\fft{2\pi N\sin(\arg\tau)}{|\tau|}\min(\{\tilde\Delta_a\},1-\{\tilde\Delta_a\})}).\label{eq:gamma:3}
\end{split}
\end{equation}
%

%%%%%
\subsubsection*{The contribution from $-\log H(\hat u;\Delta,\tau)$} \label{sec:JunhoJacobian}
%%%%%
Next we consider the contribution from the Jacobian to the index (\ref{eq:index:BA:N=4}), which has been already studied in \cite{Hong:2018viz}. In particular, section 2.2 of \cite{Hong:2018viz} yields\footnote{Note $\Delta_a^\text{there}=2\pi\Delta_a^\text{here}~(a=1,2)$ and $\theta_1(2\pi u;\tau)^\text{there}=\theta_1(u;\tau)^\text{here}$.}
\begin{equation}
\begin{split}
    -\log H(\hat u;\Delta,\tau)&=-\log N-(N-1)\log(\fft{i}{\pi}\sum_{\Delta}\partial_{\Delta}\log\theta_1(\Delta;\fft{\tau}{N}))\\
    &\quad+\log\det(I_{N-1}+\tilde H(\hat u;\Delta,\tau)),\label{eq:H:Hong}
\end{split}
\end{equation}
where $\Delta$ take values in $\{\Delta_1,\Delta_2,-\Delta_1-\Delta_2\}$ and we have defined an $(N-1)\times(N-1)$ square matrix $\tilde H$ as
\begin{subequations}
\begin{align}
    \left[\tilde H(\hat u;\Delta,\tau)\right]_{\mu\nu}&\equiv\fft{g(\mu;\Delta,\tau)-g(\mu-\nu;\Delta,\tau)}{\sum_{k=1}^Ng(k;\Delta,\tau)},\label{eq:tilde:H:component}\\
    g(j;\Delta,\tau)&\equiv\fft{i}{2\pi}\sum_{\Delta}\partial_\Delta\log\left[\theta_1(\fft{j}{N}\tau+\Delta;\tau)\theta_1(-\fft{j}{N}\tau+\Delta;\tau)\right].\label{eq:g}
\end{align}
\end{subequations}
The second term of (\ref{eq:H:Hong}) can be computed explicitly in the large-$N$ limit using the asymptotic expansion of the elliptic theta function (\ref{elliptic:theta:1:asymp}) as
\begin{equation}
	\fft{i}{\pi}\sum_\Delta\partial_\Delta\log\theta_1(\Delta;\fft{\tau}{N})=\eta\fft{N}{\tau}+\mathcal O(e^{-\fft{2\pi N\sin(\arg\tau)}{|\tau|}\min(\{\tilde\Delta_a\},1-\{\tilde\Delta_a\})})\label{eq:diagonal}
\end{equation}
under the assumption (\ref{assumption:BA}). Here we have used $\sum_\Delta\{\Delta\}_\tau=\fft{3+\eta}{2}$ from (\ref{eq:eta}) and $\Delta\in\{\Delta_1,\Delta_2,-\Delta_1-\Delta_2\}$. Substituting (\ref{eq:diagonal}) into (\ref{eq:H:Hong}) then gives
\begin{equation}
\begin{split}
	-\log H(\hat u;\Delta,\tau)&=-N\log N+(N-1)\log\fft{\tau}{\eta}-\log\det(I_{N-1}+\tilde H(\hat u;\Delta,\tau))\\
	&\quad+\mathcal O(e^{-\fft{2\pi N\sin(\arg\tau)}{|\tau|}\min(\{\tilde\Delta_a\},1-\{\tilde\Delta_a\})}).\label{eq:H}
\end{split}
\end{equation}
The final step would be therefore estimating $-\log\text{det}(I_{N-1}+\tilde H)$.

Since it is difficult to estimate $-\log\text{det}(I_{N-1}+\tilde H)$ in general, first we take the Cardy-like limit $|\tau|\ll1$. Using the asymptotic formula (\ref{elliptic:theta:1:asymp}), we can obtain the Cardy-like expansion of the $g$-function (\ref{eq:g}) under the assumption (\ref{assumption:BA}) as
\begin{equation}
	g(j;\Delta,\tau)=\fft{\eta}{\tau}+\mathcal O(e^{-\fft{2\pi\sin(\arg\tau)}{|\tau|}\min(\{\tilde\Delta_a\},1-\{\tilde\Delta_a\})}),\label{eq:g:1}
\end{equation}
where we have used $\sum_\Delta\{\Delta\}_\tau=\fft{3+\eta}{2}$ as before. Substituting (\ref{eq:g:1}) back into $[\tilde H(\hat u;\Delta,\tau)]_{\mu\nu}$ (\ref{eq:tilde:H:component}) then gives
\begin{equation}
	\tilde H_{\mu\nu}=\mathcal O(N^{-1}e^{-\fft{2\pi\sin(\arg\tau)}{|\tau|}\min(\{\tilde\Delta_a\},1-\{\tilde\Delta_a\})}).\label{eq:tilde:H:component:1}
\end{equation}
The contribution from the Jacobian (\ref{eq:H}) is then simplified as
\begin{equation}
	-\log H(\hat u;\Delta,\tau)=-N\log N+(N-1)\log\fft{\tau}{\eta}+\mathcal O(e^{-\fft{2\pi\sin(\arg\tau)}{|\tau|}\min(\{\tilde\Delta_a\},1-\{\tilde\Delta_a\})})\label{eq:H:Cardy}
\end{equation}
in the Cardy-like limit \cite{Hong:2018viz}.

We want to estimate $-\log\text{det}(I_{N-1}+\tilde H)$ in the large-$N$ limit, however, not in the Cardy-like limit. To do that, we use the Gershgorin Circle Theorem: every eigenvalue of $\tilde H$ lies within at least one of the $N-1$ Gershgorin discs $(\mu=1,2,\cdots,N-1)$ 
\begin{equation}
	D(\tilde H_{\mu\mu},\sum_{\nu=1\,(\neq\mu)}^{N-1}|\tilde H_{\mu\nu}|),
\end{equation}
where the first and the second argument of $D(\cdot,\cdot)$ denotes the center and the radius of a disk respectively. Due to (\ref{eq:tilde:H:component:1}), every Gershgorin disc can be located within the unit disk whose center is at the origin for a small enough but finite $|\tau|$, and therefore every eigenvalue of the matrix $\tilde H$ has modulus less than 1 in that regime. Hence we can estimate $-\log\text{det}(I_{N-1}+\tilde H)$ for a small enough $|\tau|$ as
\begin{equation}
	-\log\text{det}(I_{N-1}+\tilde H)=-\tr\text{log}(I_{N-1}+\tilde H)=\tr\Big(\sum_{n=1}^\infty\fft{1}{n}(-\tilde H)^n\Big)=\mathcal O(N^0).\label{eq:tilde:H:estimate}
\end{equation}
Here we have used that every eigenvalue of $\tilde H_{\mu\nu}$ has modulus less than 1 for the taylor expansion of a logarithm in the 2nd equation. The Jacobian contribution (\ref{eq:H}) is then estimated as
\begin{equation}
	-\log H(\hat u;\Delta,\tau)=-N\log N+(N-1)\log\fft{\tau}{\eta}+\mathcal O(N^0).\label{eq:H:1}
\end{equation}
for a small enough but finite $|\tau|$.

\begin{figure}[t]
	\centering
	\includegraphics[scale=.48]{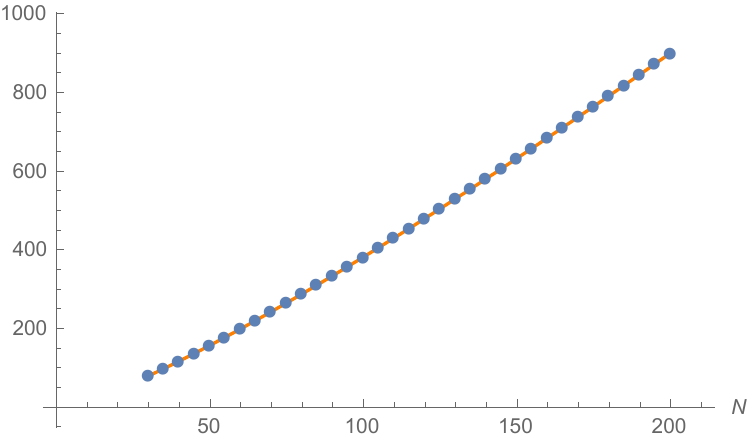}
	\includegraphics[scale=.48]{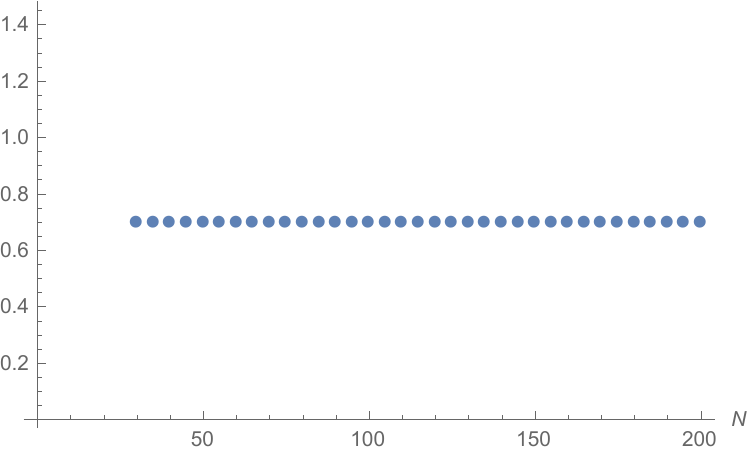}
	\caption{In the left hand side, blue dots represent numerical values of the real part of the Jacobian contribution $\Re\log H(\hat u;\Delta,\tau)$ and an orange line shows the first two leading terms read from (\ref{eq:H}), namely $N\log N-(N-1)\log|\tau|$. The figure in the right hand side shows numerical values of $\Re\log\text{det}(I_{N-1}+\tilde H)$, obtained by subtracting an orange line from blue dots in the left hand side. It converges to a certain finite value and therefore we can conclude it is of order $\mathcal O(N^0)$. \label{Jacobian}}
\end{figure}

We have not been able to estimate $-\log\text{det}(I_{N-1}+\tilde H)$ analytically for a generic finite $\tau$, which allows for some eigenvalues of $\tilde H$ to be greater than equal to $1$. Hence we move on to a numerical analysis: we investigate $-\log\text{det}(I_{N-1}+\tilde H)$ with $\Delta_1=\fft{1}{\pi}$, $\Delta_2=\fft{1}{e}$, and $\tau=2+i$ for $N=30,35,\cdots,200$ numerically. In this case the corresponding matrix $\tilde H$ (\ref{eq:tilde:H:component}) has some eigenvalues greater than $1$ so we cannot rely on the analytic argument (\ref{eq:tilde:H:estimate}). As one can see in Figure~\ref{Jacobian}, however, $-\log\text{det}(I_{N-1}+\tilde H)$ still seems to be of order $\mathcal O(N^0)$. We obtained similar results with other chemical potentials $\Delta_a$'s and $\tau$. Based on this numerical analysis, we believe that (\ref{eq:tilde:H:estimate}) and (\ref{eq:H:1}) are valid for a generic finite $\tau$ in fact.

%%%%%
%%%%%

%%%%%
\subsubsection*{The sum of all contributions}
%%%%%
Substituting (\ref{eq:kappa}), (\ref{eq:gamma:3}), (\ref{eq:H}) into (\ref{eq:index:basic}) and using the identity (\ref{pochhammer:S}), finally we have
\begin{empheq}{equation}
\begin{split}
	\log\mathcal I(\tau; \Delta)\big|_\text{Basic BA}&=-\fft{\pi iN^2}{\tau^2}\prod_{a=1}^3\left(\{\Delta_a\}_\tau-\fft{1+\eta}{2}\right)+\log N-\sum_{a=1}^3\log\widetilde\Gamma(\Delta_a;\tau) \\
	&\quad-2\log(q;q)_{\infty}+\fft{(1-\eta)\pi iN(N-1)}{2}+\fft{\pi i(6-5\eta)}{12}\\
	&\quad-\log\tau-\log\text{det}(I_{N-1}+\tilde H(\hat u;\Delta,\tau))\\
	&\quad+\mathcal O(e^{-\fft{2\pi N\sin(\arg\tau)}{|\tau|}\min(\{\tilde\Delta_a\},1-\{\tilde\Delta_a\})}).\label{eq:index:BA:N=4:1}
\end{split}
\end{empheq}
Recall that $\log\text{det}(I_{N-1}+\tilde H)$ is of order $\mathcal O(N^0)$ as we have discussed in (\ref{eq:tilde:H:estimate}) so this can be written simply as
\begin{empheq}[box=\fbox]{equation}
	\log\mathcal I(\tau; \Delta)\big|_\text{Basic BA}=-\fft{\pi iN^2}{\tau^2}\prod_{a=1}^3\left(\{\Delta_a\}_\tau-\fft{1+\eta}{2}\right)+\log N+\mathcal O(N^0)
\end{empheq}
in the large-$N$ limit, where we have neglected the pure imaginary term $\fft{(1-\eta)\pi iN(N-1)}{2}$ since it is of the form $2\pi i\mathbb Z$ due to $\eta\in\{\pm 1\}$.

If we take the Cardy-like limit after the large-$N$ limit, we can simplify \eqref{eq:index:BA:N=4:1} further using the Cardy-like Jacobian contribution (\ref{eq:H:Cardy}), the asymptotic expansion of a Pochhammer symbol (\ref{pochhammer:asymp}), and the following expansion
\begin{equation}
\begin{split}
	\sum_{a=1}^3\log\widetilde\Gamma(\Delta_a;\tau)&=-\fft{\pi i}{\tau^2}\prod_{a=1}^3\left(\{\Delta_a\}_\tau-\fft{1+\eta}{2}\right)+\fft{\pi i(\tau-2\eta)(2\tau-\eta)}{12\tau} \label{eq:gammaexp}\\
	&\quad+\mathcal O(e^{-\fft{2\pi\sin(\arg\tau)}{|\tau|}\min(\{\tilde\Delta_a\},1-\{\tilde\Delta_a\})})
\end{split}
\end{equation}
derived from (\ref{elliptic:Gamma:asymp}). The result is given as
\begin{empheq}[box=\fbox]{equation}
\begin{split}
	\log\mathcal I(\tau;\Delta)\big|_\text{Basic BA}&=-\fft{\pi i(N^2-1)}{\tau^2}\prod_{a=1}^3\left(\{\Delta_a\}_\tau-\fft{1+\eta}{2}\right)+\log N\\
	&\quad+\mathcal O(e^{-\fft{2\pi\sin(\arg\tau)}{|\tau|}\min(\{\tilde\Delta_a\},1-\{\tilde\Delta_a\})}).\label{eq:index:BA:N=4:2}
\end{split}
\end{empheq}
Similarly to what we obtained in section \ref{Sec:Saddle}, the $\fft{1}{\tau^2}$ contribution coincides with the one reported in \cite{Benini:2018ywd} and the logarithmic correction agrees perfectly with the result derived using the saddle-point approach. The origin of the $\log N$ term  can be found in the `relevant degeneracy' of the BA solutions and we will show a similar result for more generic SCFT's.

%%%%%
\subsection{Bethe Ansatz approximation for generic \texorpdfstring{${\cal N}=1$}{N=1} SCFT}\label{Sec:BAGen}
%%%%%
The goal of this section is to extend the results obtained for the SCI of $\mathcal{N} =4$ SYM to the SCI of more generic quiver gauge theories following the techniques applied in  \cite{Lezcano:2019pae,Lanir:2019abx}.  The formula for the SCI in terms of solutions to the BAEs    \cite{Closset:2017bse,Benini:2018mlo}:
\begin{eqnarray}
\mathcal{I}\left(\tau; \Delta\right) & = & \kappa_{G}\sum_{\hat{u} \in \text{BA}}\mathcal{Z}_{tot} \left(\hat{u};\Delta, \tau \right) H \left(\hat{u}; \Delta, \tau\right)^{-1}, \label{Eq:IndQuivers} \\ \nonumber
\kappa_{G} & = & \bar{\kappa}_G \times \left(\prod_{\Phi_{ab}}\widetilde{\Gamma} \left( \Delta_{ab}; \tau \right)\right)^{\text{rk}(G)}, \\ \nonumber
\mathcal{Z}_{tot}\left(u  ;\Delta, \tau \right) & = & \frac{\prod_{\Phi_{ab}}\prod_{i_a\neq j_b}\widetilde{\Gamma} \left( u_{i_a} - u_{j_b} + \Delta_{ab}; \tau \right)}{\prod_{a=1}^{n_{\text{v}}}\prod_{i_a \neq j_a}\widetilde{\Gamma} \left(u^a_{ij};\tau,\tau\right)},
\\\nonumber
H \left(u;\Delta, \tau\right) & = & \text{det}\left[\frac{1}{2 \pi i} \frac{\partial Q_{i_a}\left(u; ,\Delta, \tau\right)}{\partial u_{j_b}}\right]_{i_a j_b}.
\end{eqnarray}
Where we are interested in solutions of the BA equations, which we write as:
\begin{eqnarray}
Q_{i_a} \left(u ;\Delta, \tau\right)& = & e^{2\pi i\left( \sum_b\lambda_b - \sum_{j_b} u_{ij}^{ab}\right)}\prod_{\langle a,b \rangle}\prod_{j_b}\frac{\theta_0\left( -u_{ij}^{ab} +\Delta_{ab} ; \tau\right)}{\theta_0\left( -u_{ji}^{ba}  +\Delta_{ab}  ; \tau\right)} =1, \label{Eq:baeq1}
\end{eqnarray} 
where the notation is such that $\langle a, b \rangle$ represents the  set of chiral multiplets $\Phi_{ab}$ for a fixed value of $a$. We evaluate in BA solutions of the form: $u_{ij}^{ab} = \frac{\tau}{N}\left(i_a - j_b\right)$. These solutions appeared first in \cite{Hosseini:2016cyf} while evaluating the topologically twisted of 4d ${\cal N}=1$ theories on $T^2\times S^2$ in the high temperature limit; it was later shown in \cite{Hong:2018viz} that such configuration provides an exact solution to the BAEs. In \cite{Lezcano:2019pae, Lanir:2019abx} it was shown that this type of solution indeed satisfies the BA equations \eqref{Eq:baeq1} and provides the expected leading contribution of the form \eqref{Eq:SE}. Let us now proceed to take the large $N$ limit of \eqref{Eq:IndQuivers} keeping track of corrections of sub-leading corrections.

%%%%%%%%%%%%%%%%%%%%%%%%%%%%%%%%
\subsubsection*{The contribution from $\kappa_G$ and degeneracy} \label{sbs:kappaG}
Starting from $\kappa_G$ in \eqref{Eq:IndQuivers} we take the large $N$ limit:
\begin{eqnarray}
n_{\text{v}}N|\mathcal{W}_G|\kappa_{G} & = &  n_{\text{v}} N\left(q ; q \right)_{\infty}^{2 n_{\text{v}}(N-1)}\left(\prod_{\Phi_{ab}}\widetilde{\Gamma} \left( \Delta_{ab}; \tau\right)\right)^{N-1}, \label{Eq:KappaG} \\ \nonumber
\log n_{\text{v} }N + \log \left(|\mathcal{W}_G|\kappa_{G}\right) & = & \log n_{\text{v}} N + 2 n_{\text{v}} (N-1) \log \left(q ; q \right)_{\infty} + (N-1) \sum_{\Phi_{ab}}\log \widetilde{\Gamma} \left( \Delta_{ab}; \tau \right). 
\end{eqnarray}
The factor $n_{\text{v}} N$ in \eqref{Eq:KappaG} enters if we include the possibility of shifting each holonomy in the BA solution $N$ of times yielding inequivalent basic solutions . 

%%%%%%%%%%%%%%%%%%%%%%%%%%%%%%%%%%%%%%%%%%%%%%%%%%%%%%%%%%%%%%%%%%%%%%%%
\subsubsection*{The contribution from $\mathcal{Z}_{tot}$} \label{sbs:Ztot}
Let us now consider the expression for $\log \mathcal{Z}_{tot}$:
\begin{eqnarray}
\log \mathcal{Z}_{tot} & = & \log \left(\frac{\prod_{\Phi_{ab}}\prod_{i_a\neq j_b}\widetilde{\Gamma} \left( u_{i_a} - u_{j_b} + \Delta_{ab}; \tau \right)}{\prod_{a=1}^{n_{\text{v}}}\prod_{i_a \neq j_a}\widetilde{\Gamma} \left(u^a_{ij};\tau,\tau\right)}\right) \label{Eq:LogZquiver} \\ \nonumber
& = & \sum_{\Phi_{ab}}\sum_{i_a\neq j_b}\log\widetilde{\Gamma} \left( u_{i_a} - u_{j_b} + \Delta_{ab}; \tau \right) - \sum_{a=1}^{n_{\text{v}}}\sum_{i_a \neq j_a} \log \widetilde{\Gamma} \left(u^a_{ij};\tau\right).
\end{eqnarray}
Taking the large $N$ limit here requires just to reproduce the same calculation of section \ref{Sec:BA:N=4}, only there are $n_{\text{v}}$ contributions coming from the vector multiplets and we sum over $n_{\chi}$ chiral multiplets. Explicitly we have:
\begin{eqnarray}
\widetilde{\Gamma} \left(u_{ij}^{ab} + \Delta_{ab}; \tau\right) & = & \frac{e^{- \pi i \widetilde{\mathcal{Q}}\left(u_{ij}^{ab} + \Delta_{ab};\tau\right)}}{\theta_0 \left(\frac{u_{ij}^{ab} + \Delta_{ab}}{\tau}; - \frac{1}{\tau}\right)} \times \prod_{k =0}^{\infty}
\frac{\psi \left(\frac{k +1 +u_{ij}^{ab}}{\tau}\right)}{\psi \left(\frac{k - u_{ij}^{ab} - \Delta_{ab}}{\tau}\right)}, \label{EGF}\\ \nonumber
\widetilde{\mathcal{Q}} \left(u + \Delta; \tau, \tau\right) & = & \frac{u^3}{3 \tau^2} + u^2 \left(\frac{\Delta}{\tau^2} - \frac{2 \tau -1}{2 \tau^2} \right) + u \left(\frac{1 - 6 \tau + 5 \tau^2}{6 \tau^2} + \frac{\Delta^2}{\tau^2} - \frac{2 \tau -1}{ \tau^2} \Delta\right) - \\ \nonumber
& - & \frac{\Delta^2}{2 \tau^2}\left(2 \tau -1\right)+ \frac{\Delta}{6 \tau^2} \left(5 \tau^2 - 6 \tau + 1\right) + \frac{1}{12 \tau^2} \left(2 \tau - 1\right)\left(2 \tau - \tau^2\right) + \frac{\Delta^3}{\tau^2}.
\end{eqnarray}
Note that, the leading contribution coming from vector multiplets can be obtained from  (\ref{EGF}) by setting $\Delta_{ab} = 0$. We can write:
\begin{equation}
\begin{split}
\log \mathcal{Z}_{tot} & = \sum_{\Phi_{ab}}\sum_{i_a \neq j_b}\log \widetilde{\Gamma} \left(u_{ij}^{ab}+ \Delta_{ab}; \tau, \tau\right) -\sum_{a=1}^{n_{\text{v}}}\sum_{i_a \neq j_a}\widetilde{\Gamma} \left(u^a_{ij}, \tau, \tau\right) \\ 
& =  - i \pi  \sum_{\Phi_{ab}}\sum_{i_a\neq j_b}  \widetilde{\mathcal{Q}}\left(u_{ij}^{ab} + \Delta_{ab};\tau, \tau\right) + \sum_{\Phi_{ab}}\sum_{i_a\neq j_b} \sum_{k = 0}^{\infty} \log \frac{\psi \left(\frac{k +1 +u_{ij}^{ab}}{\tau}\right)}{\psi \left(\frac{k - u_{ij}^{ab} - \Delta_{ab}}{\tau}\right)}  \\ 
& \quad-   \sum_{\Phi_{ab}}\sum_{i_a\neq j_b} \log \theta_0 \left(\frac{u_{ij}^{ab} + \Delta_{ab}}{\tau}; - \frac{1}{\tau}\right) \\ 
& \quad+  i \pi n_{\text{v}} \sum_{i \neq j}^N\widetilde{\mathcal{Q}}\left(u_{ij}; \tau\right) +n_{\text{v}}N \log N + 2 n_{\text{v}}N \log \frac{\left(\widetilde{q}^N; \widetilde{q}^N\right)_{\infty}}{\left(\widetilde{q}; \widetilde{q}\right)_{\infty}} - \frac{i n_{\text{v}}\pi }{12}\left(N - 1 \right) \\ 
& =  - \fft{i \pi N^2}{3\tau^2} \sum_{\Phi_{ab}}B_3\left([\Delta_{ab}]_{\tau}- \tau +1\right) + i \pi \sum_{\Phi_{ab}}  \frac{\left(\left[\Delta_{ab}\right]_{\tau} - \tau +\frac{1}{2}\right)}{6}  \\ 
 & \quad-  N \sum_{\Phi_{ab}} \log \widetilde{\Gamma} \left(\Delta_{ab}; \tau, \tau\right) - n_{\text{v}} \fft{i  \pi}{3 \tau^2} N(N-1) B_3(\tau) -n_{\text{v}}\fft{i \pi }{6}(N-1) \tau \\ 
 &\quad+ n_{\text{v}}N \log N- 2 n_{\text{v}}N \log\left(\widetilde{q}; \widetilde{q}\right)_{\infty} + \mathcal{O}\left(Ne^{-\fft{N}{|\tau|}}\right), \label{eq:logia}   
 \end{split}
\end{equation}
where once again we have used the relation $\{x\}_{\tau} = [x]_{\tau} +1$. Equation \eqref{eq:logia} can be simplified making use of \eqref{eq:constraintba} in the following way
\begin{equation}
\begin{split}
\log \mathcal{Z}_{tot}  &= - \fft{i \pi N^2}{3\tau^2} \sum_{\Phi_{ab}}B_3\left([\Delta_{ab}]_{\tau}- \tau +1\right)  + \frac{i \pi}{12}\sum_{a=1}^{n_{\text{v}}}\eta_a  \\ 
 & \quad-  N \sum_{\Phi_{ab}} \log \widetilde{\Gamma} \left(\Delta_{ab}; \tau\right) - n_{\text{v}} \fft{i  \pi}{3 \tau^2} N(N-1) B_3(\tau)-n_{\text{v}}\fft{i \pi }{6}N\tau \\
 & \quad+  n_{\text{v}}N \log N - 2 n_{\text{v}}N \log\left(\widetilde{q}; \widetilde{q}\right)_{\infty} + \mathcal{O}\left(Ne^{-\fft{N}{|\tau|}}\right).
 \end{split}\label{Eq:Ztot1}
\end{equation}
If we remain within the domain of chemical potentials \eqref{domain}, which in the case of $\mathcal{N} =4$ SYM corresponds to taking $\eta =1$, and perform the shifting $\Delta_l \rightarrow \Delta_l - \fft{2 \tau}{d}$, we have
\begin{equation}
\begin{split}
\log \mathcal{Z}_{tot} & =  -\fft{i \pi N^2 }{6 \tau^2}C_{IJK}\left[\Delta_I\right]_{\tau}\left[\Delta_J\right]_{\tau}\left[\Delta_K\right]_{\tau}+ \frac{i \pi}{12}\sum_{a=1}^{n_{\text{v}}}\eta_a +\fft{i \pi n_{\text{v}}N}{3 \tau^2}B_3(\tau)-n_{\text{v}}\fft{i \pi }{6}N\tau\label{Eq:Ztot2}\\
&\quad -  N \sum_{\Phi_{ab}} \log \widetilde{\Gamma} \left(\Delta_{ab};  \tau\right) +  n_{\text{v}}N \log N- 2 n_{\text{v}}N \log\left(\widetilde{q}; \widetilde{q}\right)_{\infty}  + \mathcal{O}\left(Ne^{-\fft{N}{|\tau|}}\right).
\end{split} 
\end{equation}

\subsubsection*{The contribution from the Jacobian} \label{sbs:Jacobian}
 Let us consider the contribution from the Jacobian $H$ in \eqref{Eq:baeq1}. We need to study the large $N$ behavior of the matrix elements of the Jacobian matrix. The explicit form of the Jacobian is the following:
 \begin{eqnarray}
 H & = &  \text{det} \left[\frac{1}{2 \pi i}\frac{\partial \left(Q_{1_1},\cdots, Q_{N_1},  \cdots, Q_{1_{n_{\text{v}}}},\cdots, Q_{N_{n_{\text{v}}}}\right)}{\partial \left(u_{1_1}, \cdots, u_{N_1 -1}, \lambda_{1},\cdots, u_{1_{n_{\text{v}}}},\cdots, u_{N_{n_{\text{v}}}- 1},\lambda_{n_{\text{v}}}\right)}\right], \label{Eq:Hiajb}
 \end{eqnarray}
 where the BA operator is given as
 \begin{eqnarray}
Q_{i_a} \left(u ; \Delta, \tau \right)& = & e^{2\pi i\left( \sum_b\lambda_b - \sum_{j_b} u_{ij}^{ab}\right)}\prod_{\langle a,b \rangle}\prod_{j_b}\frac{\theta_0\left( -u_{ij}^{ab} +  \Delta_{a b } ; \tau\right)}{\theta_0\left( -u_{ji}^{ba}  + \Delta_{ba } ; \tau\right)}. \label{Eq:Qia}
\end{eqnarray} 
 Equation \eqref{Eq:Hiajb} is the determinant of an $n_{\text{v}}N \times n_{\text{v}}N  $ matrix that can be seen as $n_{\text{v}} \times n_{\text{v}}$ blocks of $N \times N$ matrices. For each fixed values of $a, b$, one can run the argument in section \ref{sec:JunhoJacobian} to show that the determinant is factorized as
 \begin{eqnarray}
 H= \text{det}H_{i_a, j_b} & = & N^{n_{\text{v}}} \text{det} H_{\mu_a \nu_b}% =  n_{\text{v}} N \exp \left( \text{Tr} \log H_{\mu_a \nu_b} \right)
  \label{Eq:FactorizationQuiver},
 \end{eqnarray}
 where Greek letters take values  $\mu_a = 1, \cdots, N-1$ for each $a\in\{1,\cdots,n_v\}$. Consider now the following matrix element evaluated in the BA solutions:
 \begin{eqnarray}
 H_{\mu_a, \nu_b}\Big{|}_{\text{BA}} & = & \frac{1}{2 \pi i}\frac{\partial \log Q_{\mu_a}}{\partial u_{\nu_b}}\Big{|}_{\text{BA}} \label{Eq:MelementQuiver} \\ \nonumber
 & = &   \sum_{\langle a, c \rangle}  \sum_{\nu_b}\left[\left(- \frac{\partial u_{\mu, \sigma}^{ac}}{\partial u_{\nu_b}}\right)\Big{|}_{\text{BA}}+\frac{1}{2 \pi i} \frac{\partial}{\partial u_{\nu_b}} \log \frac{\theta_0\left( -u_{\mu , \sigma}^{ac} +  \Delta_{a c }  ; \tau\right)}{\theta_0\left( -u_{\sigma, \mu}^{ca}  + \Delta_{ca }   ; \tau\right)}\Big{|}_{\text{BA}}\right].
 \end{eqnarray}
  As already seen in \ref{sec:JunhoJacobian}, the only important contribution in the large $N$ limit of \eqref{Eq:MelementQuiver} comes from the sum $\sum_{\nu_b}$ and it is of the form $\sim N -1$. Moreover, the structure of \eqref{Eq:MelementQuiver} thus, we have:
 \begin{equation}
\begin{split}
    -\log \text{det}H_{\mu_a, \nu_b} & =(N-1) \log \left(\prod_{a=1}^{n_{\text{v}}}\fft{i}{2\pi} \sum_{\phi_a =1}^{n_{\phi_a}} \partial_{\Delta_{\phi_a}} \log \theta_1 (\Delta_{\phi_a};\fft{\tau}{N})\right) + \mathcal{O}(N^0)  \\ 
 & =  (N-1) \log \left(\prod_{a=1}^{n_{\text{v}}}\fft{\tau}{N \eta_a}\right) + \mathcal{O}(N^0) \\
 & =-n_{\text{v}} N \log N +n_{\text{v}} \log N + n_{\text{v}}(N-1) \log \tau+ \mathcal{O}(N^0),\label{Eq:prel}
\end{split}
 \end{equation}
which is an immediate generalization of \eqref{eq:H:1}. We have used $\eta_a = 1, \hspace{2mm} \forall \hspace{2mm} a=1, \cdots,  n_{\text{v}}$, which can be proven using \eqref{eq:constraintba} and the conservation of flavor $U(1)$ charges. Inserting \eqref{Eq:prel} into the $\log$ of \eqref{Eq:FactorizationQuiver}, we obtain
 \begin{eqnarray}
 -\log H = - \log \text{det} H_{i_a j_b} & = & -n_{\text{v}} N \log N  + n_{\text{v}}(N-1) \log \tau + \mathcal{O}(N^0). \label{Eq:deth}
 \end{eqnarray}

%%%%%%%%%%%%%%%%%%%%%%%%%%%%%%%%%%%%%%%%%%%%%%%%%%%%%%%%
\subsubsection*{The sum of all contributions }
After collecting the results \eqref{Eq:KappaG}, \eqref{Eq:Ztot2}, and \eqref{Eq:deth} all together, we obtain
\begin{equation}
\begin{split}
  	\log \mathcal{I}(\tau, \Delta)\big|_\text{Basic BA}& = \log \left(n_{\text{v}}N|\mathcal{W}_G|\kappa_G\right) + \log \mathcal{Z}_{tot} -  \log H  \\ 
  	& =   -\fft{i \pi N^2 }{6 \tau^2}C_{IJK}\left[\Delta_I\right]_{\tau}\left[\Delta_J\right]_{\tau}\left[\Delta_K\right]_{\tau}  + \log n_{\text{v}}N +\frac{i \pi n_{\text{v}}}{12}\\
  	&\quad-2 n_{\text{v}} \left(\log \left(q ; q \right)_{\infty}+\fft{1}{2} \log \tau\right)-\sum_{\Phi_{ab}}\log \widetilde{\Gamma} \left( \Delta_{ab}; \tau \right)+ \mathcal{O}\left(N^0\right)\label{Eq:LogNQuiver}
\end{split}
\end{equation}
%\begin{equation}
%\begin{split}
%	\log \mathcal{I}(\tau, \Delta)\big|_\text{Basic BA}& =   \log \left(n_{\chi}N|\mathcal{W}_G|\kappa_G\right) + \log \mathcal{Z}_{tot} -  \log H \label{Eq:LogNQuiver}\\ 
%	& =  -\fft{i \pi 
%	N^2}{6 \tau^2}C_{IJK}\left[\Delta_I\right]_{\tau}\left[\Delta_J\right]_{\tau}\left[\Delta_K\right]_{\tau} -\sum_{\Phi_{ab}}\log\widetilde{\Gamma}(\Delta_{ab}; \tau) \\
%	&\quad+\log n_{\text{v}} N  - 2 n_{\text{v}}  \log \left(q ; q \right)_{\infty}+ n_{\text{v}} N \log \tau\\
%	&\quad- n_{\text{v}} (N-1) \log \left( \fft{\tau}{\sum_{a=1}^{n_{\text{v}}}\eta_a}\right) + \mathcal{O}(N^0)
%\end{split}
%\end{equation}
%
in the large-$N$ limit where we have used the identity \eqref{pochhammer:S}. 

Taking the Cardy-like limit, we can simplify further using the asymptotic expansion of the q-Pochhammer symbol  \eqref{pochhammer:asymp}  and the Elliptic Gamma function \eqref{elliptic:Gamma:asymp}, hence
 \begin{empheq}[box=\fbox]{equation}
 \begin{split}
\log \mathcal{I}(\tau; \Delta)\big|_\text{Basic BA}& = -\fft{i \pi (N^2 -1)}{6 \tau^2}C_{IJK}\left[\Delta_I\right]_{\tau}\left[\Delta_J\right]_{\tau}\left[\Delta_K\right]_{\tau} + \log n_{\text{v}}N+  \mathcal{O}(e^{-1/|\tau|}). \label{eq:Index4}
 \end{split}
 \end{empheq}
The first term in \eqref{eq:Index4} reproduces the result of \cite{Kim:2019yrz, Amariti:2019mgp, Lezcano:2019pae, Lanir:2019abx}. Note that even in the $\mathcal{N}=1$ case the coefficient of the $\log N$ term is $1$ and it appears crucially due to the degeneracy factor. 
%%%%%

%%%%%%%%%%%%%%%%%%%%%%%%%%%%%%%%%%%%%%%%%%%%%%
\section{Conclusions}\label{Sec:Conclusions}
%%%%%%%%%%%%%%%%%%%%%%%%%%%%%%%%%%%%%%%%%%%%%%%

One of the main results of this paper is the exploration of the ${\cal N}=4$ SCI beyond the leading orders in Cardy-like\,/\,large-$N$ limit respectively, which was required for a reliable estimation of logarithmic corrections. We have demonstrated by direct computation that the two main approaches to the SCI, namely the saddle point approach for the Cardy-like limit and the BA approach for the large-$N$ limit, are consistent with each other up to exponentially suppressed terms in the Cardy-like limit. This result was expected but it is highly nontrivial given the different approximation schemes involved in each computation. Our result can best be summarized as: 
\begin{equation}\label{Eq:Expansions}
    {\cal I}(\tau;\Delta) =
    \begin{cases}
    {\cal I}(\tau;\Delta)\big|_{\rm {Main\, Saddle}} + \text{contribution from other saddles},\\
    {\cal I}(\tau;\Delta)\big|_{\rm{ Basic \, BA}} + \text{contribution from other BA solutions},
\end{cases}
\end{equation}
where $\log {\cal I}(\tau;\Delta)\big|_{\rm {Main\, Saddle}}$ and $\log {\cal I}(\tau;\Delta)\big|_{\rm{ Basic \, BA}}$ are given explicitly in (\ref{Eq:ResultSP}) and (\ref{Eq:ResultBA}) respectively. Some of the contribution from other saddles and BA solutions have been studied in Appendix \ref{App:C-center}.

The nature of other contributions in the BA approach were recently discussed in \cite{ArabiArdehali:2019orz} where it was noticed that the structure of BA solutions might include entire continuous families beyond the naturally expected SL$(2,\mathbb{Z})$ type. This result indicates that the full expression for the SCI in the BA approach might require new techniques. As far as we are aware, there were no results regarding what the existence of a continuous family of BA solutions means in the saddle point approach to the SCI.  Our work in Appendix \ref{App:saddle:sublead} shows the existence of a saddle whose contribution to the SCI does not look like the contribution from any SL$(2,\mathbb Z)$ type BA solution, and therefore it is natural to expect that this saddle is related to a certain continuous family of BA solutions. Investigating this relation further may set the stage for a full understanding of the SCI. 

One of the byproducts of our analysis for the saddle point approximation is a direct window into the {\it effective} theory of the SCI. Namely, some important elements of the effective field theory approach were originally proposed in \cite{DiPietro:2014bca} and subsequently developed and extended in \cite{Ardehali:2015bla,DiPietro:2016ond}. The main paradigm is that at high temperatures the 4d theory is described by an effective 3d theory. Starting from 4d theories and taking the Cardy-like expansion, we have directly uncovered the matrix model connected with the 3d Chern-Simons theory. It would be quite interesting to systematically develop this effective field theory approach. For example, it would be quite interesting to formulate dynamical questions in terms of those degrees of freedom.  It is worth noticing  that aspects of such effective field theory approach might have powerful implications for the gravitational side as recently suggested in \cite{Larsen:2019oll,Nian:2020qsk,David:2020ems}.

Throughout our work in this project we relied heavily on numerical studies to inform us at crucial analytical turns. The ultimate product of our investigation can be entirely understood analytically.  We relegated some of our numerical discussion to various appendices because the results have been  mostly geared to confirm and motivate analytical results in the large-$N$ limit. There have been, however, two recent studies exploiting a numerical approach to the index with the main goal of better understanding finite $N$  aspects \cite{Murthy:2020rbd,Agarwal:2020zwm}. It would be interesting to explore the structure of the index in more details, in particular, to gain an understanding of the combinations governing the convergence of the expressions to aspects of black hole physics. The recent numerical experiments reported in \cite{Murthy:2020rbd,Agarwal:2020zwm} discuss finite $N$ aspects  and indicate  convergence of the numerical result to the saddle point expression  for relatively small values of $N\sim 6$. It would be quite interesting to  explore, for example,  how the convergence rates compare among the various approaches to the SCI, such analysis might possibly shed light on the expansion in Eq. (\ref{Eq:Expansions}).

The AdS/CFT correspondence posits that ${\cal N}=4$ SYM is the UV complete description of the gravity theory containing the AdS$_5$ black holes of interest. To the level of approximations described in this manuscript we have obtained a term of the form $\log N$. The precise coefficient of this term is very robust, it is present in this identical form in the generic case of 4d ${\cal N}=1$ theories with, for example,  $\text{SU}(N)^{n_{\text{v}}}$ gauge group.  This answer should pass  the litmus test of being reproduced from low energy supergravity. At the moment, the gravity side of this litmus test seems much more formidable. There are a number of difficulties along this road. For example, the  particular advantages of working in an odd-dimensional space that were crucial for clarifying the AdS$_4$ cases    \cite{Liu:2017vbl,Gang:2019uay,Benini:2019dyp} are now gone as the dual theory is 10d IIB sugra.  Nevertheless, the simplicity of the field theory answer indicates that the gravity answer might be achieved by carefully considering a small set of fields. Moreover, the fact that the result is universal for a large class of asymptotically AdS$_5\times SE_5$ black holes indicates that, most likely, the answer is the contribution of  certain zero modes. It would be quite interesting to pursue this computation fully. 

It would be quite natural to elucidate similar aspects in 3d theories. Although there is currently no BA approach to the SCI in 3d, there are  two approaches to the SCI: one using a continuum approximation for the monopole charge sums in the conformal index   \cite{Choi:2019zpz} and the other based on localization \cite{Nian:2019pxj}; they both exploit a Cardy-like limit. It would be interesting to elucidate the relation between these two approaches along the lines of the work presented here. The situation in 3d is quite peculiar because many of the computations are essentially reduced to the matrix model for the topologically twisted index, discussed in \cite{Benini:2015eyy}. For a class of matrix models describing topologically twisted indices, numerical results for the $\log N$ contribution have been worked out in \cite{Liu:2017vll,Liu:2018bac,PandoZayas:2019hdb} and precise agreement with the logarithmic entropy corrections for  magnetically charged black holes was shown in \cite{Liu:2017vbl,Gang:2019uay}. For a general class of rotating, electrically charged, black holes and their 3d SCI, agreement of the logarithmic corrections was established in \cite{Benini:2019dyp}. Quite remarkably, certain universality of the logarithmic terms in 3d partition functions was  established in  \cite{Marino:2011eh} and its dual side matched in \cite{Bhattacharyya:2012ye}; it would be quite enlightening to have such universal understanding  extended to the topologically twisted index and the 3d  SCI.

Let us finally remark that another approach to the SCI  has been proposed via doubly-periodic  extension of the index  for ${\cal N}=4$  SYM \cite{ Cabo-Bizet:2019eaf} and  in subsequent work for the generic ${\cal N}=1$ case \cite{Cabo-Bizet:2020nkr}. The leading $N^2$ term in this approach matches the other two approximations to the SCI. The methods used in this manuscript that allow to go beyond the leading order rely heavily on the analytical structure which is precisely lost in this doubly-periodic approach.  It would be quite interesting to penetrate the sub-leading structures in that approach where one might naturally hope that some aspects of modularity will be a powerful guiding principle.

%\item The {\it missing black holes problem}. Recently it has  been shown that the structures persists for a generic five-dimensional  Sasaki-Einstein manifold \cite{Benini:2020gjh}.  Some attack in the AdS$_4$ context \cite{Hong:2019wyi}

%%%%%%%%%%%%%%%%%%%%%%%%%%%%%%%%%%%%%%%%%%%%%%
\section*{Acknowledgments}
%%%%%%%%%%%%%%%%%%%%%%%%%%%%%%%%%%%%%%%%%%%%%%%

We gratefully acknowledge useful discussions with
Francesco Benini, Alejandro Cabo-Bizet, Alejandra Castro, Chandramouli Chowdhury, Marina David, Jewel Ghosh, Jun Nian,  Adu Offei-Danso, Kyriakos Papadodimas, Antonello Scardicchio and Yu Xin. This work was supported in part by the U.S. Department of Energy under grant DE-SC0007859. JH is supported in part by a Leinweber Graduate Summer Fellowship from the University of Michigan and by a Grant for Doctoral Study from the Korea Foundation for Advanced Studies.

\appendix
%%%%%

%%%%%
\section{Elliptic functions}
\label{App:elliptic:functions}
%%%%%
Here we gather definitions and useful identities of elliptic functions. 
%%%%%
\subsection{Definitions}
%%%%%
The Pochhammer symbol is defined as
\begin{equation}
	(z;q)_{\infty}=\prod_{k=0}^\infty(1-zq^k).\label{def:pochhammer}
\end{equation}
The elliptic theta functions used in this paper have the following product forms:
\begin{subequations}
\begin{align}
	\theta_0(u;\tau)&=\prod_{k=0}^\infty(1-e^{2\pi i(u+k\tau)})(1-e^{2\pi i(-u+(k+1)\tau)}),\label{eq:theta:0}\\
	\theta_1(u;\tau)&=-ie^{\fft{\pi i\tau}{4}}(e^{\pi iu}-e^{-\pi iu})\prod_{k=1}^\infty(1-e^{2\pi ik\tau})(1-e^{2\pi i(k\tau+u)})(1-e^{2\pi i(k\tau-u)})\nn\\
	&=ie^{\fft{\pi i\tau}{4}}e^{-\pi iu}\theta_0(u;\tau)\prod_{k=1}^\infty(1-e^{2\pi ik\tau}).\label{eq:theta:1}
\end{align}\label{eq:theta}%
\end{subequations}
The elliptic gamma function and the `tilde' elliptic gamma function are defined as
\begin{subequations}
\begin{align}
	\Gamma(z;p,q)&=\prod_{j,k=0}^\infty\fft{1-p^{j+1}q^{k+1}z^{-1}}{1-p^jq^kz},\label{def:gamma}\\
	\widetilde\Gamma(u;\sigma,\tau)&=\prod_{j,k=0}^\infty\fft{1-e^{2\pi i[(j+1)\sigma+(k+1)\tau-u]}}{1-e^{2\pi i[j\sigma+k\tau+u]}}.\label{def:tilde:gamma}
\end{align}
\end{subequations}
In this paper, we are mainly interested in the case with $\sigma=\tau$ and abbreviate $\Gamma(z;q,q)$ and $\widetilde\Gamma(u;\tau,\tau)$ as $\Gamma(z,q)$ and $\widetilde\Gamma(u;\tau)$ respectively. We also define a special function $\psi(u)$ as
\begin{equation}
	\psi(u)\equiv\exp[u\log(1-e^{-2\pi iu})-\fft{1}{2\pi i}\text{Li}_2(e^{-2\pi iu})].\label{def:psi}
\end{equation}
%

%%%%%
\subsection{Basic properties}
%%%%%
The elliptic theta functions have quasi-double-periodicity, namely
\begin{subequations}
\begin{align}
	\theta_0(u+m+n\tau;\tau)&=(-1)^ne^{-2\pi inu}e^{-\pi in(n-1)\tau}\theta_0(u;\tau),\label{theta0:periodic}\\
	\theta_1(u+m+n\tau;\tau)&=(-1)^{m+n}e^{-2\pi inu}e^{-\pi in^2\tau}\theta_1(u;\tau),\label{theta1:periodic}
\end{align}
\end{subequations}
for $m,n\in\mathbb Z$. The inversion formula of $\theta_0(u;\tau)$ can be written simply as
\begin{equation}
    \theta_0(-u;\tau)=-e^{-2\pi iu}\theta_0(u;\tau)\label{theta0:inversion}.
\end{equation}

The elliptic gamma function also has quasi-double-periodicity, namely
\begin{equation}
	\widetilde\Gamma(u;\sigma,\tau)=\widetilde\Gamma(u+1;\sigma,\tau)=\theta_0(u;\tau)^{-1}\widetilde\Gamma(u+\sigma;\sigma,\tau)=\theta_0(u;\sigma)^{-1}\widetilde\Gamma(u+\tau;\sigma,\tau).\label{Gamma:periodic}
\end{equation}
It also satisfies the inversion formula
\begin{equation}
    \widetilde\Gamma(u;\sigma,\tau)=\widetilde\Gamma(\sigma+\tau-u;\sigma,\tau)^{-1}.\label{Gamma:inversion}
\end{equation}

The Pochhammer symbol and the elliptic theta functions are transformed under the $S$-transformation as ($q=e^{2\pi i\tau}$, $\tilde q=e^{-\fft{2\pi i}{\tau}}$)
\begin{subequations}
\begin{align}
	(\tilde q;\tilde q)_{\infty}&=(-i\tau)^\fft12e^{\fft{\pi i}{12}(\tau+\fft{1}{\tau})}(q;q)_{\infty},\label{pochhammer:S}\\
	\theta_0(u/\tau;-1/\tau)&=e^{\fft{\pi i}{\tau}(u^2+u+\fft16)-\pi i(u+\fft12)+\fft{\pi i\tau}{6}}\theta_0(u;\tau),\label{theta:0:S}\\
	\theta_1(u/\tau;-1/\tau)&=-i(-i\tau)^\fft12e^{\fft{\pi iu^2}{\tau}}\theta_1(u;\tau).\label{theta:1:S}
\end{align}
\end{subequations}
The elliptic gamma function can be written in terms of these $S$-transformed elliptic theta functions and the $\psi$-function (\ref{def:psi}) as (see \cite{ArabiArdehali:2019tdm} for example)
\begin{equation}
	\widetilde\Gamma(\Delta_a;\tau)=\fft{e^{2\pi iQ(\{\Delta_a\}_\tau;\tau)}}{\theta_0(\fft{\{\Delta_a\}_\tau-1}{\tau};-\fft{1}{\tau})}\prod_{k=0}^\infty\fft{\psi(\fft{k+\{\Delta_a\}_\tau}{\tau})}{\psi(\fft{k+1-\{\Delta_a\}_\tau}{\tau})},\label{Gamma:S-transform}
\end{equation}
where $Q(\cdot;\cdot)$ is defined in (\ref{eq:Q:Gamma}) and repeated here as
\begin{equation}
	Q(u;\tau)\equiv-\fft{B_3(u)}{6\tau^2}+\fft{B_2(u)}{2\tau}-\fft{5}{12}B_1(u)+\fft{\tau}{12}.\label{eq:Q:Gamma:App}
\end{equation}
%

%%%%%
\subsection{Asymptotic behaviors}
%%%%%
For a small $|\tau|$ with fixed $0<\arg\tau<\pi$, the Pochhammer symbol can be approximated as
\begin{equation}
	\log(q;q)_\infty=-\fft{\pi i}{12}(\tau+\fft{1}{\tau})-\fft12\log(-i\tau)+\mathcal O(e^{-\fft{2\pi\sin(\arg\tau)}{|\tau|}}).\label{pochhammer:asymp}
\end{equation}

To study asymptotic behaviors of elliptic functions, first we introduce a $\tau$-modded value of a complex number $u$, namely $\{u\}_\tau$, as
\begin{equation}
	\{u\}_\tau\equiv u-\lfloor\Re u-\cot(\arg\tau)\Im u\rfloor\quad(u\in\mathbb C).\label{tau-modded}
\end{equation}
By definition, the $\tau$-modded value satisfies 
\begin{equation}
	\{u\}_\tau=\{\tilde u\}_\tau+\check u\tau,\qquad
	\{-u\}_\tau=\begin{cases}
	1-\{u\}_\tau & (\tilde u\notin\mathbb Z)\\
	-\{u\}_\tau & (\tilde u\in\mathbb Z),
	\end{cases}
\end{equation}
where we have defined $\tilde u,\check u\in\mathbb R$ as 
\begin{equation}
	u=\tilde u+\check u\tau.\label{u:component}
\end{equation}
Note that, for a real number $x$, a $\tau$-modded value $\{x\}_\tau$ reduces to a normal modded value $\{x\}$ defined as 
\begin{equation}
	\{x\}\equiv x-\lfloor x\rfloor\quad(x\in\mathbb R).\label{modded}
\end{equation}

Now, the elliptic theta function $\theta_0(u;\tau)$ can be approximated for a small $|\tau|$ with fixed $0<\arg\tau<\pi$ as
\begin{equation}
\begin{split}
	\log\theta_0(u;\tau)&=\fft{\pi i}{\tau}\{u\}_\tau(1-\{u\}_\tau)+\pi i\{u\}_\tau-\fft{\pi i}{6\tau}(1+3\tau+\tau^2)\\
	&\quad+\log(1-e^{-\fft{2\pi i}{\tau}(1-\{u\}_\tau)})\left(1-e^{-\fft{2\pi i}{\tau}\{u\}_\tau}\right)+\mathcal O(e^{-\fft{2\pi\sin(\arg\tau)}{|\tau|}}),\label{elliptic:theta:0:asymp}
\end{split}
\end{equation}
based on an alternative product form of $\theta_0(u;\tau)$:
\begin{equation}
\begin{split}
	\theta_0(u;\tau)&=-ie^{-\fft{\pi i}{6}(\tau+\fft{1}{\tau})}e^{\fft{\pi i}{\tau}\{u\}_\tau(1-\{u\}_\tau)}e^{\pi i\{u\}_\tau}\\
	&\quad\times\prod_{k=1}^\infty(1-e^{-\fft{2\pi i}{\tau}(k-\{u\}_\tau)})(1-e^{-\fft{2\pi i}{\tau}(k-1+\{u\}_\tau)}).
\end{split}
\end{equation}
This product form can be derived by combining (\ref{eq:theta:0}) with the $S$-transformation (\ref{theta:0:S}) and following the definition (\ref{tau-modded}). 

Similarly, the elliptic theta function $\theta_1(u;\tau)$ is approximated for a small $|\tau|$ with fixed $0<\arg\tau<\pi$ as
\begin{equation}
\begin{split}
	\log\theta_1(u;\tau)&=\fft{\pi i}{\tau}\{u\}_\tau(1-\{u\}_\tau)-\fft{\pi i}{4\tau}(1-\tau)+\pi i\lfloor\Re u-\cot(\arg\tau)\Im u\rfloor-\fft12\log\tau\\
	&\quad+\log(1-e^{-\fft{2\pi i}{\tau}(1-\{u\}_\tau)})\left(1-e^{-\fft{2\pi i}{\tau}\{u\}_\tau}\right)+\mathcal O(e^{\fft{2\pi\sin(\arg\tau)}{|\tau|}}),\label{elliptic:theta:1:asymp}
\end{split}
\end{equation}
based on an alternative product form of $\theta_1(u;\tau)$:
\begin{equation}
\begin{split}
	\theta_1(u;\tau)&=(-i\tau)^{-\fft12}e^{-\fft{\pi i}{4\tau}}e^{\pi i\lfloor\Re u-\cot(\arg\tau)\Im u\rfloor}e^{\fft{\pi
	i}{\tau}\{u\}_\tau(1-\{u\}_\tau)}\\
	&\quad\times\prod_{k=1}^\infty(1-e^{-\fft{2\pi
	i}{\tau}k})(1-e^{-\fft{2\pi i}{\tau}(k-\{u\}_\tau)})(1-e^{-\fft{2\pi
	i}{\tau}(k-1+\{u\}_\tau)}).
\end{split}
\end{equation}
This product form can be derived by combining (\ref{eq:theta:1}) with the $S$-transformation (\ref{theta:1:S}) and following the definition (\ref{tau-modded}). 

For a small $|\tau|$ with fixed $0<\arg\tau<\pi$, the elliptic gamma function can be approximated as
\begin{equation}
\begin{split}
	\log\widetilde\Gamma(u;\tau)&=2\pi i\,Q(\{u\}_\tau;\tau)+\mathcal O(|\tau|^{-1}e^{-\fft{2\pi\sin(\arg\tau)}{|\tau|}\min(\{\tilde u\},1-\{\tilde u\})}),\label{elliptic:Gamma:asymp}
\end{split}
\end{equation}
provided $\tilde u\not\to\mathbb Z$ (see \cite{ArabiArdehali:2019tdm} for example). See (\ref{eq:Q:Gamma}) or (\ref{eq:Q:Gamma:App}) for the definition of $Q(\cdot;\cdot)$.

%%%%%
\section{Contribution from \texorpdfstring{$C$-center}{C-center} solutions}\label{App:C-center}
%%%%%
In this Appendix we repeat the same procedures in \ref{Sec:Saddle:N=4} and \ref{Sec:BA:N=4} for a more general class of saddle point solutions and the BA solutions respectively. Both solutions are denoted by a \emph{finite}, positive divisor of $N$, namely $C$, and the solution with $C=1$ corresponds to what we have discussed in the main text.

The final results of this Appendix, namely (\ref{eq:index:2:saddle:finite:C:3}) and (\ref{eq:index:BA:C}), are consistent with each other for the first three terms. The remaining pure imaginary or order $\mathcal O(N^0)$ terms do not match apparently: more detailed analysis on contour deformations in the saddle point approach and on the Jacobian contribution in the Bethe Ansatz approach (see section 4.2 of \cite{Hong:2018viz} for example) would be required for a perfect match and we leave it for future research.

Another important implication of this Appendix is that 3D SU($N$) Chern-Simons theory arises from $\mathcal N=4$ SU($N$) SYM on $S^1\times S^3$ in the Cardy-like limit \emph{independently} of saddle point solutions. In the main text we have observed it for a particular saddle point  (\ref{eq:saddle:ansatz:finite}) and the following section~\ref{App:C-center:saddle} will generalize this result to $C$-center saddle points (\ref{eq:saddle:ansatz:C:finite}).

Lastly, it is worth highlighting the robustness of the universal $\log N$ term. We will demonstrate that these $C$-center solutions, which can be dominant in certain domain of chemical potentials $\Delta_a$, still contribute $\log (\frac{N}{C})$ to the SCI which is compatible with the result in the main body of the manuscript.

%%%%%
\subsection{Saddle point approach}\label{App:C-center:saddle}
%%%%%
In \ref{Cardy:finite}, we have investigated the contribution from a particular saddle point ansatz (\ref{eq:saddle:ansatz:finite}) to the index through the saddle point approximation (\ref{eq:index:1:saddle}). Here we repeat the same procedure but with a more general ansatz for $C$-center solutions \cite{ArabiArdehali:2019orz}, namely
\begin{equation}
	\hat u^{(C,m)}=\left\{u_j^{(C,m)}=\fft{m}{N}+\fft{\lfloor\fft{j-1}{N/C}\rfloor-\fft{C-1}{2}}{C}+v_j\tau\,\bigg|\,v_j\sim\mathcal O(|\tau|^0),~\sum_{j=1}^Nv_j=0\right\}\label{eq:saddle:ansatz:C:finite}
\end{equation}
with $m\in\{0,1,\cdots,\fft{N}{C}-1\}$. The range of $m$ is for the integration contour deformed from (\ref{contour}) as
\begin{equation}
	\bigcup_{\mu=1}^{N-1}(v_\mu\tau-\fft{1}{2N}-\fft{C-1}{2C},\,v_\mu\tau+1-\fft{1}{2N}-\fft{C-1}{2C}],\label{contour:deformed:C}
\end{equation}
which passes through the above saddle point $\hat u^{(C,m)}$. The $C$-center solution ansatz (\ref{eq:saddle:ansatz:C:finite}) and the corresponding deformed contour (\ref{contour:deformed:C}) reduce to the ones in the main text (\ref{eq:saddle:ansatz:finite}) and (\ref{contour:deformed}) respectively for $C=1$.

In the strict Cardy-like limit $|\tau|\to0$, the $C$-center solution ansatz (\ref{eq:saddle:ansatz:C:finite}) reduces to $C$ groups of holonomies, where each group has equal number\,($N/C$) of condensed holonomies and is separated from adjacent groups by $1/C$ along the domain $(0,1]$ with 0 identified with 1. The name `$C$-center' comes from its symmetry breaking pattern $\mathbb Z_N\to\mathbb Z_C$ \cite{ArabiArdehali:2019orz}.

Following \ref{Cardy:finite} and using the following identity of Bernoulli polynomials
\begin{equation}
	\sum_{J=0}^{C-1}B_n(\{\fft{J}{C}+u\}_\tau)=\fft{1}{C^{n-1}}B_n(\{Cu\}_\tau)\quad(u\in\mathbb C),\label{Bernoulli:identity}
\end{equation}
we simplify the effective action (\ref{eq:Seff:2:finite}) \underline{near} the $C$-center Ansatz (\ref{eq:saddle:ansatz:C:finite}) up to exponentially suppressed terms as
\begin{equation}
\begin{split}
	N^2S_\text{eff}(\hat u;\Delta,\tau)&\sim\sum_{I=0}^{C-1}\left(-\fft{\pi i\eta_CN}{C\tau^2}\sum_{i=1}^{N/C}(u_{I,i}-\bar u_I)^2+\sum_{i\neq j}^{N/C}\log(2\sin\fft{\pi(u_{I,i}-u_{I,j})}{\tau})\right)\\
	&\quad-\fft{\pi i}{2\tau^2}\fft{N^2}{C^2}\sum_{I,J=0}^{C-1}\xi_{I-J}(\bar u_{IJ})^2-\fft{\pi iN^2}{C^3\tau^2}\prod_{a=1}^3\left(\{C\Delta_a\}_\tau-\fft{1+\eta_C}{2}\right)\\
	&\quad+\fft{\pi i}{\tau^2}\prod_{a=1}^3\left(\{\Delta_a\}_\tau-\fft{1+\eta_1}{2}\right)-\fft{5\pi i\eta_CN^2}{12C}+\fft{\pi iN}{2}-\fft{\pi i(6-5\eta_1)}{12}\\
	&\quad-(N-1)\log\tau,\label{eq:Seff:2:finite:near-ansatz:C}
\end{split}
\end{equation}
where we have introduced $u_{I,i}$ and $\bar u_I$ as
\begin{equation}
\begin{split}
	u_i&=\fft{m}{N}+\fft{I-\fft{C-1}{2}}{C}+u_{I,i-(N/C)I}\quad(I=\left\lfloor\fft{i-1}{N/C}\right\rfloor,~i=1,\cdots,N),\\
	\bar u_I&=\fft{1}{N/C}\sum_{i=1}^{N/C}u_{I,i}.\label{C:grouping}
\end{split}
\end{equation}
Note that $\sum_{I=0}^{C-1}\bar u_I=0$ from the SU($N$) constraint. We have also defined $\xi_J$ and $\eta_C$ as
\begin{equation}
\begin{split}
	\sum_{a=1}^3\{\fft{J}{C}+\Delta_a\}_\tau&=2\tau+\fft{3+\xi_J}{2},\\
	\sum_{a=1}^3\{C\Delta_a\}_\tau&=2C\tau+\fft{3+\eta_C}{2},\label{eq:eta:C}
\end{split}
\end{equation}
which are related with each other as $\eta_C=\sum_{J=0}^{C-1}\xi_J\in\{\pm1\}$ under the assumption $\{C\tilde\Delta_a\}\neq0$. Note that $\eta_1$ is equivalent to the $\eta$ introduced in the main text (\ref{eq:eta}).    

Substituting the effective action (\ref{eq:Seff:2:finite:near-ansatz:C}) into the saddle point approximation (\ref{eq:index:1:saddle}) gives the SCI as
\begin{equation}
\begin{split}
	\mathcal I(\tau; \Delta)&\sim\sum_{m=0}^{N/C-1}\fft{\mathcal A}{((N/C)!)^C}\int_{D_{\hat u^{(C,m)}}}\prod_{\mu=1}^{N-1}du_\mu\,e^{N^2S_\text{eff,$u$-dept}(\hat u;\Delta_a,\tau)}\\
	&\quad+(\text{contribution from the other saddles}),\label{eq:index:2:saddle:finite:C:1}
\end{split}
\end{equation}
where $D_{\hat u^{(C,m)}}$ denotes a small neighborhood of a saddle point solution $\hat u^{(C,m)}$ (\ref{eq:saddle:ansatz:C:finite}) on the contour (\ref{contour:deformed:C}), namely
\begin{equation}
	\bigcup_{\mu=1}^{N-1}(v_\mu\tau+\fft{m}{N}+\fft{\lfloor\fft{\mu-1}{N/C}\rfloor-\fft{C-1}{2}}{C}-\epsilon,v_\mu\tau+\fft{m}{N}+\fft{\lfloor\fft{\mu-1}{N/C}\rfloor-\fft{C-1}{2}}{C}+\epsilon]\label{contour:saddle:C:u}
\end{equation}
with some small positive number $\epsilon$. The $u$-dependent part of the effective action, namely $N^2S_\text{eff,$u$-dept}(\hat u;\Delta,\tau)$, denotes the first three terms of (\ref{eq:Seff:2:finite:near-ansatz:C}) and the prefactor $\mathcal A$ is related to the remaining $u$-independent part of (\ref{eq:Seff:2:finite:near-ansatz:C}) as
\begin{equation}
\begin{split}
	\mathcal A&=\text{exp}\left[-\fft{\pi iN^2}{C^3\tau^2}\prod_{a=1}^3\left(\{C\Delta_a\}_\tau-\fft{1+\eta_C}{2}\right)+\fft{\pi i}{\tau^2}\prod_{a=1}^3\left(\{\Delta_a\}_\tau-\fft{1+\eta_1}{2}\right)\right.\\
	&\kern3em~\left.-\fft{5\pi i\eta_CN^2}{12C}+\fft{\pi iN}{2}-\fft{\pi i(6-5\eta_1)}{12}-(N-1)\log\tau\right].\label{eq:C:A}
\end{split}
\end{equation}
Note that we have $((N/C)!)^C$ instead of the original $N!$ in the denominator of (\ref{eq:index:2:saddle:finite:C:1}) taking an extra factor of $\fft{N!}{((N/C)!)^C}$ into account, which corresponds to the number of distributing $N$ holonomies into $C$ groups as (\ref{C:grouping}).

Introducing new integration variables $\lambda_{I,i}$ and $\bar\lambda_I$ as
\begin{equation}
\begin{split}
	-i\lambda_{I,i}\tau&=u_{I,i}-\bar u_I\qquad(I=0,\cdots,C-1~\text{and}~i=1,\cdots,N/C-1),\\
	-i\bar\lambda_I\tau&=\bar u_I\qquad(I=0,\cdots,C-2),
\end{split}
\end{equation}
whose Jacobian is given as
\begin{equation}
	\left|\fft{\partial(u_1,\cdots,u_{N-1})}{\partial(\lambda_{0,1},\cdots,\lambda_{0,N/C-1},\bar\lambda_0,\lambda_{1,1},\cdots,\lambda_{C-1,N/C-1})}\right|=e^{-\fft{\pi i(N-1)}{2}}\left(\fft{N}{C}\right)^{C-1}\tau^{N-1},
\end{equation}
the index (\ref{eq:index:2:saddle:finite:C:1}) can be rewritten as
\begin{equation}
\begin{split}
	\mathcal I(\tau; \Delta)&=\fft{N}{C}\tau^{N-1}\mathcal A\,e^{-\fft{\pi i(N^2/C-1)}{2}}\left(Z_{\text{SU}(N/C)}^{CS}\right)^C\int\prod_{I=0}^{C-2}d\bar\lambda_I\,e^{\fft{\pi i}{2}\sum_{I,J=0}^{C-1}\xi_{I-J}(\bar\lambda_{IJ})^2}\\
	&\quad+(\text{contribution from other saddles}).\label{eq:index:2:saddle:finite:C:2}
\end{split}
\end{equation}
Here we have assumed smooth deformations of contours as we have done from (\ref{contour:saddle:lambda}) to real lines in the main text. Note that the original SU($N$) group breaks down into $C$ copies of SU($N/C$) groups and the remaining $C-1$ copies of U(1) groups: accordingly, we obtained $C$ copies of the SU($N/C$) Chern-Simons partition function together with the extra $(C-1)$-dimensional integral for U(1) terms. We denote the latter simply as $Z_{\text{U}(1)\text{'s}}$. Finally, substituting the partition function of SU($N$) CS theory (\ref{eq:Z:SU(N)}) with $N\to N/C$ into (\ref{eq:index:2:saddle:finite:C:2}) gives
\begin{equation}
\begin{split}
	\mathcal I(\tau; \Delta)&=\fft{N}{C}\exp\left[-\fft{\pi iN^2}{C^3\tau^2}\prod_{a=1}^3\left(\{C\Delta_a\}_\tau-\fft{1+\eta_C}{2}\right)+\fft{\pi i}{\tau^2}\prod_{a=1}^3\left(\{\Delta_a\}_\tau-\fft{1+\eta_1}{2}\right)\right.\\
	&\kern4em~~\left.+\fft{5\pi i(\eta_1-C\eta_C)}{12}\right]\times Z_{\text{U}(1)\text{'s}}\\
	&\quad+(\text{contribution from other saddles}).\label{eq:index:2:saddle:finite:C:3}
\end{split}
\end{equation}
%

%%%%%
\subsection{Bethe Ansatz approach}\label{App:C-center:BA}
%%%%%
In \ref{Sec:BA:N=4}, we have investigated the contribution from basic solutions  (\ref{BAE:sol:basic:many}) to the index through the Bethe Ansatz formula (\ref{eq:index:BA}). Here we generalize it with a larger set of solutions denoted by a positive divisor of $N$, namely $C$, as
\begin{equation}
\begin{split}
	\hat u_C&=\left\{u_i=\bar u+\fft{I}{C}+\fft{i-(N/C)I}{N/C}\tau+m_i+n_i\tau\,\Big|\,I=\left\lfloor\fft{i}{N/C}\right\rfloor,~i=1,\cdots,N\right\}\label{BAE:sol:C:many}
\end{split}
\end{equation}
with arbitrary integers $m_i$'s and $n_i$'s. Note that this solution is equivalent to the $\{C,N/C,0\}$ solution in \cite{Hong:2018viz} and the $(C,N/C)$ saddle in \cite{Cabo-Bizet:2019eaf}. Since the calculation is parallel to the one in the main text, we summarize the key intermediate results only.

First, the degeneracy gives $\log N!+\log N$ by the same token we discussed in the beginning of \ref{Sec:BA:N=4}. The prefactor contribution also remains the same as (\ref{eq:kappa}). Calculating the contribution from $\log\mathcal Z(\hat u_C;\Delta,\tau)$ is more involved but does not require extra techniques other than using (\ref{eq:gamma:2}) and (\ref{Bernoulli:identity}). The result is given as
\begin{equation}
\begin{split}
	\log\mathcal Z(\hat u_C;\Delta,\tau)&=C\sum_{J=0}^{C-1}\sum_{i,j=0}^{N/C-1}\sum_{a=1}^3\log\tilde\Gamma(\fft{i-j}{N/C}\tau+\fft{J}{C}+\Delta_a;\tau)-N\sum_{a=1}^3\log\tilde\Gamma(\Delta_a;\tau)\\
	&\quad-C\sum_{J=1}^{C-1}\sum_{i,j=0}^{N/C-1}\log\tilde\Gamma(\fft{i-j}{N/C}\tau+\fft{J}{C};\tau)-C\sum_{i,j=0\,(i\neq j)}^{N/C-1}\log\tilde\Gamma(\fft{i-j}{N/C}\tau;\tau)\\
	&=-\fft{\pi iN^2}{C^3\tau^2}\prod_{a=1}^3\left(\{C\Delta_a\}_\tau-\fft{1+\eta_C}{2}\right)+\fft{\pi i(1-\eta_C)N^2}{2C}+\fft{\pi i(1-3\tau+\tau^2)N}{6\tau}\\
	&\quad+\fft{\pi i\eta_CC}{12}-N\sum_{a=1}^3\log\tilde\Gamma(\Delta_a;\tau)+N\log\fft{N}{C}-2N\log(\tilde q;\tilde q)_{\infty}\\
	&\quad+\mathcal O(Ne^{-\fft{2\pi N\sin(\arg\tau)}{|\tau|}\min(\{\fft{J}{C}+\tilde\Delta_a\},1-\{\fft{J}{C}+\tilde\Delta_a\}|\,J=0,1,\cdots,C-1)}),\\
\end{split}
\end{equation}
where we have introduced the quantities $\eta_C$ as in (\ref{eq:eta:C}). The contribution from the Jacobian $-\log H(\hat u_C;\Delta,\tau)$ can also be obtained by following the procedure in \ref{Sec:BA:N=4} and the detailed discussion on the Jacobian matrix in \cite{Hong:2018viz} as
\begin{equation}
\begin{split}
	-\log H(\hat u_C;\Delta,\tau)&=-\log N-(N-1)\log(\fft{i}{\pi}\sum_{\Delta}\partial_\Delta\log\theta_1(C\Delta;\fft{\tau}{N/C^2}))-\log\det(I_{N-1}+\tilde H)\\
	&=-N\log N+(N-1)\log\fft{C\tau}{\eta_C}-\log\det(I_{N-1}+\tilde H)\\
	&\quad+\mathcal O(e^{-\fft{2\pi N\sin(\arg\tau)}{|\tau|}\min(\{\fft{J}{C}+\tilde\Delta_a\},1-\{\fft{J}{C}+\tilde\Delta_a\}|\,J=0,1,\cdots,C-1)}).
\end{split}
\end{equation}

Substituting all the contributions to the BA formula (\ref{eq:index:BA}) and using (\ref{pochhammer:S}), finally we have the contribution from the BA solutions (\ref{BAE:sol:C:many}):
\begin{equation}
\begin{split}
	\log\mathcal I(\tau; \Delta)\big|_{\{C,N/C,0\}}&=-\fft{\pi iN^2}{C^3\tau^2}\prod_{a=1}^3\left(\{C\Delta_a\}_\tau-\fft{1+\eta_C}{2}\right)+\log\fft{N}{C}+\fft{\pi i(1-\eta_C)N^2}{2C}\\
	&\quad+\fft{\pi i\eta_CC}{12}-\fft{\pi i(1-\eta_C)(N-1)}{2}\\
	&\quad-\sum_{a=1}^3\log\tilde\Gamma(\Delta_a;\tau)-2\log(q;q)_{\infty}-\log\tau-\log\det(I_{N-1}+\tilde H)\\
	&\quad+\mathcal O(Ne^{-\fft{2\pi N\sin(\arg\tau)}{|\tau|}\min(\{\fft{J}{C}+\tilde\Delta_a\},1-\{\fft{J}{C}+\tilde\Delta_a\}|\,J=0,1,\cdots,C-1)}).
\end{split}
\end{equation}
In the Cardy-like limit that imposes $|\tau|\ll1$, this reduces to
\begin{equation}
\begin{split}
	\log\mathcal I(\tau; \Delta)\big|_{\{C,N/C,0\}}&\sim-\fft{\pi iN^2}{C^3\tau^2}\prod_{a=1}^3\left(\{C\Delta_a\}_\tau-\fft{1+\eta_C}{2}\right)+\fft{\pi i}{\tau^2}\prod_{a=1}^3\left(\{\Delta_a\}_\tau-\fft{1+\eta_1}{2}\right)\\
	&\quad+\log\fft{N}{C}+\fft{\pi i(1-\eta_C)N^2}{2C}-\fft{\pi i(1-\eta_C)(N-1)}{2}+\fft{\pi i\eta_CC}{12}\\
	&\quad-\fft{\pi i(6-5\eta_1)}{12}-\log\det(I_{N-1}+\tilde H).\label{eq:index:BA:C}
\end{split}
\end{equation}
up to exponentially suppressed terms.

%%%%%
\section{Saddle point solutions of 3D Chern-Simons theory}\label{App:saddle}
%%%%%
In this Appendix we investigate the saddle point equation (\ref{eq:saddle:2:finite}) from the effective action of the $\mathcal N=4$ SU($N$) SYM theory in the Cardy-like expansion (\ref{eq:Seff:2:finite:near-ansatz}), namely
\begin{equation}
	i\eta\,v_j=\fft{1}{N}\sum_{k=1\,(\neq j)}^N\cot\pi v_{jk}\quad(i=1,\cdots,N).
\end{equation}
This equation is in fact equivalent to the saddle point equation of 3D Chern-Simons theory with a 't~Hooft coupling $t$ \cite{Aganagic:2002wv,Halmagyi:2003ze},
\begin{equation}
	\fft{1}{t}u_j=\fft1N\sum_{k=1\,(\neq j)}^N\coth\fft{u_{jk}}{2},\label{xxeq:saddle:CS}
\end{equation}
under $v_j\to {i}u_j/{2\pi}$ and $t={2\pi i}/{\eta}$. We solve this saddle point equation in the planar limit, or equivalently in the large-$N$ limit.

The partition function of 3D Chern-Simons theory on $S^3$ can be written as \cite{Aganagic:2002wv,Halmagyi:2003ze}
\begin{equation}
    Z=\fft1{N!}\int\prod_idu_i\prod_{i<j}\left(2\sinh\fft{u_{ij}}2\right)^2\exp\left(-\fft1{2g_s}\sum_iu_i^2\right),
\label{eq:3DCS}
\end{equation}
where $g_s=2\pi i/\hat k$ and $\hat k$ is the effective Chern-Simons level.  As we have seen in (\ref{eq:Seff:2:finite:near-ansatz}), the fluctuations around the dominant saddle point of the $\mathcal N=4$ SYM theory are described by such a Chern-Simons theory, provide we make the identification $t=2\pi i/\eta$ where $t=g_sN$ is the `t~Hooft coupling.  Although this partition function can be evaluated directly \cite{Kapustin:2009kz}, as detailed in Appendix~\ref{App:CS:partition:function}, it is important to note that our starting point is a saddle point evaluation of the $\mathcal N=4$ SYM index.  Hence, in principle, we should seek a saddle point evaluation of the 3D Chern-Simons partition function.  As we demonstrate in this Appendix, the saddle point result coincides with the exact partition function in the large-$N$ limit, so in practice this distinction is immaterial.  However, we highlight an interesting observation that there are, in fact, multiple saddle point solutions to the Chern-Simons model and that it is important to properly identify the dominant saddle in order to find agreement.

%%%%%
\subsection{The dominant saddle point}\label{App:saddle:lead}
%%%%%

The saddle point equation obtained by varying the action in (\ref{eq:3DCS}) takes the form
\begin{equation}
	\fft{1}{t}u_j=\fft1N\sum_{k=1\,(\neq j)}^N\coth\fft{u_{jk}}{2}.
\label{eq:saddle:CS}
\end{equation}
As in \cite{Marino:2011nm}, it is convenient to introduce the exponentiated eigenvalues $X_j=e^{u_j}$, so that the saddle point equation becomes
\begin{equation}
	\log X_j=\fft{t}{N}\sum_{k=1\,(\neq j)}^N\left(-1+\fft{2X_j}{X_j-X_k}\right).
\label{eq:saddle:CS:exp}
\end{equation}
As usual, in the large-$N$ limit, we assume the eigenvalues condense along a single cut, $x\in[a,b]$ on the real axis, provided the `t~Hooft parameter $t$ is real.  (Later on we will analytically continue to complex $t$.)  We then introduce the density of eigenvalues $\rho(x)$ such that
\begin{equation}
    \sum_i f(x_i)\quad\longrightarrow\quad N\int_a^bdx\,\rho(x)f(x).
\end{equation}

The important properties of the matrix partition function are now encoded in the eigenvalue density.  In order to determine $\rho(x)$, we introduce the resolvent
\begin{equation}
	\omega(X)\equiv-t+2t\int_a^bdy\,\rho(y)\fft{X}{X-Y}\qquad(X\in\mathbb C\setminus\mathcal C).
\label{eq:omega}
\end{equation}
This function is analytic in the complex $X$ plane except for a cut $\mathcal C$ from $[e^a,e^b]$ on the positive real axis.  By studying $\omega(X)$ on both sides of the cut, we can reproduce the saddle point equation
\begin{equation}
    \log X=\fft12[\omega_+(X)+\omega_-(X)]\qquad(X\in\mathcal C),
\label{eq:saddle:omega}
\end{equation}
and also recover the eigenvalue density
\begin{equation}
    \rho(x)=-\fft1{4\pi it}	[\omega_+(X)-\omega_-(X)]\qquad(X\in\mathcal C).
\label{eq:discontinuity}
\end{equation}
Here we have defined
\begin{equation}
    \omega_\pm(X)=\omega(e^{x\pm i\epsilon})=\omega(X\pm i\epsilon)\qquad(X\in\mathcal C).
\end{equation}

Following \cite{Marino:2011nm}, we can use the following trick to derive the resolvent $\omega(X)$. Recall that $\omega(X)$ is analytic on $X\in\mathbb C\setminus\mathcal C$. Then it is straightforward to check that the function $g(X)$ defined as
\begin{equation}
	g(X)\equiv e^{\omega(X)/2}+Xe^{-\omega(X)/2}\qquad(X\in\mathbb C\setminus\mathcal C),
\label{eq:g(X)}
\end{equation}
can be analytically continued to the entire complex plane including $\mathcal C$ since
\begin{equation}
	g_+(X)=e^{\omega_+(X)/2}+Xe^{-\omega_+(X)/2}\\
	=Xe^{-\omega_-(X)/2}+e^{\omega_-(X)/2}=g_-(X)\quad(X\in\mathcal C),
\end{equation}
where the middle equality corresponds to the saddle point equation, (\ref{eq:saddle:omega}). Furthermore, using the asymptotic behavior of (\ref{eq:omega})
\begin{equation}
	\lim_{X\to0}\omega(X)=-t,\qquad \lim_{|X|\to\infty}\omega(X)=t,
\label{omega:asymptotic}
\end{equation}
we deduce the form of $g(X)$ as
\begin{equation}
	g(X)=e^{-t/2}(X+1)\qquad(X\in\mathbb C).
\label{eq:g(X):1}
\end{equation}
Substituting this into (\ref{eq:g(X)}) then gives
\begin{equation}
	e^{\omega(X)/2}=\fft12\left(g(X)\pm\sqrt{g(X)^2-4X}\right).
\label{eq:omega:g}
\end{equation}
Consistency of this solution demands that the branch cut of the square root is along $\mathcal C$.  In particular, note that the branch points of the square root are given by
\begin{equation}
    X_\pm=2e^t-1\pm2(e^{2t}-e^t)^\fft12,
\label{eq:endpoints}
\end{equation}
with the product $X_+X_-=1$.

%%%%%
\subsubsection{The solution for \texorpdfstring{$t>0$}{t>0}}\label{App:saddle:lead:t>0}
%%%%%

Although we have assumed that the eigenvalues condense along the real line, the endpoints $X_\pm$ are only real for real $t>0$.  Assuming this to be the case, the resolvent (\ref{eq:omega:g}) can be written as
\begin{equation}
	e^{\omega(X)/2}=\fft12\left(e^{-t/2}(X+1)-e^{-t/2}(X-X_+)^\fft12(X-X_-)^\fft12\right),
\label{eq:omega:lead:t>0}
\end{equation}
where the principal branch is taken for both square roots.  The eigenvalue density can then be recovered from the discontinuity across the cut using (\ref{eq:discontinuity}), with the result \cite{Marino:2011nm}
\begin{equation}
	\rho(x)=\fft{1}{\pi t}\tan^{-1}\fft{\sqrt{e^t-\cosh^2\fft{x}{2}}}{\cosh\fft{x}{2}}\quad(x\in[a,b]),
\label{eq:rho:1}
\end{equation}
where the endpoints are given by $-a=b=2\cosh^{-1}(e^{t/2})$.

Substituting this eigenvalue density into the saddle point action is non-trivial, but can be shown to give the genus-zero free energy (see \textit{e.g.}\ Appendix~A of \cite{Halmagyi:2003ze})
\begin{equation}
    \log Z=N^2\left(\frac{\zeta(3)-\Li_3(e^{-t})}{t^2}+\fft{t}6-\fft{\pi^2}{6t}\right)+o(N^2).
\label{eq:ZCSF}
\end{equation}
This has a simple expansion in the large-$t$ limit
\begin{equation}
    \log Z/N^2\sim\fft{t}6-\fft{\pi^2}{6t}+\fft{\zeta(3)}{t^2}+\mathcal O(e^{-t})\qquad(t\gg1),
\end{equation}
but remains valid for real $t>0$.  For small $t$, it has an expansion
\begin{equation}
    \log Z/N^2=\fft12\log t-\fft34+\fft{t}{12}+\fft{t^2}{288}+\cdots\qquad(t\to0^+),
\end{equation}
which diverges logarithmically as $t\to0$.

%%%%%
\subsubsection{The solution for \texorpdfstring{$t={2\pi i}/{\eta}$}{t=2Pi I/eta} with \texorpdfstring{$\eta=\pm1$}{eta = +/- 1}}\label{App:saddle:lead:t=2PiI}
%%%%%

While we have worked with real $t$ above, in order to connect to $\mathcal N=4$ SYM, we want to analytically continue to a purely imaginary value $t={2\pi i}/{\eta}$ where $\eta=\pm1$.  However, this continuation is subtle, since $\eta=\pm1$ turns out to be the endpoints of a singular region of the Chern-Simons matrix model.  In particular, there is a divergence for $t={2\pi i}/{\eta}$ with $-1<\eta<1$ \cite{Morita:2011cs}.  This subtlety can also be seen by noting that the endpoints of the cut,  $X_\pm$ in (\ref{eq:endpoints}), collapse to $X_\pm=1$ when $\eta=\pm1$.

%%%%%
\begin{figure}[t]
	\centering
	\includegraphics[scale=.65]{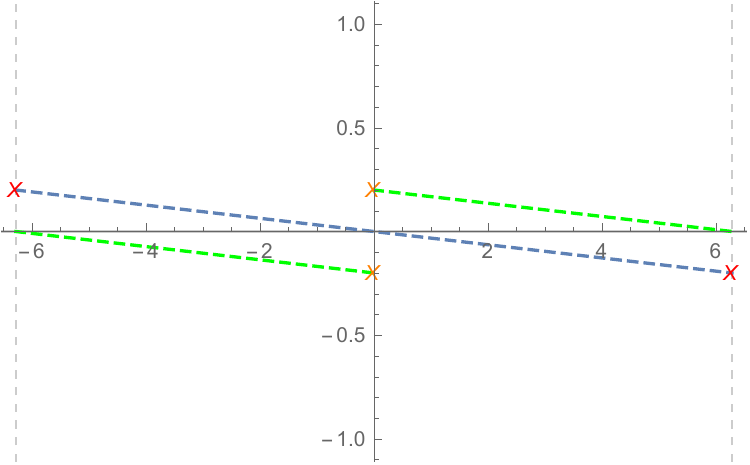}
	\caption{Orange (red) crosses are branch points and green (blue) lines are branch cuts of $h_+(x)$ and $h_-(x)$, respectively. Here we chose $\epsilon={1}/{10}$ for presentation. \label{Branch-cuts}}
\end{figure}
%%%%%

To avoid this singularity issue for $\eta=\pm1$, we take $t={2\pi i}/{\eta}+\epsilon^2$ where $\epsilon$ is a small positive number.  Although we have assumed real $t$ above, it was not strictly needed in order to obtain the resolvent (\ref{eq:omega:g}).  We thus start from there and analytically continue to imaginary eigenvalues, $x\to ix$.  In particular, we take $X=e^{ix}$, in which case the resolvent takes the form
\begin{equation}
	e^{\omega(X)/2}=\fft12\left(e^{-t/2}(e^{ix}+1)+(e^{-t/2}(e^{ix}+1)+2e^{ix/2})^\fft12(e^{-t/2}(e^{ix}+1)-2e^{ix/2})^\fft12\right).
\label{eq:omega:lead:t=2PiI:1}
\end{equation}
For $t=\pm{2\pi i}+\epsilon^2$, the square root factors have the following branch cuts:
\begin{equation}
\begin{split}
	h_+(x)\equiv(e^{-t/2}(e^{ix}+1)+2e^{ix/2})^\fft12&:~\bigcup_{n\in\mathbb Z}[(4n+2)\pi-x_\ast,(4n+2)\pi+x_\ast],\\
	h_-(x)\equiv(e^{-t/2}(e^{ix}+1)-2e^{ix/2})^\fft12&:~\bigcup_{n\in\mathbb Z}[4n\pi-x_\ast,4n\pi+x_\ast],
\end{split}
\end{equation}
where $x_\ast=2\pi-2i\epsilon+\mathcal O(\epsilon^2)$ (see Figure~\ref{Branch-cuts}).  Using\footnote{This Taylor expansion becomes subtle as $x\to2\pi\mathbb Z$ where the leading order vanishes. So we focus on the bulk and ignore this subtle issue near the endpoints $x=2\pi\mathbb Z$.}
\begin{equation}
	h_\pm(x)=(-(1\mp e^{ix/2})^2)^\fft12+\mathcal O(\epsilon^2),
\end{equation}
we can write down $h_\pm(x)$ more explicitly with the above specified branch cuts as
\begin{subequations}
	\begin{align}
	h_+(x)&=\begin{cases}
	\pm i(1-e^{ix/2})+\mathcal O(\epsilon^2) & (\text{above the cuts of }h_+(x))\\
	\mp i(1-e^{ix/2})+\mathcal O(\epsilon^2) & (\text{below the cuts of }h_+(x)),
	\end{cases}\\
	h_-(x)&=\begin{cases}
	\pm i(1+e^{ix/2})+\mathcal O(\epsilon^2) & (\text{above the cuts of }h_-(x))\\
	\mp i(1+e^{ix/2})+\mathcal O(\epsilon^2) & (\text{below the cuts of }h_-(x)).
	\end{cases}
	\end{align}\label{eq:h}%
\end{subequations}

We now rewrite the resolvent (\ref{eq:omega:lead:t=2PiI:1}) using (\ref{eq:h}) within the strip $\Re x\in(-2\pi,2\pi)$ explicitly as
\begin{subequations}
	\begin{align}
	e^{\omega(X)/2}&=\begin{cases}
	-1+\mathcal O(\epsilon^2) & (\text{above the cuts of }h_\pm(x))\\
	-e^{ix}+\mathcal O(\epsilon^2) & (\text{between the cuts of }h_\pm(x))\\
	-1+\mathcal O(\epsilon^2) & (\text{below the cuts of }h_\pm(x)),
	\end{cases}\label{eq:omega:lead:t=2PiI:2}\\
	\to\quad\omega(X)&=\begin{cases}
	-\fft{2\pi i}{\eta}+\mathcal O(\epsilon^2) & (\text{above the cuts of }h_\pm(x))\\
	-\fft{2\pi i}{\eta}(1-\fft{x}{\pi})+\mathcal O(\epsilon^2) & (\text{between the cuts of }h_\pm(x),~\Re x\in[0,2\pi))\\
	\fft{2\pi i}{\eta}(1+\fft{x}{\pi})+\mathcal O(\epsilon^2) & (\text{between the cuts of }h_\pm(x),~\Re x\in(-2\pi,0))\\
	\fft{2\pi i}{\eta}+\mathcal O(\epsilon^2) & (\text{below the cuts of }h_\pm(x)).\end{cases}\label{eq:omega:lead:t=2PiI:3}
	\end{align}
\end{subequations}
Since (\ref{eq:omega:lead:t=2PiI:2}) determines $\omega(X)$ only up to $4\pi i\mathbb Z$, we have used the asymptotic conditions from (\ref{omega:asymptotic}) along with continuity outside of the branch cuts to fix $\omega(X)$.  Finally, the eigenvalue density can be obtained by substituting (\ref{eq:omega:lead:t=2PiI:3}) into (\ref{eq:discontinuity})
\begin{equation}
	\rho(x)=\begin{cases}
	\fft{1}{2\pi}\left(1-\fft{x}{2\pi}\right)+\mathcal O(\epsilon^2) & x\in[0,x_\ast)\\
	\fft{1}{2\pi}\left(1+\fft{x}{2\pi}\right)+\mathcal O(\epsilon^2) & x\in(-x_\ast,0).
	\end{cases}
\label{eq:rho:1:2PiI}
\end{equation}
Taking the limit $\epsilon\to0$ then gives the simple expression
\begin{equation}
    \rho(x)=\fft{1}{2\pi}\left(1-\fft{|x|}{2\pi}\right)\qquad x\in(-2\pi,2\pi),
\label{eq:rholeadim}
\end{equation}
which satisfies the normalization condition $\int_{-2\pi}^{2\pi}dx\,\rho(x)=1$ as expected.  Recall that, since we have analytically continued, the actual eigenvalues $u=ix$ are now distributed between $\pm2\pi i$ along the imaginary axis.

The genus-zero free energy can be obtained by evaluating the saddle point action
\begin{equation}
    S_{\mathrm{eff}}/N^2=\left[-\fft1{2t}\int dx\rho(x)u^2+\fft12\int\rho(x)\rho(\tilde x)dx\,d\tilde x\log\left(4\sinh^2\fft{u-\tilde u}2\right)^2\right]_{u=ix,\,\tilde u=i\tilde x}
\label{eq:CSSeff}
\end{equation}
on the solution given by (\ref{eq:rho:1:2PiI}).  Here some care must be taken in keeping the $\epsilon$ regulator while integrating the log term because of branch issues.  The result is simply
\begin{equation}
    \left.\log Z/N^2\right|_{t=\pm2\pi i}=\fft{5\pi i}{12}\eta,
\label{eq:domsad}
\end{equation}
which is purely imaginary.  This result can also be obtained directly by analytic continuation, namely by inserting $t=2\pi i/\eta$ into (\ref{eq:ZCSF}) but our careful analysis provides some direct insight into the structure of eigenvalues.

%%%%%
\subsection{The sub-leading saddle point}\label{App:saddle:sublead}
%%%%%

In deriving the resolvent, (\ref{eq:omega:g}), we assumed a one-cut solution with the cut extending along $[X_-,X_+]$.  The function $g(X)$ defined in (\ref{eq:g(X)}) is then argued to be analytic in the complex plane.  For $t>0$, this picture is evident as the cut is on the positive real axis in the $X$ plane.  However, for $t=\pm2\pi i$, the cut starts at $1-2\epsilon$, wraps twice along the unit circle, and ends at $1+2\epsilon$, where $\epsilon$ prevents the cut from overlapping with itself.

This picture of a cut wrapping twice around the unit circle in the $X$ plane suggests the possibility of another solution where the cut extends only once around the circle.  We have in fact identified such a solution where the cut starts at $X=-1$, goes around the circle, and ends again at $X=-1$.  What is special about this solution is that the double endpoint $X=-1$ may be singular, and this allows for $g(X)$ defined in (\ref{eq:g(X)}) to have a pole at $X=-1$. In particular, we find that
\begin{equation}
	g(X)=e^{-t/2}(X+1)+e^{t/2}\fft{X}{X+1}\qquad(X\in\mathbb C\setminus\{-1\}),
\label{eq:g:2}
\end{equation}
is consistent with analyticity except for a pole at $X=-1$.  The regular (first) term is identical to that of the standard solution, (\ref{eq:g(X):1}), while the pole (second) term is new but does not modify the asymptotic conditions (\ref{omega:asymptotic}).

%%%%%
\subsubsection{The solution for \texorpdfstring{$t>0$}{t>0}}\label{App:saddle:sublead:t>0}
%%%%%

\begin{figure}[t]
	\centering
	\includegraphics[scale=.6]{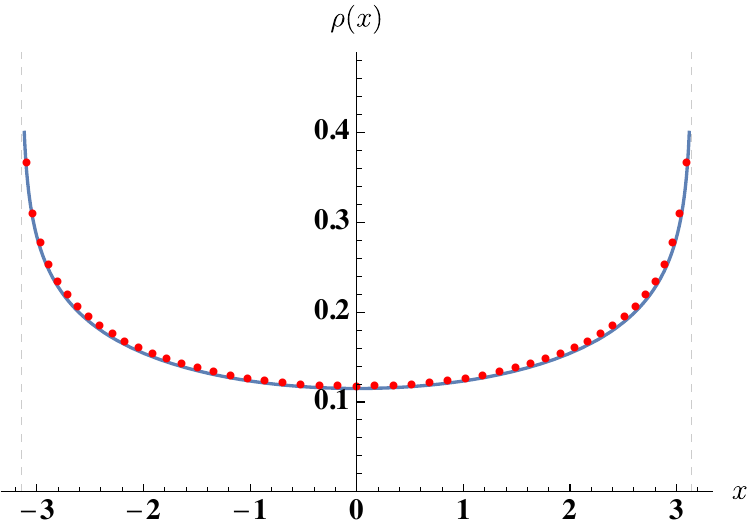}
	\caption{The numerically determined eigenvalue density, $\rho(x)$, for $N=50$ and $t=5$ (red dots) along with the large-$N$ analytic solution (blue line), (\ref{eq:rho:sublead:t>0}).  The numerical density is obtained by finite differencing.
\label{Subdominant-rho}}
\end{figure}

For $t>0$, we choose the cut to lie along the unit circle, starting and ending at the singular point $X=-1$.  Using (\ref{eq:omega:g}), we obtain the resolvent
\begin{equation}
	\omega(X)=\begin{cases}
	-t+2\log(1+X) & (|X|<1)\\
	t-2\log(1+1/X) & (|X|>1),
	\end{cases}
\label{eq:omega:sublead:t>0}
\end{equation}
where the principal branch is taken for the log.  Here the `inside' and `outside' solutions are chosen to satisfy the asymptotic conditions (\ref{omega:asymptotic}).  In this case, the matrix eigenvalues are imaginary and lie in the interval $(-i\pi,i\pi)$.  The eigenvalue density is obtained from (\ref{eq:discontinuity}), and is given by
\begin{equation}
	\rho(x)=\fft{1}{2\pi}\left(1-\fft1t\log(4\cos^2\fft{x}{2})\right)\qquad(x\in(-\pi,\pi)),
\label{eq:rho:sublead:t>0}
\end{equation}
and the eigenvalues themselves are $u=ix$.  Although the 't~Hooft coupling multiplies the log term, it averages to zero over the interval $(-\pi,\pi)$, so the normalization condition is satisfied with an average eigenvalue density of $1/2\pi$.  This sub-leading solution is somewhat unusual as $\rho(x)$ diverges logarithmically at the endpoints, as highlighted in Figure~\ref{Subdominant-rho}.

%%%%%
%%%%%

The genus-zero free energy can be obtained by using the above eigenvalue density in (\ref{eq:CSSeff}), with the result
\begin{equation}
    \log Z/N^2=\fft{\zeta(3)}{t^2}+\fft{\pi^2}{6t}+(\mbox{$t$-independent imaginary term}),
\label{eq:subleadF}
\end{equation}
where we have not been careful enough to keep track of the log branch issues that go into computing the imaginary term.  Note that, even though here we have taken real $t>0$, the saddle point free energy is complex since this sub-leading saddle itself is complex.

%%%%%
\subsubsection{The solution for \texorpdfstring{$t={2\pi i}/{\eta}$}{t=2 Pi I/eta} with \texorpdfstring{$\eta=\pm1$}{eta = +/- 1}}\label{App:saddle:sublead:t=2PiI}
%%%%%

\begin{figure}[t]
	\centering
	\includegraphics[scale=.6]{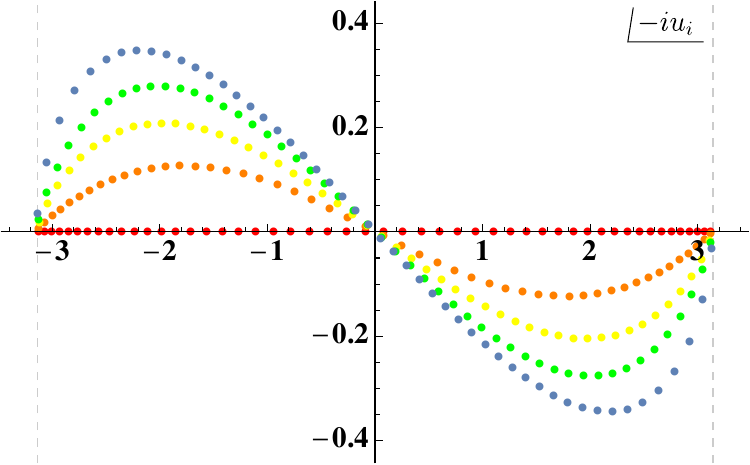}
	\caption{The numerically determined eigenvalues, $-iu_j$ for $N=50$.  The family of solutions correspond to $t=5\,(\text{red})$, $t=5+{\pi i}/{2}\,(\text{orange})$, $t=5+2\pi i\,(\text{yellow})$, $t=3+2\pi i\,(\text{green})$, and $t=2\pi i\,(\text{blue})$, respectively.
\label{Subdominant-saddle}}
\end{figure}

For connection to the $\mathcal N=4$ SYM saddle, we are interested in analytically continuing to $t={2\pi i}/{\eta}$ with $\eta=\pm1$.  While in the previous cases the eigenvalues either lie entirely on the real or imaginary axis, this is no longer the case for the sub-leading saddle with $t=2\pi i/\eta$.  Instead, from numerical observations, the eigenvalues lie along a curve connecting $u\in (-i\pi,i\pi)$.  We have been unable to obtain an analytic form of this curve.  However, it can be examined numerically, as shown in Figure~\ref{Subdominant-saddle}, where the 't~Hooft coupling is analytically continued from $t=5$ to $t=2\pi i$.

%%%%%
%%%%%

The genus-zero free energy for the sub-leading saddle with $t=\pm2\pi i$ may be obtained by analytic continuation of (\ref{eq:subleadF})
\begin{equation}
    \left.\log Z/N^2\right|_{t=\pm2\pi i}=-\fft{\zeta(3)}{4\pi^2}+(\mbox{imaginary}).
\end{equation}
Since this has a negative real part, it is always sub-dominant to the leading saddle whose free energy, (\ref{eq:domsad}), has vanishing real part.

%%%%%
\subsection{Saddle point solutions of \texorpdfstring{$\mathcal N=4$}{N=4} SYM from direct numerical evaluation}\label{App:saddle:compare}
%%%%%

We now return to the original problem at hand, namely the saddle point evaluation of the $\mathcal N=4$ SYM index in the Cardy-like limit.  As we have shown in (\ref{eq:index:2:saddle:finite:2}), the effective action reduces to that of 3D SU($N$) Chern-Simons theory.  As a result, we may simply apply the saddle point solution of the latter theory to the $\mathcal N=4$ SYM index.  However, it is instructive to see how this works in practice.  To do so, {\it  we have numerically solved the saddle point equation arising from the effective action in (\ref{eq:Seff:1})}.  This was performed using FindRoot in Mathematica, where the elliptic gamma function was approximated by truncating its product representation, (\ref{def:gamma}).

We find that numerical solutions to the saddle point equation for the $\mathcal N=4$ SYM index are sensitive to the initial trial configuration for the eigenvalues.  Based on large-$N$ investigations of the index that suggest the eigenvalues are distributed along the `thermal' circle \cite{Cabo-Bizet:2019eaf,ArabiArdehali:2019orz}, it is natural to start with an initial configuration distributed uniformly along the interval $(-\tau/2,\tau/2)$.  This starting point, however, converges to the sub-leading saddle point solution corresponding to that discussed in section~\ref{App:saddle:sublead}.  In order to find the dominant saddle point corresponding to section~\ref{App:saddle:lead}, we have to instead start with an initial configuration mirroring (\ref{eq:rholeadim}) of the 3D Chern-Simons theory.  Here the initial eigenvalues go twice around the `thermal' circle, and are distributed non-uniformly in the interval $(-\tau,\tau)$.

%%%%%
\begin{figure}[t]
	\centering
	\leavevmode\raise6.5mm\hbox{\includegraphics[width=.48\textwidth]{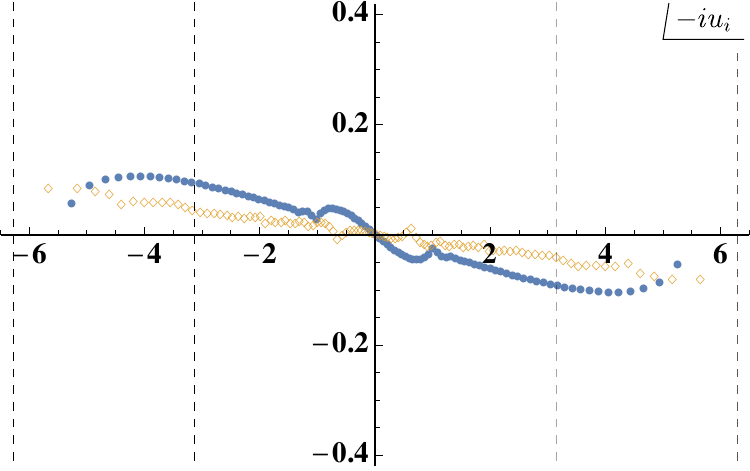}}
	\includegraphics[width=.48\textwidth]{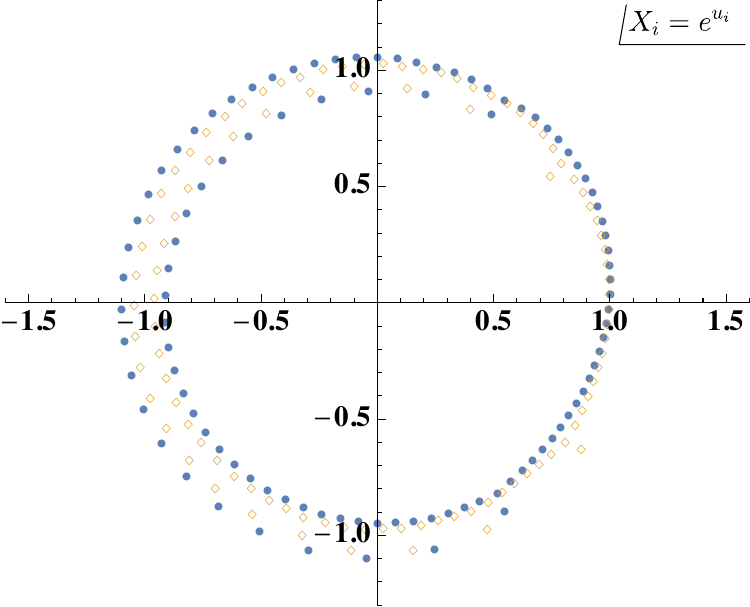}
	\caption{Comparison between the $\mathcal N=4$ SYM (blue dots) and 3D Chern-Simons (orange diamonds) solutions for the dominant saddle point.  Here we have taken $N=100$ along with $\tau=ie^{i\pi/6}$ and $\Delta_a=(2/3,2/3,2/3+2\tau)$, which maps to $t=2\pi i$ in the Chern-Simons theory.  As seen in the figure on the right, the exponentiated eigenvalues go twice around the circle.  The 3D Chern-Simons eigenvalues $u_i$ are given as in (\ref{eq:3DCS}), while the $\mathcal N=4$ SYM eigenvalues $\tilde u_i$ are mapped according to $u_i=2\pi i\tilde u_i/\tau$.
\label{fig:dom-saddles}}
\end{figure}
%%%%%

\begin{figure}[t]
	\centering
	\leavevmode\raise4.5mm\hbox{\includegraphics[width=.48\textwidth]{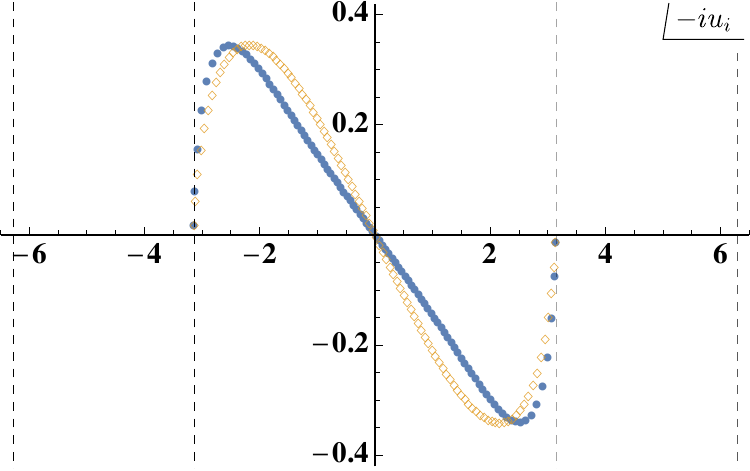}}
	\includegraphics[width=.48\textwidth]{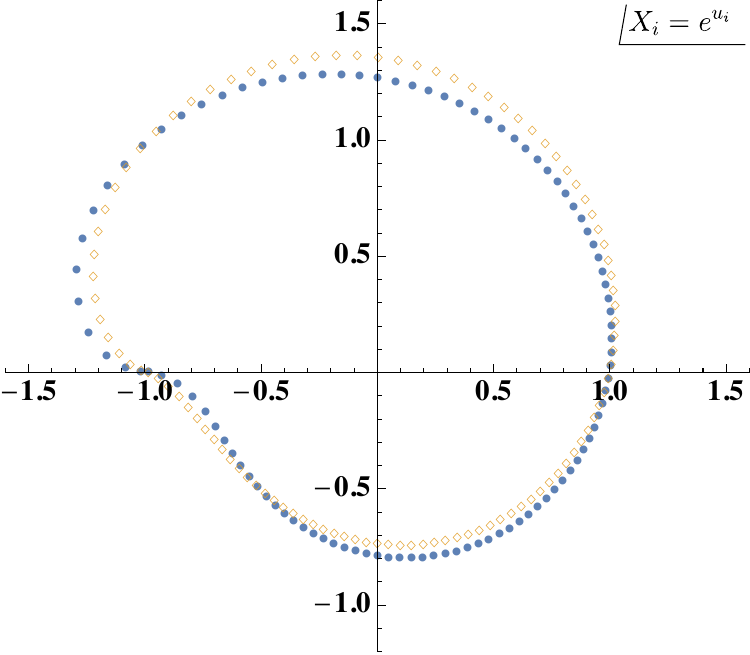}
	\caption{Comparison between the $\mathcal N=4$ SYM (blue dots) and 3D Chern-Simons (orange diamonds) solutions for the sub-leading saddle point.  The parameters are the same as in Figure~\ref{fig:dom-saddles}, but for the sub-leading saddle the exponentiated eigenvalues go only once around the (distorted) circle.
	\label{fig:sub-saddles}}
\end{figure}

As an example, we compare the numerical solution to the $\mathcal N=4$ SYM saddle point equations with those from the 3D Chern-Simons theory in Figure~\ref{fig:dom-saddles} for the leading saddle and Figure~\ref{fig:sub-saddles} for the sub-dominant saddle.  For $\mathcal N=4$ SYM, we take $\tau=ie^{i\pi/6}$ and chemical potentials such that $\eta=1$, so that $t=2\pi i$ in the Chern-Simons theory.  Since $|\tau|=1$, the numerical results are not taken in the Cardy-like limit.  Nevertheless, the similarity of the full SYM solution with that of the corresponding Chern-Simons theory is apparent.  We have observed numerically that the sub-leading saddle point solution becomes indistinguishable from that of the Chern-Simons theory in the Cardy-like limit.  However, the leading order saddle is more sensitive to $1/N$ effects arising from the repulsion between eigenvalues on the inner and outer circles of Figure~\ref{fig:dom-saddles}.  In any case, the distinction between $\mathcal N=4$ SYM and 3D Chern-Simons solutions is small compared to the difference between the dominant and sub-leading saddles which is clearly evident when comparing Figure~\ref{fig:dom-saddles} with Figure~\ref{fig:sub-saddles}.

%%%%%
%%%%%

%%%%%
\section{The \texorpdfstring{$S^3$}{S3} partition function of \texorpdfstring{SU($N$)}{SU(N)} Chern-Simons theory}
\label{App:CS:partition:function}
%%%%%
Here we compute the $S^3$ partition function of SU($N$) Chern-Simons theory, namely
\begin{equation}
	Z_{\text{SU}(N)}^{CS}=\fft{1}{N!}\int_{-\infty}^\infty(\prod_{j=1}^{N-1}d\lambda_j)\,e^{-ik\pi\sum_{j=1}^N\lambda_j^2}\prod_{i\neq j}2\sinh\pi\lambda_{ij}
\end{equation}
with the constraint $\sum_{j=1}^N\lambda_j=0$, where $k=-\eta N~(\eta=\pm1)$. 

Recall that the $S^3$ partition function of U($N$) Chern-Simons theory is given in Appendix B of \cite{Kapustin:2009kz} as
\begin{equation}
\begin{split}
	Z_\text{U(N)}^{CS}&=\fft{1}{N!}\int_{-\infty}^\infty(\prod_{j=1}^{N}d\lambda_j)\,e^{-ik\pi\sum_{j=1}^N\lambda_j^2}\prod_{i\neq j}2\sinh\pi\lambda_{ij}\\
	&=\fft{(-1)^{\fft{N(N-1)}{2}}e^{-\fft{\pi iN(N-1)}{4}}e^{-\fft{\pi i}{6k}N(N^2-1)}}{(ik)^{N/2}}\prod_{m=1}^{N-1}\left(2\sin\fft{\pi m}{k}\right)^{N-m}.\label{eq:Z:U(N)}
\end{split}
\end{equation}
Under the change of variables $\lambda_\mu\to\lambda_\mu+\sum_{j=1}^N\lambda_j~(\mu=1,\cdots,N-1)$, whose Jacobian is given as
\begin{equation}
	\prod_{j=1}^Nd\lambda_j\quad\to\quad N\prod_{j=1}^Nd\lambda_j,
\end{equation}
the U(N) partition function (\ref{eq:Z:U(N)}) can be rewritten as
\begin{equation}
\begin{split}
	Z_\text{U(N)}^{CS}&=\fft{1}{(N-1)!}\int_{-\infty}^\infty(\prod_{j=1}^Nd\lambda_j)\,e^{-ik\pi\sum_{\mu=1}^{N-1}(\lambda_\mu+\sum_{j=1}^N\lambda_j)^2-ik\pi\lambda_N^2}\prod_{\mu\neq\nu}^{N-1}2\sinh\pi\lambda_{\mu\nu}\\
	&\kern5em\times\prod_{\mu=1}^{N-1}2\sinh\pi(\lambda_\mu+\Sigma_j\lambda_j-\lambda_N)2\sinh\pi(\lambda_N-\lambda_\mu-\Sigma_j\lambda_j)\\
	&=\fft{1}{(N-1)!}\int_{-\infty}^\infty(\prod_{\mu=1}^{N-1}d\lambda_\mu)\,\prod_{i\neq j}^N2\sinh\pi\lambda_{ij}\Big|_{\lambda_N=-\sum_{\mu=1}^{N-1}\lambda_\mu}\\
	&\kern5em\times\int_{-\infty}^\infty d\lambda_N\,e^{-ik\pi N(\lambda_N+\sum_{\mu=1}^{N-1}\lambda_\mu)^2}e^{-ik\pi(\sum_{\mu=1}^{N-1}\lambda_\mu^2+(\sum_{\mu=1}^{N-1}\lambda_\mu)^2)}\\
	&=\left(\fft{N}{ik}\right)^\fft12Z_{\text{SU}(N)}^{CS}.\label{eq:Z:U(N):SU(N)}
\end{split}
\end{equation}
For $k=-\eta N$ with $\eta=\pm1$, substituting the identity
\begin{equation}
	\prod_{m=1}^{N-1}\left(2\sin\fft{\pi m}{N}\right)^{N-m}=N^{N/2}
\end{equation}
into the U($N$) partition function (\ref{eq:Z:U(N)}) and using the relation (\ref{eq:Z:U(N):SU(N)}), we have
\begin{equation}
\begin{split}
	Z_{\text{SU}(N)}^{CS}&=e^{-\fft{\pi i\eta}{4}}\fft{e^{\fft{\pi iN(N-1)}{2}}e^{-\fft{\pi iN(N-1)}{4}}e^{\fft{\pi i\eta}{6}(N^2-1)}}{e^{-\fft{\pi i\eta N}{4}}}e^{\fft{\pi i(1+\eta)N(N-1)}{4}}\\
	&=\exp[\fft{\pi iN(N-1)}{2}+\fft{5\pi i\eta(N^2-1)}{12}].\label{eq:Z:SU(N)}
\end{split}
\end{equation}
%

%%%%%
\renewcommand{\emph}[1]{\textit{#1}}
\bibliographystyle{JHEP}
\bibliography{BHLocalization}
\end{document}